\documentclass[iop,apj,tighten,twocolappendix,numberedappendix]{emulateapj}
\usepackage{natbib}
\usepackage{amsmath,amssymb}
\usepackage{apjfonts}
\usepackage{xspace}
\usepackage[maxfloats=50]{morefloats}
\usepackage[plainpages=false, colorlinks=true, anchorcolor=blue, linkcolor=blue, citecolor=blue, bookmarks=false]{hyperref}
\shorttitle{NANOGrav Nine-year Data Set}
\shortauthors{Z.~Arzoumanian et al.}

\def\be{\begin{equation}}
\def\ee{\end{equation}}
\newcommand{\lp}{\left(}
\newcommand{\rp}{\right)}
\newcommand{\bb}{\begin{bmatrix}}
\newcommand{\eb}{\end{bmatrix}}
\newcommand{\msJ}{\mathcal{J}}

\newcommand{\tempo}{{\sc tempo}\xspace}
\newcommand{\tempotwo}{{\sc tempo2}\xspace}
\newcommand{\psrchive}{{\sc psrchive}\xspace}

\begin{document}

\title{The NANOGrav Nine-year Data Set: Observations, Arrival Time
Measurements, and Analysis of 37 Millisecond Pulsars}

\author{%
The NANOGrav Collaboration:
Zaven~Arzoumanian\altaffilmark{1},
Adam~Brazier\altaffilmark{2},
Sarah~Burke-Spolaor\altaffilmark{3},
Sydney~Chamberlin\altaffilmark{4},
Shami~Chatterjee\altaffilmark{2},
Brian~Christy\altaffilmark{5,6},
James~M.~Cordes\altaffilmark{2},
Neil~Cornish\altaffilmark{7},
Kathryn~Crowter\altaffilmark{8},
Paul~B.~Demorest\altaffilmark{3,$\dagger$},
Timothy~Dolch\altaffilmark{2,9},
Justin~A.~Ellis\altaffilmark{10,26},
Robert~D.~Ferdman\altaffilmark{11},
Emmanuel~Fonseca\altaffilmark{8},
Nathan~Garver-Daniels\altaffilmark{12},
Marjorie~E.~Gonzalez\altaffilmark{8,13},
Fredrick~A.~Jenet\altaffilmark{14},
Glenn~Jones\altaffilmark{15},
Megan~L.~Jones\altaffilmark{12},
Victoria~M.~Kaspi\altaffilmark{11},
Michael~Koop\altaffilmark{16},
T.~Joseph~W.~Lazio\altaffilmark{10},
Michael~T.~Lam\altaffilmark{2},
Lina~Levin\altaffilmark{12,17},
Andrea~N.~Lommen\altaffilmark{5},
Duncan~R.~Lorimer\altaffilmark{12},
Jing~Luo\altaffilmark{14},
Ryan~S.~Lynch\altaffilmark{18},
Dustin~Madison\altaffilmark{2},
Maura~A.~McLaughlin\altaffilmark{12},
Sean~T.~McWilliams\altaffilmark{12},
David~J.~Nice\altaffilmark{19},
Nipuni~Palliyaguru\altaffilmark{12},
Timothy~T.~Pennucci\altaffilmark{20},
Scott~M.~Ransom\altaffilmark{21},
Xavier~Siemens\altaffilmark{4},
Ingrid~H.~Stairs\altaffilmark{8},
Daniel~R.~Stinebring\altaffilmark{22},
Kevin~Stovall\altaffilmark{23},
Joseph~K.~Swiggum\altaffilmark{12},
Michele~Vallisneri\altaffilmark{10},
Rutger~van~Haasteren\altaffilmark{10,26},
Yan~Wang\altaffilmark{24,14},
and
Weiwei~Zhu\altaffilmark{8,25}
}

\altaffiltext{$\dagger$}{pdemores@nrao.edu}
\altaffiltext{1}{Center for Research and Exploration in Space Science and Technology and X-Ray Astrophysics Laboratory, NASA Goddard Space Flight Center, Code 662, Greenbelt, MD 20771, USA}
\altaffiltext{2}{Department of Astronomy, Cornell University, Ithaca, NY 14853, USA}
\altaffiltext{3}{National Radio Astronomy Observatory, P.~O.~Box O, Socorro, NM, 87801, USA}
\altaffiltext{4}{Center for Gravitation, Cosmology and Astrophysics, Department of Physics, University of Wisconsin-Milwaukee, P.O. Box 413, Milwaukee, WI 53201, USA}
\altaffiltext{5}{Department of Physics and Astronomy, Franklin and Marshall College, P.O. Box 3003, Lancaster, PA 17604, USA}
\altaffiltext{6}{Department of Mathematics, Computer Science, and Physics, Notre Dame of Maryland University 4701 N Charles St, Baltimore, MD 21210, USA}
\altaffiltext{7}{Department of Physics, Montana State University, Bozeman, MT, 59717, USA}
\altaffiltext{8}{Department of Physics and Astronomy, University of British Columbia, 6224 Agricultural Road, Vancouver, BC V6T 1Z1, Canada}
\altaffiltext{9}{Department of Physics, Hillsdale College, 33 E. College Street, Hillsdale, Michigan 49242, USA}
\altaffiltext{10}{Jet Propulsion Laboratory, California Institute of Technology, 4800 Oak Grove Dr. Pasadena CA, 91109, USA}
\altaffiltext{11}{Department of Physics, McGill University, 3600 rue Universite, Montreal, QC H3A 2T8, Canada}
\altaffiltext{12}{Department of Physics, West Virginia University, P.O.  Box 6315, Morgantown, WV 26505, USA}

\begin{abstract}

We present high-precision timing observations spanning up to nine years
for 37 millisecond pulsars monitored with the Green Bank and Arecibo
radio telescopes as part of the North American Nanohertz Observatory for
Gravitational Waves (NANOGrav) project.  We describe the observational
and instrumental setups used to collect the data, and methodology
applied for calculating pulse times of arrival; these include novel
methods for measuring instrumental offsets and characterizing low
signal-to-noise ratio timing results.  The time of arrival data are fit
to a physical timing model for each source, including terms that
characterize time-variable dispersion measure and frequency-dependent
pulse shape evolution.  In conjunction with the timing model fit, we
have performed a Bayesian analysis of a parameterized timing noise model
for each source, and detect evidence for excess low-frequency, or
``red,'' timing noise in 10 of the pulsars.  For 5 of these cases this
is likely due to interstellar medium propagation effects rather than
intrisic spin variations.  Subsequent papers in this series will present
further analysis of this data set aimed at detecting or limiting the
presence of nanohertz-frequency gravitational wave signals.

\end{abstract}

\section{Introduction}
\label{sec:intro}

The era of gravitational-wave astronomy is expected to begin within the
next decade. It will be heralded by the first direct detection of
gravitational waves as perturbations in the spacetime metric due to
acceleration of massive objects. Anticipated gravitational wave sources
include merging systems of supermassive black hole binaries (SMBHBs) and
neutron-star binaries, as well as inflation-era relics
\citep[e.g.,][]{gri05} and cosmic strings \citep[e.g.,][]{vs94}.
Several major experiments are underway in order to detect and
characterize gravitational waves.  One type of experiment is a pulsar
timing array (PTA), in which a collection of radio pulsars is monitored,
providing sensitivity to gravitational radiation in the nanohertz region
of the spectrum \citep{haa+10}.

\footnotetext[13]{Department of Nuclear Medicine, Vancouver Coastal Health Authority, Vancouver, BC V5Z 1M9, Canada}
\footnotetext[14]{Center for Gravitational Wave Astronomy, University of Texas at Brownsville, Brownsville, TX 78520, USA}
\footnotetext[15]{Department of Physics, Columbia University, 550 W. 120th St. New York, NY 10027, USA}
\footnotetext[16]{Department of Astronomy and Astrophysics, Pennsylvania State University, University Park, PA 16802, USA}
\footnotetext[17]{Jodrell Bank Centre for Astrophysics, School of Physics and Astronomy, The University of Manchester, Manchester M13 9PL, UK}
\footnotetext[18]{National Radio Astronomy Observatory, P.~O.~Box 2, Green Bank, WV, 24944, USA}
\footnotetext[19]{Department of Physics, Lafayette College, Easton, PA 18042, USA}
\footnotetext[20]{University of Virginia, Department of Astronomy, P.~O.~Box 400325 Charlottesville, VA 22904-4325, USA}
\footnotetext[21]{National Radio Astronomy Observatory, 520 Edgemont Road, Charlottesville, VA 22903, USA}
\footnotetext[22]{Department of Physics and Astronomy, Oberlin College, Oberlin, OH 44074, USA}
\footnotetext[23]{Department of Physics and Astronomy, University of New Mexico, Albuquerque, NM, 87131, USA}
\footnotetext[24]{School of Physics, Huazhong University of Science and Technology, 1037 Luoyu Road, Wuhan, Hubei Province 430074, China}
\footnotetext[25]{Max-Planck-Institut f{\" u}r Radioastronomie, Auf dem H{\" u}gel 69, D- 53121, Bonn, Germany}
\footnotetext[26]{Einstein Fellow}
\setcounter{footnote}{0}

Millisecond pulsar rotation is very stable, and pulsars in relativistic
binary systems have already been used to place the most stringent
experimental constraints on strong-field gravity so far. The orbital
decay observed in such binary systems serves as compelling indirect
evidence of gravitational radiation, as the observed orbital decay rates
match the expected rates due to loss of energy and angular momentum via
emission of gravitational waves \citep{fst14,ksm+06,wnt10}.
\citet{saz78} and \citet{det79} were the first to suggest that pulsar
signals can be used to directly measure gravitational waves,
particularly those produced by SMBHB mergers. \citet{hd83} extended this
view and developed the notion that gravitational waves produce pulse
time-of-arrival (TOA) shifts that are correlated amongst a set of
pulsars.  In principle, this allows the gravitational wave signal to be
unambiguously separated from other astrophysical phenomena affecting
measured TOAs -- these would be specific to each pulsar, thus
uncorrelated between different objects.

The North American Nanohertz Observatory for Gravitational Waves
(NANOGrav)\footnote{\url{http://www.nanograv.org}} is one of several PTA
programs across the globe that collectively form the International
Pulsar Timing Array \citep[IPTA;][]{haa+10}. These collaborations
regularly monitor the most stable members of the millisecond pulsar
(MSP) population distributed across the sky in order to achieve the
highest sensitivity possible towards gravitational wave detection.
While no direct detection has been made so far, individual PTA programs
have yielded upper limits on the amplitude $A_{\rm gw}$ of the characteristic
gravitational-wave strain $h_c$ due to a stochastic background in the
nanohertz regime \citep{vlj+11,dfg+13,src+13,ltm+15}, where  
\begin{equation}
  h_c(f) = A_{\rm gw}\bigg(\frac{f}{1\textrm{ yr}^{-1}}\bigg)^{\alpha}
\end{equation}
and $f$ is the observed gravitational wave frequency.   A number of
predictions for the expected strength of the SMBHB gravitational wave
background have been made \citep[e.g.,][]{jb03,svc08,ses13,mop14}, with
$\alpha=-2/3$ and expected values for $A_{\rm gw}$ ranging from
$\sim5\times10^{-16}$ to $4\times10^{-15}$.  Currently, the best
published experimental limit is $A_{\rm gw}<2.7\times10^{-15}$
\citep{src+13}.  In addition to measuring the gravitational wave
background, PTA measurements can be used to attempt to measure other
gravitational wave signals, such as periodic gravitational waves from
individual sources \citep{svv09,abb+14} and permanent deformations in
spacetime, referred to as gravitational-wave ``memory''
\citep[e.g.,][]{mcc14,whc+15}.
PTA data can also be used for ancillary studies of pulsars, binary
systems, and the interstellar medium.

In this study, we extend the data set analyzed by \citet{dfg+13} and
present high-precision timing observations of the updated NANOGrav PTA.
The data comprise measurements of 37 MSPs and span nine years of
observation. In Section~\ref{sec:obs}, we provide information regarding
the telescopes, methods and pulsar backends used for data collection. In
Section~\ref{sec:toa}, we describe the general procedure for data
reduction, flux and polarization calibration, TOA determination and data
excision. In Section~\ref{sec:timing}, we outline the strategy used for
generating updated timing models for each pulsar and discuss the results
of the model fitting. In Section~\ref{sec:noise}, we describe the
models used for characterizing noise in our timing data. In
Section~\ref{sec:conclusions}, we summarize the results and implications
of this work. Raw and processed data products presented here are
publicly available as of the date this work is published
(\S\ref{sec:timing}).

\section{Observations}
\label{sec:obs}

\begin{deluxetable*}{cccccccc}
\tablecaption{Observing frequencies and bandwidths}
\tablehead{
                   & \multicolumn{3}{c}{ASP/GASP} && \multicolumn{3}{c}{PUPPI/GUPPI} \\[2pt]
\cline{2-4}\cline{6-8}
\rule{0pt}{10pt}
Telescope      &                             & Frequency                         &  Usable                    &&                            & Frequency                         &   Usable                 \\
Receiver       &  Data Span\tablenotemark{a} &        Range\tablenotemark{b} & Bandwidth\tablenotemark{c} && Data Span\tablenotemark{a} &        Range\tablenotemark{b} & Bandwidth\tablenotemark{c}  \\
               &            &     (MHz)        &  (MHz)    &&           &    (MHz)         &   (MHz)    
}
\startdata
\multicolumn{8}{c}{Arecibo} \\[2pt]
\hline
\rule[0pt]{0pt}{11pt}%
   327   &  2005.0-2012.0 & $315-339$   &  34    && 2012.2-2013.8 &  $302-352$        &  \phn 50    \\
   430   &  2005.0-2012.3 & $422-442$   &  20    && 2012.2-2013.8 &  $421-445$        &  \phn 24    \\
  L-wide &  2004.9-2012.3 & $1380-1444$ &  64    && 2012.2-2013.8 & $1147-1765$           &      603    \\ 
  S-wide &  2004.9-2012.6 & $2316-2380$ &  64    && 2012.2-2013.8 & \tablenotemark{\phantom{d}}$1700-2404$\tablenotemark{d}  &   460  \\ 
\hline
\multicolumn{8}{c}{\rule[0pt]{0pt}{11pt}GBT} \\[2pt]
\hline
\rule[0pt]{0pt}{11pt}%
Rcvr\_800 &  2004.6-2011.0 & $822-866$   &  64    && 2010.2-2013.8 &  $722-919$       &   186  \\
Rcvr1\_2 &  2004.6-2010.8 & $1386-1434$ &  48    && 2010.2-2013.8 &  $1151-1885$      &   642  
\enddata
\tablenotetext{a}{Dates of instrument use.  Observation dates of individual pulsars vary; see Figure~\ref{fig:obs:epochs}.}
\tablenotetext{b}{Most common values; some observations differed. Some frequencies unusable due to radio frequency interference.}
\tablenotetext{c}{Nominal values after excluding narrow subbands with radio frequency interference.}
\tablenotetext{d}{Non-contiguous usable bands at $1700-1880$ and $2050-2404$~MHz.}
\label{tab:obs:observing_systems}
\end{deluxetable*}

This paper reports on observations of an array of millisecond pulsars
made over a 9-year span from 2004 to 2013.  Pulsars were chosen for this
project based on expectations of high time-of-arrival precision,
reliable detection across a wide range of frequencies, and lack of
unpredictable timing fluctuations from astrophysical effects (for
example, no eclipsing binary pulsars have been included).  The array
initially included 15 pulsars, and it grew to 37 pulsars over the course
of the project.  The growth came from the discovery of new millisecond
pulsars and the advent of wide-band data acquisition systems, which
allowed observation of some sources previously deemed too weak or
unreliable.  The first five years of data on 17 of the pulsars were
previously reported by \citet{dfg+13}, however all data have been
reprocessed as described in following sections.

Pulsars with declinations in the range $0^\circ<\delta<39^\circ$ were observed
with the 305\,m William E.\ Gordon Telescope of the Arecibo Observatory.
Sources outside this declination range were observed with the 100\,m Robert
C.\ Byrd Green Bank Telescope (GBT) of the National Radio Astronomy Observatory.
Two sources were observed at both telescopes, PSRs J1713+0747 and B1937+21.

\begin{figure*}
\begin{center}
\includegraphics[width=6.0in]{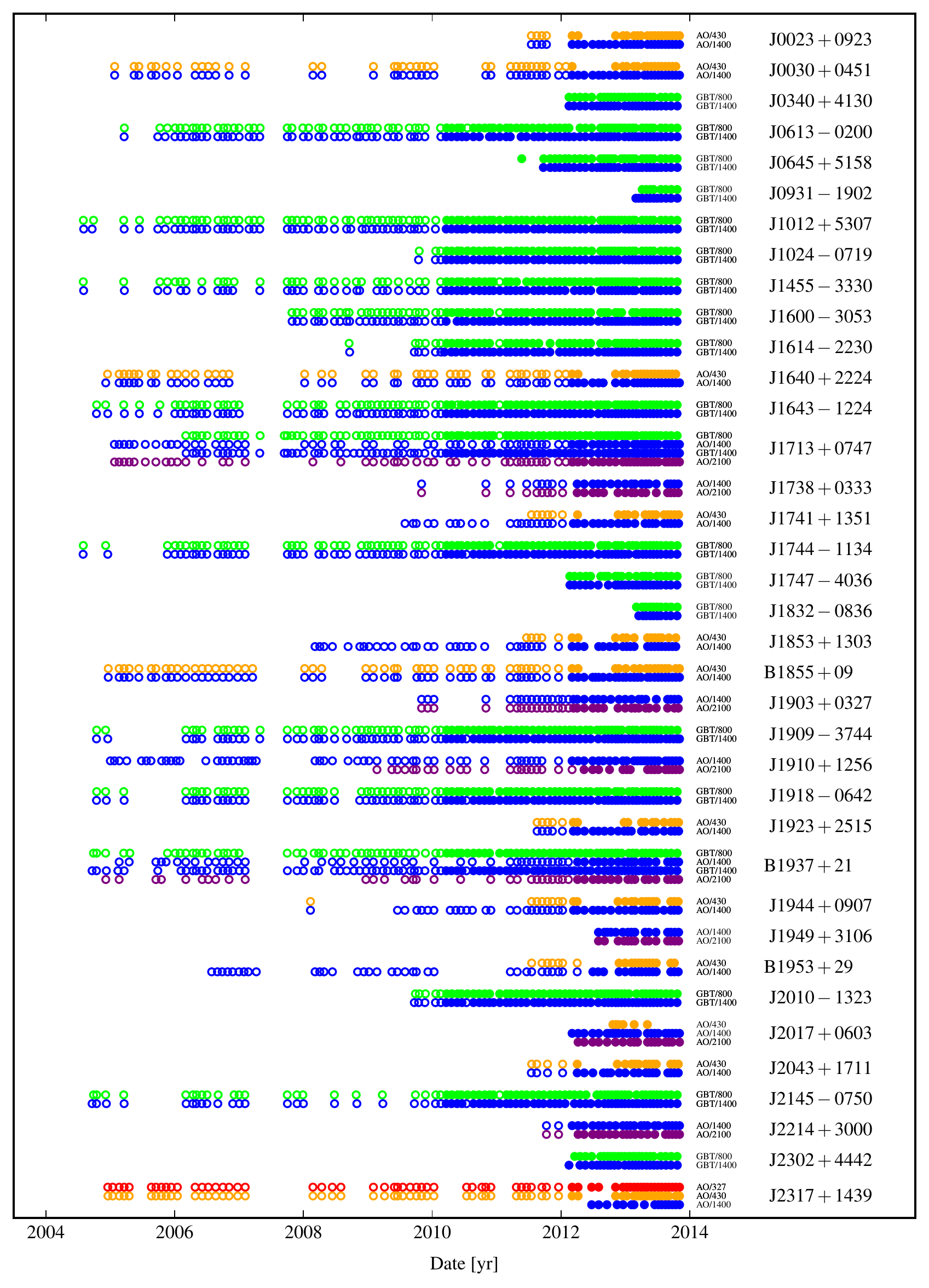}
\caption{\label{fig:obs:epochs}Epochs of all observations in the data
set.  Marker type indicates data acquisition system: open circles are
ASP or GASP; closed circles are PUPPI or GUPPI.  Colors indicate radio
frequency band, at either telescope: red is 327~MHz; orange is 430~MHz;
green is 820~MHz; blue is 1.4~GHz; and purple is 2.1~GHz.}
\end{center}
\end{figure*}

Table~\ref{tab:obs:observing_systems} summarizes the radio frequencies and
data acquisition systems used for this project; these are discussed in 
more detail below.
Observation time spans for individual sources are
listed in Table~\ref{tab:psr_params}, and observation dates are displayed in 
Figure~\ref{fig:obs:epochs}.

Sources were observed at approximately monthly intervals through most
of this program, with denser observations, every three weeks, in 2013.
Scheduling of individual observing epochs
varied depending on telescope operational considerations and
sometimes deviated from regular cadences.
Observations at both telescopes were interrupted in 2007 due to telescope
painting (Arecibo) and azimuth track refurbishment (GBT).

Each pulsar was observed using radio receivers at two separate
frequencies throughout this program in order to measure and remove
frequency-dependent dispersive effects.  Exceptions were Arecibo
observations of five sources that were made at a single frequency before
2009 or 2011 (depending on the source), and certain Arecibo observations
during 2012, when technical issues impeded use of the 430 MHz receiver.

At Arecibo, observations of a given pulsar using two receivers were made
in immediate succession within $\sim$1 hour.  At the GBT, observations
using two receivers were typically separated by a few days due to the
need for a physical receiver change at that telescope.  Observations
without complementary data from the other receiver taken within 14~days
were excluded from the data set.  Exceptions to this rule were made for
early Arecibo single-receiver observations of several sources mentioned
above, and for wide-band 1400 MHz data, in which the wide frequency
range of one receiver partially made up for the lack of data from a
second receiver.  The typical observation duration was about 25 minutes,
with some variations over the course of the program.

All receivers are sensitive to dual linear polarizations,  with the
exception of the Arecibo 430 MHz receiver, which measures dual circular
polarizations.  Polarization cross-products were recorded so that full
Stokes parameters could be recovered.  However, for the present work, we
only use total-intensity measurements, obtained by summing the
calibrated signals from pairs of orthogonal polarizations.

Two sets of data acquisition systems were used.  Early observations
(through 2012.3 at Arecibo and through 2011.0 at Green Bank) were
recorded by the nearly-identical Astronomical Signal Processor (ASP) and
Green Bank Astronomical Signal Processor (GASP) data acquisition systems
\citep{dem07}.  Later observations (beginning 2012.2 at Arecibo and
2010.2 at Green Bank) were recorded using the nearly-identical Puerto
Rican Ultimate Pulsar Processing Instrument (PUPPI) and the Green Bank
Ultimate Pulsar Processing Instrument \citep[GUPPI;][]{drd+08,fdr10}.  
Each of these systems digitized incoming baseband radio telescope
voltage signals at the appropriate Nyquist rate, channelized them into
subbands, performed real-time coherent dedispersion, calculated self-
and cross-products to record full polarization information, and folded
the data with the dynamically-calculated pulsar period using a
pre-computed ephemeris.  The end product of each instrument was folded
pulse profiles (2048 bins) with self- and cross-products recorded over a
series of frequency channels and integrated over short subintervals over
the course of an observation.

The ASP and GASP systems processed up
to 64 MHz of bandwidth, recording in contiguous 4~MHz subbands.  Data were usually recorded in consecutive
60-second subintervals over the course of an observation.  

The PUPPI and GUPPI systems processed 100, 200, or 800 MHz of bandwidth, depending
on the mode of operation.  In each case, data were recorded in
contiguous subbands of width 1.5625~MHz.
Data were usually recorded in consecutive 10-second subintervals, with
1-second subintervals used for some Arecibo 1400 MHz observations to aid interference
excision.  

In some cases, particularly with PUPPI and GUPPI, bandwidth was limited
by telescope receivers rather than the data acquisition instrument (Table~\ref{tab:obs:observing_systems}).
In post-processing, narrowband radio frequency noise was removed and adjacent subbands were summed before 
arrival times were calculated (see \S\ref{sec:toa:calib}).

Each pulsar observation was preceded or followed by measurement of a pulsed noise signal using a
setup identical to the pulsar observation in order to calibrate the signal levels.  The pulsed noise signals
themselves were calibrated in on- and off-source observations of unpolarized continuum radio
sources on a monthly basis.

For the timing analysis in this paper, only the polarization self-products were used.
Data were summed in time and polarization (\S\ref{sec:toa:calib}).
Simultaneous observations between ASP and PUPPI at Arecibo, and between GASP and GUPPI
at the GBT, were used to measure the time offset between these pairs of instruments 
(Appendix~\ref{sec:offset}).

\section{Calibration and TOA Determination}
\label{sec:toa}

The results of the observations described in Section~\ref{sec:obs} are
``raw'' pulse profiles.  This section describes the procedures employed
to turn the raw profiles into usable pulse times of arrival (TOAs).
These included: RFI excision, polarization and flux calibration,
additional averaging in time and frequency, derivation of template
profiles, and finally TOA determination.  All data processing operations
described in this section were carried out using the \psrchive software
package
\citep{hvm04,vdo12}.\footnote{\url{http://psrchive.sourceforge.net};
\psrchive version \texttt{2015-01-15 b4826eb} was used for this work.}
These were organized into a processing pipeline via a set of scripts
that are available
online.\footnote{\url{http://github.com/demorest/nanopipe}} Overall, the
calibration and processing strategy used here is similar to
\cite{dfg+13}, and is based on standard methods for pulsar data
analysis.

\subsection{Calibration and averaging}
\label{sec:toa:calib}

The polarization calibration procedure used for these data sets was
identical to that described by \cite{dfg+13}:  A locally-generated
broadband noise source is pulsed at 25~Hz, split into two copies, and
coupled in to the two polarization signal paths.  Before each pulsar
observation, a short ($\sim$1 minute) observation of the pulsed noise
signal is recorded by the backend systems.  This correlated noise source
observation is used to calibrate the two leading polarization terms:
differential gain and phase between the two hands of polarization.  The
noise source power is not assumed to be equal in each hand -- its power
in each polarization is measured separately at each observing epoch by
observing the noise source while the telescope is pointed on and off a
bright, unpolarized quasar (B1442$+$101 at Green Bank; J1413$+$1509 at
Arecibo).  For purposes of this paper, we used this calibration to
balance the gains of orthogonal polarizations before summing to produce
total intensity profiles used for pulse timing.  While we have not
solved for a complete polarization calibration solution here, the
calibration data can also be used to create full-polarimetry profiles
and to flux-calibrate the pulse profiles using the known flux densities
of the quasars used as calibration sources.

Following calibration, excision of data corrupted by radio frequency
interference (RFI) was performed in two steps.  First, a set of
consistently bad frequency ranges for each telescope receiver was
determined manually and was removed from all data sets.  The interfering
signals primarily responsible for these cuts are satellite transmissions
near $\sim$1.6~GHz and radar signals near $\sim$1.2~GHz.  This step
resulted in removal of up to 15\% of the full bandwidth recorded by
GUPPI and PUPPI; the narrow-band ASP and GASP instruments were tuned to
avoid these strong signals.  The remaining usable bandwidth for each
receiver and data acquisition system is listed in
Table~\ref{tab:obs:observing_systems}.  Following this initial cut,
remaining RFI was removed via the median filter algorithm in \psrchive.
In each 20-channel wide frequency window, the median off-pulse variation
was computed, and any channels exceeding 4 times this value were
removed.  Finally, prior to the final averaging described below, all
profiles were normalized to have constant off-pulse variance.  This step
acted to down-weight any remaining corrupted data.

As the TOA-determination procedure described in the next section begins
to fail at very low signal-to-noise ratios (see Appendix~\ref{sec:snr}),
it is advantageous to average as much data as possible into each profile
before measuring a TOA.  The final time and frequency resolution that
should be used is ultimately limited by the need to resolve TOA shifts
as a function of time (for example, from orbital motion) or frequency
(from profile frequency evolution or variable dispersion measure).  We
chose to average profiles in time up to a maximum of 30 minutes or 2\%
of the pulsar's binary period, whichever is shorter.  Data from each
observing session were divided equally -- for example, a 40-minute
observation would be averaged into two 20-minute sections.  The shortest
averaging was for PSR~J0023+0923, which has an orbital period of
200~minutes (see $P_b$ column in Table~\ref{tab:psr_params}).  For
frequency averaging, we adopted a slightly different strategy for each
instrument.  For ASP and GASP, no frequency averaging was done, and the
final profile data remain at the instrumental resolution of 4~MHz.  For
GUPPI and PUPPI, the data were averaged to different final frequency
resolutions depending on which receiver system was in use: 1.5625~MHz for
327~MHz and 430~MHz data, 3.125~MHz for 0.82~GHz, and 12.5~MHz for all
frequencies above 1~GHz.   

To summarize, the profile data set for any given
observation consists of calibrated total-intensity profiles
collected simultaneously across many subbands (typically between 5 and 60)
divided into one or more subintervals (typically 20--30 minutes).

\subsection{Measuring times-of-arrival}
\label{sec:toa:toa}

We calculated a pulse time-of-arrival (TOA) from each averaged profile
resulting from the procedure described in the previous section.  Thus
any given observation results in a large number of TOAs, computed from
data collected simultaneously in different subbands.

We calculated TOAs using the Fourier-domain algorithm of \cite{tay92} as
implemented in the \psrchive program \verb+pat+.  This method determined
each TOA and its uncertainty via a least-squares fit for the pulse phase
shift between an observed total-intensity pulse profile and an ideal
template profile.  Template determination was done using the same
procedure as \cite{dfg+13}.  In short, for each pulsar and each
receiver, we made a signal-to-noise-weighted sum of all GUPPI/PUPPI
profile data.  We then de-noised these profiles via wavelet
decomposition and thresholding of the wavelet coefficients (as
implemented in the \psrchive program \verb+psrsmooth+).  The same
template profiles were used to calculate TOAs from both GUPPI/PUPPI data
and from GASP/ASP data.  All templates were aligned so that phase zero
occurs at the peak of the pulse profile.

The pairs of data acquisition systems used at each telescope (GASP and
GUPPI or ASP and PUPPI) had different signal path lengths and different
internal latencies, which led to systematic TOA offsets.  These must be
measured and removed in order to avoid corrupting the pulsar timing
results.  This has typically been done in the past using the pulsar TOA
measurements themselves -- a time offset between two systems is fit for,
either as a term in the overall timing model fit (see
Section~\ref{sec:timing}), or separately using a subset of
contemporaneous TOAs \citep[e.g.,][]{tw89}.  More recently,
\cite{mhb+13} applied a method where a locally-generated timing signal
was injected, measured and used to derive per-backend timing offsets.
For the present work, we developed a new method that analyzes the noise
in the pulsar profiles collected simultaneously with a given pair of
data acquisition systems.  In simultaneous data the noise is correlated,
and cross-correlating the pulse profile data from the two systems
provided much higher-precision offset measurements than could be made
from TOAs, where by design most of the noise was filtered out by the
template-matching process.  The results were offsets between GASP and
GUPPI (also ASP and PUPPI), with typical value $\sim$1~$\mu$s and
uncertainty $\sim$5~ns, that were applied directly to the TOAs.
Additional details are presented in Appendix~\ref{sec:offset}. 

\subsection{Editing time-of-arrival data sets}
\label{sec:toa:edit}

After all TOAs were generated as described above, several
cuts were made on the set of TOAs from each pulsar in order to arrive at
the final set of data used in the analysis described in the next
section.
\begin{enumerate}

  \item For observations where simultaneous data were recorded with both
  sets of instrumentation, the ASP/GASP TOAs were removed.

  \item In order to meaningfully determine a time-variable dispersion
  measure, data from observing epochs with low fractional bandwidth,
  $\nu_{\rm high}/\nu_{\rm low} < 1.1$, were removed.  In practice, this
  criterion removed TOAs for any pulsar that did not have, within any
  given 14-day window, either TOAs measured using two separate receivers
  or TOAs measured using one wide-band receiver with a wide-band data
  acquisition system.

  \item As TOA measurement uncertainties become both underestimated and
  significantly non-Gaussian at low signal-to-noise ratio,
  TOAs from profiles with signal-to-noise ratio less 
  than 8 were removed.  See Appendix~\ref{sec:snr} for further analysis
  of this cut.

  \item A small number of outlier TOAs were manually
  identified and removed during the timing analysis.  Typically these
  were due to low-S/N data that were just above the S/N cutoff 
  or to RFI not excised by the algorithms described above.  The
  TOAs removed by this process comprise $\sim$1\% of the full data set.

\end{enumerate}

All TOA data presented in this paper are publicly
available.\footnote{\url{http://data.nanograv.org}}  TOAs removed from
the analysis as described above are included as supplementary files
along with the main data set.  The TOAs are given in \tempotwo format,
with additional flags specifying relevant meta-data (e.g., backend,
receiver, profile template file, etc).  This data format can be read
using both \tempo\footnote{\url{http://tempo.sourceforge.net}; \tempo
version {\tt 2014-11-20 76b8375} was used for this work.} and
\tempotwo\footnote{\url{http://tempo2.sourceforge.net}; \tempotwo
version {\tt 2014.11.1} was used for this work.} pulsar timing analysis
software.  Clock correction data needed to reference the TOAs to the
TT(BIPM) timescale (see Section~\ref{sec:timing:models}) are distributed
along with the TOA data, in the standard formats used by \tempo and
\tempotwo.

\begin{table*}[tp]
\caption{\label{tab:psr_params} Basic pulsar parameters and TOA statistics}
\begin{center}
\begin{tabular}{c|rrrr|rl|rl|rl|rl|rl|r}
\hline
Source & $P$  & $dP/dt$      & DM           & $P_b$ 
         & \multicolumn{10}{|c|}{Median scaled TOA uncertainty$^a$ ($\mu$s) / Number of epochs} 
         & Span \\
       & (ms) & ($10^{-20}$) & (pc~cm$^{-3}$) & (d)  
         & \multicolumn{2}{|c|}{327~MHz}
         & \multicolumn{2}{|c|}{430~MHz}
         & \multicolumn{2}{|c|}{820~MHz}
         & \multicolumn{2}{|c|}{1.4~GHz}
         & \multicolumn{2}{|c|}{2.3~GHz} 
         & (yr) \\
\hline
J0023$+$0923 & 3.05 & 1.14 & 14.3 & 0.1& \multicolumn{2}{|c|}{-} & 0.151 & 22 & \multicolumn{2}{|c|}{-} & 0.179 & 28 & \multicolumn{2}{|c|}{-} & 2.3\\
J0030$+$0451 & 4.87 & 1.02 & 4.3 & -& \multicolumn{2}{|c|}{-} & 0.265 & 53 & \multicolumn{2}{|c|}{-} & 0.261 & 60 & \multicolumn{2}{|c|}{-} & 8.8\\
J0340$+$4130 & 3.30 & 0.71 & 49.6 & -& \multicolumn{2}{|c|}{-} & \multicolumn{2}{|c|}{-} & 0.782 & 25 & 1.595 & 27 & \multicolumn{2}{|c|}{-} & 1.7\\
J0613$-$0200 & 3.06 & 0.96 & 38.8 & 1.2& \multicolumn{2}{|c|}{-} & \multicolumn{2}{|c|}{-} & 0.116 & 88 & 0.450 & 90 & \multicolumn{2}{|c|}{-} & 8.6\\
J0645$+$5158 & 8.85 & 0.49 & 18.2 & -& \multicolumn{2}{|c|}{-} & \multicolumn{2}{|c|}{-} & 0.269 & 29 & 0.985 & 32 & \multicolumn{2}{|c|}{-} & 2.4\\
J0931$-$1902 & 4.64 & 0.41 & 41.5 & -& \multicolumn{2}{|c|}{-} & \multicolumn{2}{|c|}{-} & 0.804 & 8 & 1.554 & 11 & \multicolumn{2}{|c|}{-} & 0.6\\
J1012$+$5307 & 5.26 & 1.71 & 9.0 & 0.6& \multicolumn{2}{|c|}{-} & \multicolumn{2}{|c|}{-} & 0.377 & 96 & 0.518 & 99 & \multicolumn{2}{|c|}{-} & 9.2\\
J1024$-$0719 & 5.16 & 1.86 & 6.5 & -& \multicolumn{2}{|c|}{-} & \multicolumn{2}{|c|}{-} & 0.582 & 50 & 0.846 & 53 & \multicolumn{2}{|c|}{-} & 4.0\\
J1455$-$3330 & 7.99 & 2.43 & 13.6 & 76.2& \multicolumn{2}{|c|}{-} & \multicolumn{2}{|c|}{-} & 0.868 & 82 & 1.766 & 80 & \multicolumn{2}{|c|}{-} & 9.2\\
J1600$-$3053 & 3.60 & 0.95 & 52.3 & 14.3& \multicolumn{2}{|c|}{-} & \multicolumn{2}{|c|}{-} & 0.267 & 71 & 0.202 & 77 & \multicolumn{2}{|c|}{-} & 6.0\\
J1614$-$2230 & 3.15 & 0.96 & 34.5 & 8.7& \multicolumn{2}{|c|}{-} & \multicolumn{2}{|c|}{-} & 0.336 & 51 & 0.424 & 66 & \multicolumn{2}{|c|}{-} & 5.1\\
J1640$+$2224 & 3.16 & 0.28 & 18.5 & 175.5& \multicolumn{2}{|c|}{-} & 0.076 & 61 & \multicolumn{2}{|c|}{-} & 0.082 & 67 & \multicolumn{2}{|c|}{-} & 8.9\\
J1643$-$1224 & 4.62 & 1.85 & 62.4 & 147.0& \multicolumn{2}{|c|}{-} & \multicolumn{2}{|c|}{-} & 0.301 & 93 & 0.481 & 93 & \multicolumn{2}{|c|}{-} & 9.0\\
J1713$+$0747 & 4.57 & 0.85 & 16.0 & 67.8& \multicolumn{2}{|c|}{-} & \multicolumn{2}{|c|}{-} & 0.100 & 90 & 0.050 & 175 & 0.025 & 68 & 8.8\\
J1738$+$0333 & 5.85 & 2.41 & 33.8 & 0.4& \multicolumn{2}{|c|}{-} & \multicolumn{2}{|c|}{-} & \multicolumn{2}{|c|}{-} & 0.316 & 29 & 0.301 & 27 & 4.0\\
J1741$+$1351 & 3.75 & 3.02 & 24.2 & 16.3& \multicolumn{2}{|c|}{-} & 0.155 & 20 & \multicolumn{2}{|c|}{-} & 0.233 & 42 & \multicolumn{2}{|c|}{-} & 4.2\\
J1744$-$1134 & 4.07 & 0.89 & 3.1 & -& \multicolumn{2}{|c|}{-} & \multicolumn{2}{|c|}{-} & 0.114 & 89 & 0.203 & 91 & \multicolumn{2}{|c|}{-} & 9.2\\
J1747$-$4036 & 1.65 & 1.32 & 153.0 & -& \multicolumn{2}{|c|}{-} & \multicolumn{2}{|c|}{-} & 0.895 & 22 & 1.112 & 25 & \multicolumn{2}{|c|}{-} & 1.7\\
J1832$-$0836 & 2.72 & 0.87 & 28.2 & -& \multicolumn{2}{|c|}{-} & \multicolumn{2}{|c|}{-} & 0.577 & 11 & 0.422 & 10 & \multicolumn{2}{|c|}{-} & 0.6\\
J1853$+$1303 & 4.09 & 0.87 & 30.6 & 115.7& \multicolumn{2}{|c|}{-} & 0.369 & 18 & \multicolumn{2}{|c|}{-} & 0.369 & 50 & \multicolumn{2}{|c|}{-} & 5.6\\
B1855$+$09 & 5.36 & 1.78 & 13.3 & 12.3& \multicolumn{2}{|c|}{-} & 0.155 & 74 & \multicolumn{2}{|c|}{-} & 0.148 & 84 & \multicolumn{2}{|c|}{-} & 8.9\\
J1903$+$0327 & 2.15 & 1.88 & 297.6 & 95.2& \multicolumn{2}{|c|}{-} & \multicolumn{2}{|c|}{-} & \multicolumn{2}{|c|}{-} & 0.444 & 35 & 0.437 & 35 & 4.0\\
J1909$-$3744 & 2.95 & 1.40 & 10.4 & 1.5& \multicolumn{2}{|c|}{-} & \multicolumn{2}{|c|}{-} & 0.041 & 89 & 0.102 & 98 & \multicolumn{2}{|c|}{-} & 9.1\\
J1910$+$1256 & 4.98 & 0.97 & 38.1 & 58.5& \multicolumn{2}{|c|}{-} & \multicolumn{2}{|c|}{-} & \multicolumn{2}{|c|}{-} & 0.239 & 75 & 0.275 & 37 & 8.8\\
J1918$-$0642 & 7.65 & 2.57 & 26.6 & 10.9& \multicolumn{2}{|c|}{-} & \multicolumn{2}{|c|}{-} & 0.317 & 86 & 0.547 & 92 & \multicolumn{2}{|c|}{-} & 9.0\\
J1923$+$2515 & 3.79 & 0.96 & 18.9 & -& \multicolumn{2}{|c|}{-} & 0.442 & 17 & \multicolumn{2}{|c|}{-} & 0.535 & 24 & \multicolumn{2}{|c|}{-} & 2.2\\
B1937$+$21 & 1.56 & 10.51 & 71.0 & -& \multicolumn{2}{|c|}{-} & \multicolumn{2}{|c|}{-} & 0.007 & 93 & 0.012 & 154 & 0.007 & 47 & 9.1\\
J1944$+$0907 & 5.19 & 1.73 & 24.3 & -& \multicolumn{2}{|c|}{-} & 0.365 & 24 & \multicolumn{2}{|c|}{-} & 0.403 & 47 & \multicolumn{2}{|c|}{-} & 5.7\\
J1949$+$3106 & 13.14 & 9.96 & 164.1 & 1.9& \multicolumn{2}{|c|}{-} & \multicolumn{2}{|c|}{-} & \multicolumn{2}{|c|}{-} & 1.414 & 22 & 1.349 & 15 & 1.2\\
B1953$+$29 & 6.13 & 2.97 & 104.5 & 117.3& \multicolumn{2}{|c|}{-} & 0.475 & 19 & \multicolumn{2}{|c|}{-} & 0.558 & 51 & \multicolumn{2}{|c|}{-} & 7.2\\
J2010$-$1323 & 5.22 & 0.48 & 22.2 & -& \multicolumn{2}{|c|}{-} & \multicolumn{2}{|c|}{-} & 0.370 & 53 & 0.733 & 55 & \multicolumn{2}{|c|}{-} & 4.1\\
J2017$+$0603 & 2.90 & 0.80 & 23.9 & 2.2& \multicolumn{2}{|c|}{-} & 0.237 & 6 & \multicolumn{2}{|c|}{-} & 0.238 & 27 & 0.243 & 20 & 1.7\\
J2043$+$1711 & 2.38 & 0.52 & 20.7 & 1.5& \multicolumn{2}{|c|}{-} & 0.121 & 24 & \multicolumn{2}{|c|}{-} & 0.170 & 32 & \multicolumn{2}{|c|}{-} & 2.3\\
J2145$-$0750 & 16.05 & 2.98 & 9.0 & 6.8& \multicolumn{2}{|c|}{-} & \multicolumn{2}{|c|}{-} & 0.213 & 71 & 0.535 & 73 & \multicolumn{2}{|c|}{-} & 9.1\\
J2214$+$3000 & 3.12 & 1.47 & 22.6 & 0.4& \multicolumn{2}{|c|}{-} & \multicolumn{2}{|c|}{-} & \multicolumn{2}{|c|}{-} & 0.399 & 26 & 0.399 & 22 & 2.1\\
J2302$+$4442 & 5.19 & 1.38 & 13.7 & 125.9& \multicolumn{2}{|c|}{-} & \multicolumn{2}{|c|}{-} & 1.018 & 26 & 1.592 & 26 & \multicolumn{2}{|c|}{-} & 1.7\\
J2317$+$1439 & 3.45 & 0.24 & 21.9 & 2.5& 0.070 & 75 & 0.070 & 74 & \multicolumn{2}{|c|}{-} & 0.202 & 19 & \multicolumn{2}{|c|}{-} & 8.9\\
\hline
\multicolumn{5}{r|}{Nominal scaling factor$^b$ (ASP/GASP)} 
  & \multicolumn{2}{|c|}{0.6}
  & \multicolumn{2}{|c|}{0.4}
  & \multicolumn{2}{|c|}{0.8}
  & \multicolumn{2}{|c|}{0.8}
  & \multicolumn{2}{|c|}{0.8}
  & \\
\multicolumn{5}{r|}{Nominal scaling factor$^b$ (GUPPI/PUPPI)} 
  & \multicolumn{2}{|c|}{0.7}
  & \multicolumn{2}{|c|}{0.5}
  & \multicolumn{2}{|c|}{1.4}
  & \multicolumn{2}{|c|}{2.5}
  & \multicolumn{2}{|c|}{2.1}
  & \\
\hline
\end{tabular}

\vspace{0.5em}

{$^a$ For this table, the original TOA uncertainties were scaled by their
bandwidth-time product $\left( \frac{\Delta \nu}{100~\mathrm{MHz}}
\frac{\tau}{1800~\mathrm{s}} \right)^{1/2}$ to remove variation due to
different instrument bandwidths and integration time.}

\vspace{0.5em}

{$^b$ TOA uncertainties can be rescaled to the nominal full instrumental
bandwidth as listed in Table~\ref{tab:obs:observing_systems} by dividing by the
scaling factors given here.}

\end{center}
\end{table*}

\section{Timing Analysis}
\label{sec:timing}

We fit the TOA set for each pulsar to a timing model using standard
procedures as described by \citet{lk05}, supplemented by novel
methods to compensate for frequency-dependent pulse shape variations
and to model noise in the TOA data sets.  In this section, we start by 
summarizing our use of standard timing models, and then we describe
the new algorithm for handling frequency-dependent pulse shape variations.
In the next section of the paper, we describe the noise model.

\subsection{Timing models}
\label{sec:timing:models}

All TOAs were initially measured
using a local, topocentric time provided by hydrogen-maser clocks located
at the observatories.  Observatory clock corrections, determined by
daily monitoring of the maser offsets compared to times determined
using Global Positioning System (GPS) receivers, were used to
transform the TOAs to Coordinated Universal Time (UTC).  The times
were further transformed to Terrestrial Time (TT) as published by the
Bureau International des Poids et Mesures, TT(BIPM), after
accounting for the effects of the Earth's varying rotation rate as
published by the International Earth Rotation and Reference Systems
Service (IERS). Finally, relativistic corrections are applied to
convert the times to Barycentric Dynamical Time (TDB)\footnote{See
\url{http://www.iausofa.org/2015\_0209\_C/sofa/sofa\_ts\_c.pdf} for a detailed
discussion of the various timescale transformations.}. Propagation delays in
the solar system, used to project the TOAs to the Solar-System Barycenter
(SSB) are calculated using the JPL DE421 planetary
ephemeris\footnote{\url{http://naif.jpl.nasa.gov/naif/}},
rotated into an ecliptic reference frame using the 2010 IAU value
of the obliquity of the ecliptic.

We derived timing models by enumerating all rotations of each pulsar and
accounting for the various physical processes, discussed below, that can
cause observed timing delays.  This modelling was done in conjunction
with a parameterized model for noise in the arrival-time data, described in
detail in Section~\ref{sec:noise} and Appendix~\ref{sec:noise_app}.  In
effect, the noise model determined a separate ``weight'' for each data subset,
defined by the combination of receiver and backend system used, along
with a measurement of temporally correlated ``red'' noise.  We used the
\tempo
and \tempotwo
pulsar-timing analysis programs, making use of recently implemented
generalized least-squares (GLS) fitting procedures which take into
account correlations in the TOA noise when determining timing model
parameter values and their uncertainties \citep[e.g.,][]{chc+11}.

The code-bases for \tempo and \tempotwo are different but not fully
independent.  With appropriate timing model options such that the two
programs employed the same algorithms and conventions (using the same
clock standards, employing the same solar system ephemeris, using the
same obliquity of the ecliptic, excluding a solar wind model (see
below), etc.) the fit results were nearly identical between these two
programs.  The vast majority of all fit parameters agreed to
$\lesssim$10\% of their 1-$\sigma$ uncertainties, and the ephemeris
files we provide are able to be used in both programs.  PSR~J1713$+$0747
is an exception, as its complicated timing model includes time-varying
orbital geometry terms that are handled slightly differently by each
program.  In this paper we report only the \tempo results for all the
pulsars.

The number of fit parameters in each timing model depends on the
observed spin, astrometric, and environmental properties of the given
pulsar.  We used ecliptic coordinates to fit for all astrometric
parameters in order to reduce parameter covariances, and we fit for
parallax for all pulsars, regardless of whether the resulting fit value
was physically meaningful (i.e.,~positive) or significant.  We also fit
for proper motion for all pulsars except for the two (PSRs~J0931$-$1902
and J1832$-$0836) which had observing timespans less than one year. The
timing models contain fits for spin frequency and its first
time-derivative, with higher-order spin noise, if present, being
parametrized by the red noise model.  We fit five Keplerian binary
parameters for all binary pulsars using either the \citet[``DD'']{dd85}
or \citet[``ELL1'']{lcw+01} binary models.  The former is a generally
applicable, fully relativistic description of the pulsar's orbit, while
the latter is an alternate parameterization that improves numerical
stability for very low-eccentricity orbits.  We introduced
post-Keplerian parameters \citep[e.g.,][]{dt92} when model accuracy was
significantly improved as determined by an $F$-test significance value
of 0.0027 (i.e.,~3-$\sigma$ significant). 
Information on the timing models, noise models and residual statistics
is presented in Table~\ref{tab:timingpar}.

We incorporated timing model parameters to describe dispersive delays in
the TOAs from the time-variable integrated column of ionized gas
between the observatory and each pulsar.  These delays are primarily due to the
turbulent interstellar medium (ISM) but also include smaller
contributions from the solar wind and the Earth's ionosphere. The
delays are characterized by the time-dependent
dispersion measure (DM) for each pulsar which is directly proportional
to the integrated electron column density. The expected TOA delay for
a broadband radio pulse is $\Delta t_{\rm DM} \propto$ DM$\nu^{-2}$,
where $\nu$ is the observing frequency.

We measured a value of DM at nearly every observing epoch for each
pulsar, enumerated using the \tempo/\tempotwo parameter ``DMX''.   Since
dual-receiver observations were sometimes separated by several days, we
allowed a single constant DM value to apply to a window of up to 14 days
of observations.  These measured DM values include the effects of
ionospheric, interstellar, and solar wind dispersion (i.e., the solar
wind DM model typically applied by default in \tempo/\tempotwo was
disabled for this analysis).  The best-fit DMX values are shown in the
timing summary figures for each pulsar below.  The average DM value for
each pulsar is highly covariant with profile shape evolution versus
frequency (\S\ref{sec:timing:fd}).  However, DM {\it variation} can be
easily distinguished from the constant-in-time profile shape terms.  The
DM error bars shown in the summary figures represent the uncertainty on
the mean-subtracted values ($\mathrm{DMX}_i - \left< \mathrm{DMX}
\right>$), removing the large covariance affecting the mean DM.  These
uncertainties are determined via an appropriate linear transformation of
the original post-fit parameter covariance matrix.

Variations in DM over time are primarily attributed to an evolving view
of the line-of-sight electron column due to the relative motion of the
Earth, the pulsar, and the ISM \citep[e.g.][]{rdb+06}.  Many pulsars in
our data set exhibit slow, long-term DM trends that are at least
qualitatively in agreement with the expectation of turbulent electron
density structure in the ISM.  For several pulsars, annual variations in
DM are also apparent.  Those at low ecliptic latitude likely have a
significant solar wind contribution to their DM -- notable examples are
J1614$-$2230 and J2010$-$1323 with ecliptic latitudes of $-$1.2 and 6.5
degrees respectively.  These pulsars show a sharp increase in DM at the
times of year that they pass behind the Sun.  Less sharp, more
sinusoidal annual trends are seen in some cases, notably J0613$-$0200
and J1643$-$1224.  A possible explanation for this is line-of-sight
motion due to Earth's orbit projected onto an ISM density gradient, as
discussed by \citet{kcs+13}.  Using an independent data set,
\citet{kcs+13} also detect significant annual DM modulation in several
of the same sources where it is apparent in our results.  Finally,
isolated DM ``events'' such as the 2008--2009 dip shown by J1713$+$0747
are occasionally visible.  This event is not yet well explained, but
indicates the need for additional study.  A more detailed astrophysical
analysis of all DM results from this data set is currently underway.

Regardless of the cause, since each pulsar's DM is evolving with time,
the fact that our dual-frequency measurements are separated by up to 14
days will result in a DM estimation error due to the DM being slightly
different at the time of each of the two observations.  This effect was
studied in detail by \citet{lcc+15}, who conclude that 14-day
separations can result in timing errors on the order of $\sim$50~ns,
depending on the turbulent spectrum of electron density fluctuations in
the ISM.  In practice this effect will be significantly smaller as the
majority of our paired observations are separated by $\lesssim$3 days
\citep[e.g.,][Figure~4]{lcc+15}.

\subsection{Compensation for frequency evolution of pulsar profiles}
\label{sec:timing:fd}

\begin{figure}[tp]
\centering
\includegraphics[width=\columnwidth]{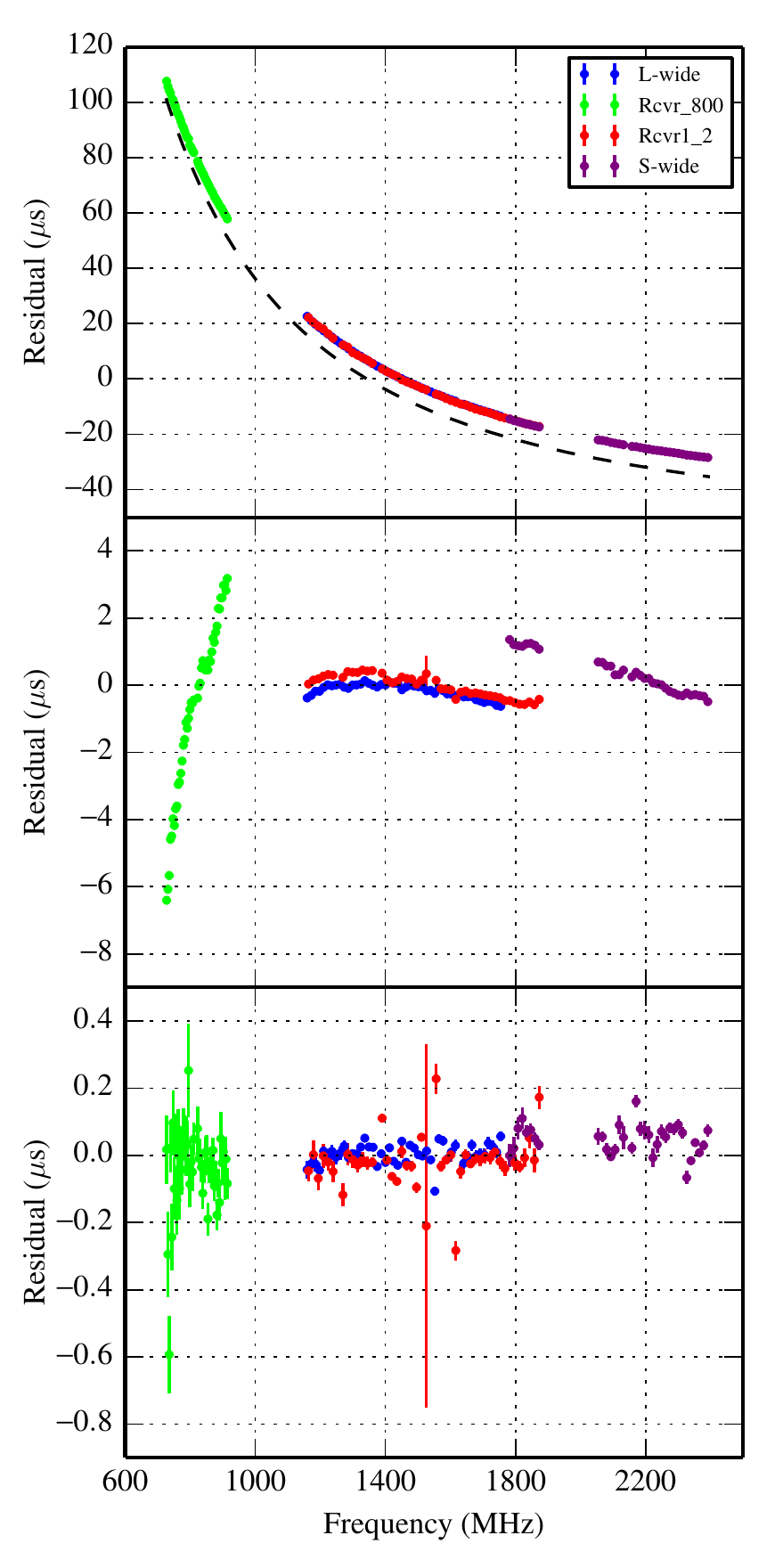}
\caption{Average timing residual versus radio frequency for
PSR~J1713$+$0747.  {\it Upper panel:} Non-dispersive frequency dependent (FD)
residuals with all other timing model parameters held fixed at their best-fit
values.  The dashed line indicates the best-fit FD model from
Eqn.~\ref{eq:FD}, offset for plot clarity.  {\it Middle panel:}
Residuals when FD model is not included in the fit. {\it Lower panel:}
final residuals when FD model is included in the fit.  Note the
different y-axis scales in each panel.  See also a similar presentation
of these data by \citet{zsd+15}.  \label{fig:fd}}
\end{figure}

For each pulsar, we calculated TOAs for all frequency channels recorded
with a given receiver using a single standard template profile.  Because
pulse shapes vary with frequency, this produced small systematic
frequency-dependent perturbations in the TOAs in addition to the
$\nu^{-2}$ offsets due to dispersion.  An example of such common
behavior in our data set is in Figure~\ref{fig:fd}.  In previous work,
we compensated for these frequency-dependent perturbations for each
pulsar by fitting arbitrary time offsets to each spectral channel, thus
adding a large number of free parameters to the timing model
\citep{dfg+13}.

For this analysis, we developed a heuristic approach to remove as much of
the frequency-dependent (``FD'') bias from our timing residuals as
possible by incorporating an additional timing delay $\Delta t_{\rm
FD}$ to all timing models, where
\begin{equation}
	\Delta t_{\rm FD} = \sum_{i=1}^n c_i {\rm log}\bigg(\frac{\nu}{1 { \rm GHz}}\bigg)^i
	\label{eq:FD}
\end{equation}
and the coefficients $c_i$ are fit parameters in the timing models.  For
any given pulsar, the number of terms needed was determined by an
$F$-test significance value of 0.0027, the same criterion used for other
timing model parameters.  The number of parameters used ranged from 0 to
5 (Table~\ref{tab:timingpar}), and reflects the degree to which profile
evolution is important for each pulsar. An illustrative example of the
application of our FD model for bias removal is shown in
Figure~\ref{fig:fd} \citep[see also discussion of this approach
by][]{zsd+15}.

This algorithm significantly reduced the number of free parameters in
the timing model compared to our previous work, but it remains an {\it
ad hoc} procedure -- it is applied only after TOAs are calculated and
does not directly utilize the additional pulse shape information
available in the profile data.  In future work we plan to explore
additional approaches to solving this problem, including TOAs derived
from a broadband profile-fitting approach that directly incorporates
profile evolution versus frequency \citep{pdr14}.

\subsection{Timing summary}
\label{sec:timing:summary}

A basic summary of the timing model fit results, including number of
TOAs, number of fit parameters, noise model results, and basic
statistics of the residuals is presented in Table \ref{tab:timingpar}.
The full set of best-fit timing model parameter values, and their
associated uncertainties, are publicly available and are distributed
along with the TOA data presented here.  The models are provided in the
standard ``par file'' format understood by \tempo and \tempotwo.  These
results potentially contain a signficant amount of astrophysical
information about these pulsars, their orbits and binary companions, and
the ISM properties along their line-of-sight, however we have postponed
a detailed interpretation of this to future work (see
Section~\ref{sec:conclusions} for a partial list).  It should be noted
that while the models presented here provide an accurate description of
the time of arrival data for the purposes of gravitational wave
detection, certain model parameters (in particular, parallax and Shapiro
delay) may require a more sophisticated uncertainty analysis before
astrophysically meaningful conclusions can be drawn from their values.

\begin{table*}[tp]
\begin{center}
\caption{\label{tab:timingpar} Summary of timing model fits}
\begin{tabular}{c|r|rrrrrr|cc|ccr|c}
\hline
Source
  & Number
  & \multicolumn{6}{|c|}{Number of Fit Parameters$^a$}
  & \multicolumn{2}{|c|}{RMS$^b$ ($\mu$s)}
  & \multicolumn{3}{|c|}{Red Noise$^c$}
  & Figure \\

  & of TOAs
  & S
  & A
  & B
  & DM
  & FD
  & J
  & Full
  & White
  & $A_{\mathrm{red}}$
  & $\gamma_{\mathrm{red}}$
  & log$_{10}B$
  & Number\\
\hline
J0023+0923 & 4598 & 3 & 5 & 5 & 28 & 1 & 1 & 0.320  & - & - & - & 1.40  & \ref{fig:summary-J0023+0923} \\
J0030+0451 & 2468 & 3 & 5 & 0 & 60 & 0 & 1 & 0.723  & 0.212  & 0.014\phantom{$^d$} & $-$4.8 & 4.99  & \ref{fig:summary-J0030+0451} \\
J0340+4130 & 3008 & 3 & 5 & 0 & 27 & 1 & 1 & 0.385  & - & - & - & 0.01  & \ref{fig:summary-J0340+4130} \\
J0613$-$0200 & 7651 & 3 & 5 & 7 & 90 & 2 & 1 & 0.592  & 0.165  & 0.093\phantom{$^d$} & $-$2.9 & 3.71  & \ref{fig:summary-J0613-0200} \\
J0645+5158 & 2896 & 3 & 5 & 0 & 33 & 2 & 1 & 0.052  & - & - & - & $-$0.08  & \ref{fig:summary-J0645+5158} \\
J0931$-$1902 & 719 & 3 & 3 & 0 & 11 & 0 & 1 & 0.381  & - & - & - & $-$0.07  & \ref{fig:summary-J0931-1902} \\
J1012+5307 & 11995 & 3 & 5 & 5 & 95 & 1 & 1 & 1.197  & 0.355  & 0.669\phantom{$^d$} & $-$1.0 & 14.31  & \ref{fig:summary-J1012+5307} \\
J1024$-$0719 & 5073 & 3 & 5 & 0 & 53 & 2 & 1 & 0.280  & - & - & - & 0.08  & \ref{fig:summary-J1024-0719} \\
J1455$-$3330 & 5122 & 3 & 5 & 6 & 81 & 1 & 1 & 0.694  & - & - & - & 0.05  & \ref{fig:summary-J1455-3330} \\
J1600$-$3053 & 8174 & 3 & 5 & 8 & 74 & 2 & 1 & 0.197  & - & - & - & $-$0.01  & \ref{fig:summary-J1600-3053} \\
J1614$-$2230 & 7517 & 3 & 5 & 7 & 54 & 1 & 1 & 0.189  & - & - & - & 0.02  & \ref{fig:summary-J1614-2230} \\
J1640+2224 & 2565 & 3 & 5 & 9 & 65 & 2 & 1 & 0.158  & - & - & - & $-$0.03  & \ref{fig:summary-J1640+2224} \\
J1643$-$1224 & 7119 & 3 & 5 & 6 & 91 & 2 & 1 & 2.057  & 0.331  & 1.231$^d$ & $-$1.7 & 18.33  & \ref{fig:summary-J1643-1224} \\
J1713+0747 & 15830 & 3 & 5 & 8 & 106 & 4 & 3 & 0.116  & - & - & - & 0.01  & \ref{fig:summary-J1713+0747} \\
J1738+0333 & 2711 & 3 & 5 & 5 & 28 & 1 & 1 & 0.308  & - & - & - & 0.01  & \ref{fig:summary-J1738+0333} \\
J1741+1351 & 1600 & 3 & 5 & 8 & 27 & 0 & 1 & 0.103  & - & - & - & $-$0.02  & \ref{fig:summary-J1741+1351} \\
J1744$-$1134 & 9020 & 3 & 5 & 0 & 88 & 2 & 1 & 0.334  & - & - & - & 0.25  & \ref{fig:summary-J1744-1134} \\
J1747$-$4036 & 2778 & 3 & 5 & 0 & 25 & 1 & 1 & 0.531  & - & - & - & 0.12  & \ref{fig:summary-J1747-4036} \\
J1832$-$0836 & 1136 & 3 & 3 & 0 & 10 & 0 & 1 & 0.121  & - & - & - & $-$0.04  & \ref{fig:summary-J1832-0836} \\
J1853+1303 & 1411 & 3 & 5 & 6 & 26 & 0 & 1 & 0.235  & - & - & - & $-$0.02  & \ref{fig:summary-J1853+1303} \\
B1855+09 & 4071 & 3 & 5 & 7 & 72 & 3 & 1 & 1.339  & 0.505  & 0.017\phantom{$^d$} & $-$4.9 & 2.87  & \ref{fig:summary-B1855+09} \\
J1903+0327 & 1887 & 3 & 5 & 8 & 36 & 2 & 1 & 1.949  & 0.327  & 0.851$^d$ & $-$2.5 & 2.87  & \ref{fig:summary-J1903+0327} \\
J1909$-$3744 & 10697 & 3 & 5 & 8 & 88 & 1 & 1 & 0.079  & - & - & - & 0.72  & \ref{fig:summary-J1909-3744} \\
J1910+1256 & 2690 & 3 & 5 & 6 & 45 & 1 & 1 & 1.449  & 0.587  & 0.801$^d$ & $-$1.9 & 5.39  & \ref{fig:summary-J1910+1256} \\
J1918$-$0642 & 10035 & 3 & 5 & 7 & 87 & 3 & 1 & 0.340  & - & - & - & $-$0.02  & \ref{fig:summary-J1918-0642} \\
J1923+2515 & 939 & 3 & 5 & 0 & 24 & 1 & 1 & 0.266  & - & - & - & $-$0.06  & \ref{fig:summary-J1923+2515} \\
B1937+21 & 9966 & 3 & 5 & 0 & 102 & 5 & 3 & 1.549  & 0.104  & 0.197\phantom{$^d$} & $-$2.4 & 96.48  & \ref{fig:summary-B1937+21} \\
J1944+0907 & 1724 & 3 & 5 & 0 & 28 & 2 & 1 & 2.442  & 0.331  & 0.860$^d$ & $-$2.8 & 2.35  & \ref{fig:summary-J1944+0907} \\
J1949+3106 & 1416 & 3 & 5 & 7 & 16 & 0 & 1 & 0.647  & - & - & - & 0.03  & \ref{fig:summary-J1949+3106} \\
B1953+29 & 1329 & 3 & 5 & 6 & 24 & 2 & 1 & 4.149  & 0.531  & 0.015$^d$ & $-$6.7 & 2.14  & \ref{fig:summary-B1953+29} \\
J2010$-$1323 & 8068 & 3 & 5 & 0 & 55 & 1 & 1 & 0.312  & - & - & - & 0.08  & \ref{fig:summary-J2010-1323} \\
J2017+0603 & 1589 & 3 & 5 & 7 & 24 & 0 & 2 & 0.073  & - & - & - & 0.01  & \ref{fig:summary-J2017+0603} \\
J2043+1711 & 1394 & 3 & 5 & 7 & 23 & 1 & 1 & 0.108  & - & - & - & 0.02  & \ref{fig:summary-J2043+1711} \\
J2145$-$0750 & 7369 & 3 & 5 & 6 & 73 & 2 & 1 & 0.371  & - & - & - & 0.20  & \ref{fig:summary-J2145-0750} \\
J2214+3000 & 2624 & 3 & 5 & 5 & 25 & 1 & 1 & 0.319  & - & - & - & 0.06  & \ref{fig:summary-J2214+3000} \\
J2302+4442 & 3044 & 3 & 5 & 6 & 27 & 1 & 1 & 0.708  & - & - & - & 0.47  & \ref{fig:summary-J2302+4442} \\
J2317+1439 & 2650 & 3 & 5 & 8 & 68 & 3 & 2 & 0.267  & - & - & - & 0.04  & \ref{fig:summary-J2317+1439} \\
\hline
\end{tabular}

\vspace{0.5em}

{$^a$ Fit parameters: S=spin; B=binary; A=astrometry; DM=dispersion measure;
FD=frequency dependence; J=jump}

\vspace{0.5em}

{$^b$ Weighted root-mean-square of epoch-averaged post-fit timing residuals.
For sources with red noise, the ``Full'' RMS value includes the red noise
contribution, while the ``White'' RMS does not.}

\vspace{0.5em}

{$^c$ Red noise parameters: $A_{\mathrm{red}}$ = amplitude of red noise
spectrum at $f$=1~yr$^{-1}$ measured in $\mu$s yr$^{1/2}$;
$\gamma_{\mathrm{red}}$ = spectral index; $B$ = Bayes factor.  See
Eqn.~\ref{eqn:rn_spec} and Appendix~\ref{sec:noise_app} for details.}

\vspace{0.5em}

{$^d$ For these sources, the detected red noise is likely due to unmodeled
interstellar medium propagation effects rather than intrinsic spin noise; see
text for details.}

\end{center}
\end{table*}

\section{Noise Characterization}
\label{sec:noise}

\subsection{Noise Model}
\label{sec:noise-model}

The noise model used in the analysis is a parameterized one that
includes the effects of several noise sources that produce different
correlations of TOAs obtained in non-overlapping time blocks and
frequency channels. 
For instance, the template matching errors due to radiometer noise are 
uncorrelated in both time and frequency, but pulse-jitter noise
\citep{cs10} appears to affect all TOAs obtained simultaneously in
different frequency channels.  Correlated timing noise with a red power
spectrum occurs to varying degree in different pulsars.  Spin noise is
achromatic and is much smaller in MSPs compared to objects with stronger
magnetic fields and longer spin periods.    Chromatic red noise due to
propagation through intervening plasmas (ISM, interplanetary medium, and
ionosphere) may also be present if dispersive delays are not removed
perfectly or if scattering and refraction effects contribute
significantly.    Jitter noise appears to be highly correlated across
hundreds of MHz for those MSPs that have been analyzed in detail
\citep{2014MNRAS.443.1463S, 2014ApJ...794...21D}.  
An approach to noise modeling has been discussed extensively in
\citet{vhl13, e13, vhv14, vhv15, abb+14,e14}. For more details about the
specific noise model and implementation for the current study see Appendix
\ref{sec:noise_app}. Here we will summarize the parameterization
used for this data release.

Our model for noise starts with the measurement error on each TOA, $\sigma_{i,k}$, determined by the template-matching TOA calculation algorithm;  here $i$ is the TOA number and the subscript $k$ denotes the backend/receiver system.  Because such measurement errors may be underestimated, we allow for them to be increased by systematic quadrature and scaling factors, determined on a system-by-system basis,
\be
\sigma_{i,k} \rightarrow  E_{k} \left( \sigma_{i,k}^{2} + Q_{k}^{2}  \right)^{1/2},
\ee
where $E_k$ and $Q_k$ are the EFAC and EQUAD parameters used in the \tempo/\tempotwo timing code.

In addition to the template-fitting errors,
we also include TOA errors that are uncorrelated in time but completely correlated between TOAs obtained at different frequencies measured simultaneously, termed ``short-term correlated noise'' in Appendix  \ref{sec:noise_app}. 
 The strength of this process is characterized 
by the ECORR parameter. This term could include true pulse phase jitter
\citep{cs10}, known to be present in some pulsars,
but can also include other similarly-correlated components. 
 Lastly, we model the red noise as a stationary Gaussian process that is parameterized by a power spectrum of the form
\be
\label{eqn:rn_spec}
P(f) = A_{\rm red}^2\left( \frac{f}{f_{\rm yr}} \right)^{\gamma_{\rm red}},
\ee
where $A_{\rm red}$ is the amplitude of the red noise in $\mu$s~${\rm
yr}^{1/2}$, $\gamma_{\rm red}$ is the spectral index, and $f_{\rm
yr}=1{\rm yr}^{-1}$.

These noise parameters are included in a joint likelihood that contains
all timing model parameters. For the purposes of this paper, we
analytically marginalize over the linear timing model parameters and
explore the space of noise parameters via Markov Chain Monte-Carlo
(MCMC). We then use the MCMC results to determine the maximum likelihood
noise parameters, which are subsequently used as inputs to
\tempo/\tempotwo GLS fitting routines. For each pulsar we always include the EFAC, EQUAD, and ECORR parameters for each backend/receiver system. We only include red noise when it is preferred by the data. The red noise model selection is performed with MultiNest \citep{fhb09} using a Bayes factor threshold of 100 to determine whether red noise is included in the final model. The applicable red noise amplitudes, spectral indices, and Bayes factors are shown in Table \ref{tab:timingpar}.

\subsection{Noise Analysis}
\label{sec:noise-analysis}

Here we discuss some of the major features that our noise model
reveals. As mentioned above, we only include red noise in the noise
model when the data favor its inclusion. In our analysis, ten pulsars
meet this criterion and will now be discussed further. Intrinsic pulsar spin noise and its effects have been studied in 
\citep{bnr84, cd85, antt94} and \citep[][hereafter SC10]{sc10}. Using a sample of both canonical pulsars (CPs) and MSPs, SC10 parameterize the post-fit (after quadratic subtraction) timing noise rms as
\be
\label{eq:scrms}
\hat{\sigma}_{\rm TN,2} = C_{2}\nu^{\alpha}|\dot{\nu}_{-15}|^{\beta}T_{\rm yr}^{\gamma},
\ee 
where $\nu$, $\dot{\nu}_{-15}$, and $T_{\rm yr}$ are the spin frequency,
spin frequency derivative in units of $10^{-15}$ s$^{-2}$, and time span of the data set in years. The best fit values of the free parameters were found to be $\ln(C_{2})=1.6 \pm 0.4$, $\alpha = -1.4 \pm 0.1$, $\beta = 1.1 \pm 0.1$, and $\gamma = 2.0 \pm 0.2$. A fifth parameter, $\delta$, was used to take into account the 
empirical scatter about the mean relation in Eq.~\ref{eq:scrms} 
and was estimated to be $\delta = 1.6 \pm 0.1$ in $\ln\sigma_{\rm TN,2}$. First we note that the best fit value of $\gamma$  in Eq.~\eqref{eq:scrms} corresponds to a power spectral density index of $\gamma_{\rm red} = -(2\gamma+1) =  -5 \pm 0.4$ in Eq.~\eqref{eqn:rn_spec}. 
 We can estimate $\sigma_{\rm TN,2}$ from our noise model by
\be
\begin{split}
\label{eq:sigmaTN2}
\sigma_{\rm TN,2} &= 
	\lp\int_{1/T}^{\infty} P(f) df\rp^{1/2}\\
	& = 2.05\,{\rm ns}\, \, 
	\lp1-\gamma_{\rm red}\rp^{-1/2} \lp  
	\frac{A_{\rm red}}{3\times 10^{-3} 
	{\rm \mu s \, yr^{1/2}}}\rp 
	T_{\rm yr}^{(1-\gamma_{\rm red})/2},
\end{split}
\ee
where the lower integration limit of $1/T$ serves as a filter for quadratic subtraction. Furthermore, we can produce a distribution of $\sigma_{\rm TN,2}$ by evaluating Eq.~\eqref{eq:sigmaTN2} for all values of $A_{\rm red}$ and $\gamma_{\rm red}$ from our MCMC analysis. This will represent our uncertainty in the red noise variance by incorporating the full posterior distributions of  red noise parameters as opposed to just the maximum likelihood values.

In essence, SC10 make two predictions for intrinsic pulsar timing spin noise: (\textit{i}) the red noise spectral index is steep with a value $\sim$ $-5$, and (\textit{ii}) the red noise rms follows Eq. \eqref{eq:scrms}
to within a factor of $\exp{\left(\pm \delta\right)}$.
\begin{figure*}[ht!]
\centering
\includegraphics[scale=1.0]{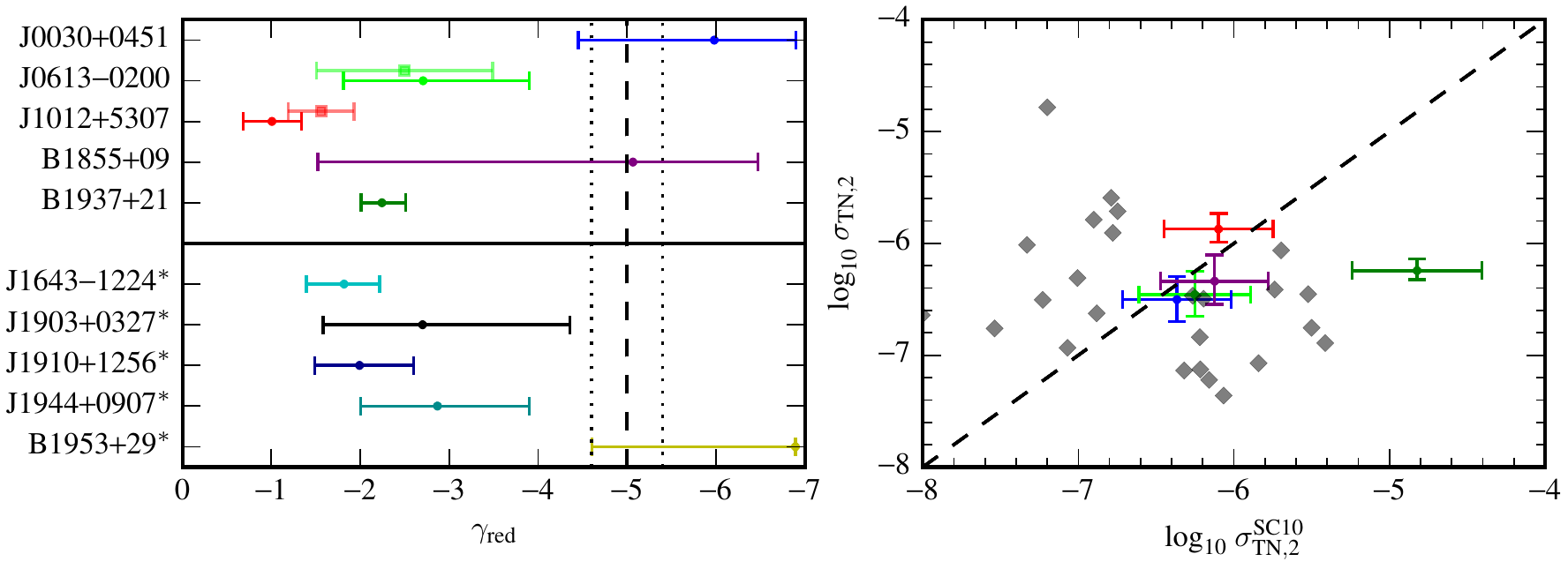}
\caption{{\it Left panel:} Maximum a-posteriori value and 68\% credible
interval on the red noise spectral index, $\gamma_{\rm red}$ for all
pulsars that display significant red noise. The dashed and dotted black
lines represent the mean and one-sigma predictions on the spectral index
from SC10. The points with square markers are the spectral index values
presented in \cite{ltm+15}. The pulsars marked with an asterisk indicate those
pulsars for which the red noise is likely due to unmodeled interstellar medium 
propagation effects, rather than intrinsic spin noise. {\it Right panel:} Measured value for red
noise RMS for all red pulsars (see text for details of this calculation)
vs. predictions for SC10. The uncertainties are the one-sigma and 68\%
credible region for the predicted and measured values, respectively. The
gray points are the 95\% upper limits from the predicted and measured
values of $\sigma_{\rm TN,2}$ for all other pulsars. The color scheme is
the same as the left panel. We do not include the pulsars marked with an asterisk
in this figure.}
\label{fig:red-comp}
\end{figure*}
In Figure \ref{fig:red-comp} we show the maximum a-posteriori value and 68\% credible interval for the red noise spectral index, $\gamma_{\rm red}$, for all pulsars that display significant red noise. 

We see that our noise analysis yields a much more shallow
spectral index in general than the predicted value of SC10. In fact, of
the 10 pulsars that display red noise, only 3 (PSRs J0030$+$0451,
B1953$+$29, and B1855$+$09) have spectral indices consistent with $-5$,
the others are consistent with $\sim$ $-2$. If we assume that this red
noise is due to a random walk in one of the quadratic spin down
parameters, then our analysis suggests a random walk in the pulsar
phase\footnote{As stated in SC10, random walks in the pulsar phase,
period and period derivative lead to underlying power spectral indices
of $-2$, $-4$, and $-6$, respectively.}. However, it is more likely that in many 
cases (pulsars marked with an asterisk in Figure \ref{fig:red-comp})
this behavior is due to un-modeled ISM effects as we will discuss. In
the right panel of Figure \ref{fig:red-comp} we see that our
measurements of $\sigma_{\rm TN,2}$ are close to one-sigma consistent
with the predictions of SC10 with the exception of PSR B1937$+$21 which
exhibits much weaker red noise than predicted. The gray points show that
the 95\% upper limits on $\sigma_{\rm TN,2}$ are not consistent for some pulsars.
Overall we can state with confidence that our noise analysis is
inconsistent with the predictions of SC10 both for the spectral index
and the overall red noise rms. To explore this more closely, we will now
look into each pulsar in more detail.

PSRs J0030$+$0451 and B1855$+$09 are consistent with the spin noise predictions of a steep red noise process. From inspection of Figures \ref{fig:summary-J0030+0451} and \ref{fig:summary-B1855+09} we see that both pulsars are timed for the full nine years and have dual frequency data and DM$(t)$ corrections for all observing epochs. Furthermore, each set of residuals displays a cubic low frequency term that is characteristic of the predicted steep red process. This appears to be the first evidence of red noise in these pulsars as they have white residuals for both five and six year datasets presented from NANOGrav \citep{dfg+13} and the Parkes Pulsar Timing Array \citep[PPTA][]{mhb+13}. 

PSR B1953+29 is also consistent with the spin noise predictions of a steep red noise process.  However, as shown in Figure \ref{fig:summary-B1953+29}, this pulsar lacks dual-frequency data early in the timing campaign, which is likely a strong contributor to the measure red noise.

Both PSRs J1643$-$1224 and  J1910$+$1256 were identified as displaying strong evidence of red noise in \cite{dfg+13}, and PSR J1643$-$1224 had a significant $\ddot{\nu}$ in \cite{mhb+13}. In the case of PSR J1643$-$1224 the shallow red noise process may be  
due to uncorrected ISM effects that include scattering and refraction  
\citep[e.g.,][]{ric90, fc90, cs10} --
as can be seen in Figure \ref{fig:summary-J1643-1224}, there is a clear
dependence of the noise on radio frequency. The red noise present in the
residuals of PSR J1910$+$1256 is likely caused by  DM variations due to
the fact that we only have single-frequency observations for the first
four years of the data set.  DM variations for a Kolmogorov spectrum
would give an $f^{-8/3}$ spectrum of TOA variations but this can be
altered by linear changes in DM from large-scale structures or changes
in pulsar distances from their line-of-sight motions.  This is further
indicated by inspection of Figure \ref{fig:summary-J1910+1256} where the
timing residuals appear relatively white after dual frequency observations had begun.

PSR B1937$+$21 displays the strongest red noise in our sample,
consistent with previous work that shows a large amount of red noise
\citep[e.g.,][]{sc10, 2013ApJ...766....5S}.  Unlike previous work, which indicates a steep, red spectrum
our analysis shows a shallower spectrum. 
Although it has a large cubic trend, the shallow red noise spectral
index measurement indicates that there are still high frequency trends
in the data. In fact, in Figure \ref{fig:summary-B1937+21} we can see
that the red noise seems to track the DM changes around 2011.5 --
2012.5. This feature, along with the large DM (71 pc cm$^{-3}$), suggest that unmodeled ISM effects may contribute to the observed red noise, particularly at high frequencies, resulting in a lower measured spectral index. An
additional explanation for the much shallower spectral index is that we
are analyzing only the most recent nine years of data on this pulsar
whereas previous analyses have used a much longer time span (SC10 uses
up to 24 years of data) not encompassing this new data. This indicates
that the noise could be non-stationary in nature. PSR J1903$+$0327 has a
similar feature around the same time in which a large drop in the DM
coincides with a peak in the red noise. Once again, this effect, in
combination with the very large DM (297.6 pc cm$^{-3}$) indicate ISM
effects as opposed to intrinsic instability as the cause of the red
noise. 

The measured red noise in PSR J1944$+$0907 is, likely due
to unmodeled DM or scattering/refraction effects since there is only
single-frequency data for a large portion of the data set.

As shown in Figure \ref{fig:red-comp}, PSRs J0613$-$0200 and
J1012$+$5307 also display low spectral-index red noise; however, although there are clear high frequency fluctuations in the residuals (Figs \ref{fig:summary-J0613-0200}, \ref{fig:summary-J1012+5307}) there are no obvious radio frequency dependent features present. Therefore, it is difficult to assess the cause of this measured red noise for these pulsars. Nonetheless, these results are very consistent with \cite{ltm+15} (square marker in Figure \ref{fig:red-comp}) where a nearly identical noise model was used for these two pulsars observed with EPTA telescopes. This at the very least can rule out any instrumental effects.

It was also pointed out in \cite{mhb+13} that PSR J1909$-$3744 displayed
some evidence of red noise for the PPTA and EPTA data sets. While our analysis of PSR J1909-3744 did note find sufficient red noise to classify it as a detection (Bayes factor greater than 100; Table \ref{tab:timingpar}), the posterior probability distributions for this pulsar hint at the presence of weak red noise, again with a shallow spectral index. This is interesting in that PSR J1909$-$3744 has very good timing precision and is ideal for GW detection prospects. Future longer data sets will test whether this pulsar truly displays red noise.

Finally, we point out PSRs J1600$-$3053 and J1747$-$4036. While not
displaying evidence of red noise, we do see non-white features in the
residuals that are radio frequency dependent. Time varying DM corrections were included for the full range of both of these datasets indicating that while the noise is radio frequency dependent, it is not likely a DM effect. 

The noise model used for this data release provides a good fit to the data for most of the pulsars in our data set. However, the model does not accommodate time-variable, chromatic phenomena other than DM variations with their $\nu^{-2}$ dependence on radio frequency.  Such phenomena may include frequency dependent dispersion measures  \citep{css15}  or scattering, and presently such phenomena are imprecisely absorbed by the red-noise and short-term correlated noise models.  As the timespan of wide-band millisecond pulsar data sets grows, it will become practical to incorporate such phenomena into the noise model.

\section{Summary and Conclusions}
\label{sec:conclusions}

In summary, we have obtained, reduced, analyzed, and made public pulse
times of arrival for 37 millisecond pulsars, using two telescopes, over
a time span of up to nine years.  A major upgrade in backend
instrumentation occurred midway through the data set; we developed a
novel method for measuring the instrumental offset between these systems
that removes the need to fit this effect using the TOA data.  We have
continued to develop and refine methods for characterizing time-variable
dispersion measure and frequency-depenent pulse shape evolution, while
fitting phyiscal timing models to these data.  A significant new
development is the parameterized noise model presented in this paper,
and its inclusion in the timing model fit via a generalized least squares
procedure.  Our noise modelling has indicated the presence of
time-correlated, or red, noise in 10 of these pulsars;  we suspect a
combination of propagation effects in the interstellar medium and
intrinsic spin noise both contribute to these detections, with levels of
each varying on a case-by-case basis.

The primary scientific motivation for this project is to detect or limit
the presence of nanohertz-frequency gravitational radiation by looking
for correlated timing fluctuations amongst this set of pulsars.  While
the analysis presented here deals with each pulsar separately,
subsequent papers in this series will perform correlation analyses to
look for the effect of different gravitational wave signal types.  These
include the stochastic background from supermassive black hole binaries
and/or cosmic strings; continuous-wave emission from individual binary
systems; and gravitational wave bursts with memory following merger
events.

In addition to the gravitational wave analyses just mentioned, a number
of additional topics are planned to be addressed in future papers,
including:  Detailed investigation of pulse jitter and other sources of
noise in these data; measurement of orbital parameters, pulsar and
companion masses, and relativistic orbital effects; the effect of
scattering on the timing results; pulsar astrometry, distance
measurements and kinematics; analysis of the polarization properties of
the pulse profiles; flux densities and population analysis; and
reanalysis of these data using wide-band timing methodologies.  Future
improvements to the data set presented here include ongoing increase in
the number of pulsars measured, increased cadence on several of the best
pulsars in the set, and inclusion of archival Arecibo data covering up
to 20 years total timespan.

\acknowledgements 
{\it Author contributions.}  An alphabetical-order author list was used
for this paper in recognition of the fact that a large, decade-timescale
project such as this is necessarily the result of the work of many
people. All authors contributed to the activities of the NANOGrav
collaboration leading to the work presented here, and reviewed the
manuscript text and figures prior to the paper's submission.  Additional
specific contributions to this paper are as follows.
ZA, KC, PBD, TD, RDF, EF, MEG, GJ, MJ, MTL, LL, MAM, DJN, TTP, SMR, IHS,
KS, JKS, and WWZ made observations for this project and developed timing
models (\S\ref{sec:timing}).
PBD co-developed the GUPPI and PUPPI instruments, wrote observing
proposals, coordinated GBT observations, performed the calibration and
TOA generation, developed and implemented the offset-measurement
analysis, developed and implemented the FD profile evolution model,
analyzed TOA accuracy in the low-signal-to-noise-regime, compiled the
data files for public release, coordinated the writing of the paper, and
contributed substantially to the text, figures, and tables.
JAE developed and implemented the noise model, used it to analyze the
present data set, analyzed the red-noise results, co-developed the
daily-average residual statistic and implemented it for the residual
figures, wrote substantial amounts of the text, and contributed tables
and figures.
DJN co-authored observing proposals, coordinated the NANOGrav Timing
Working Group, wrote portions of the text, and contributed a figure and
table.
TTP undertook careful analysis of the residuals leading to the
implementation of the signal-to-noise cutoff.
IHS coordinated the Arecibo observations and wrote portions of the text.
SMR co-developed the GUPPI and PUPPI instruments and wrote portions of
the text.
RvH contributed to the noise model and analysis and verified it using
independent code and co-developed the daily-average residual statistic.
EF wrote portions of the text.
ZA, JMC, TD, and others edited the manuscript text and/or provided
significant input about its content.

The NANOGrav project receives support from National Science Foundation
(NSF) PIRE program award number 0968296 and NSF Physics Frontier Center
award number 1430284.  NANOGrav research at UBC is supported by an NSERC
Discovery Grant and Discovery Accelerator Supplement and the Canadian
Institute for Advanced Research.  Part of this research was carried out
at the Jet Propulsion Laboratory, California Institute of Technology,
under a contract with the National Aeronautics and Space Administration.
TTP is a student at the National Radio Astronomy Observatory.  The
National Radio Astronomy Observatory is a facility of the NSF operated
under cooperative agreement by Associated Universities, Inc.  The
Arecibo Observatory is operated by SRI International under a cooperative
agreement with the NSF (AST-1100968), and in alliance with Ana G.
M\'{e}ndez-Universidad Metropolitana, and the Universities Space
Research Association. Some computational work was performed on the Nemo
cluster at UWM supported by NSF grant No. 0923409. JAE and RvH
acknowledge support by NASA through Einstein Fellowship grants
PF4-150120 and PF3-140116, respectively.\\

\appendix

\section{Timing Offset Determination}
\label{sec:offset}

A common problem in long-term pulsar timing studies is to connect the
timing results between multiple generations of instrumentation at a
given telescope.  Single-dish pulsar timing data is generally
timestamped either at the point where the radio signal is digitized, or
somewhat later at the output of a filterbank, when the data are received
by software systems.  Prior to this point, the signal accumulates
additional latency as it passes through various telescope electronics
systems.  These can include analog cable delays (tens to hundreds of
ns), transmission over long fiber optic links (tens of $\mu$s), and
filter latencies (up to a few $\mu$s).  Since only variations in pulse
phase -- not its absolute value -- are physically meaningful in timing
analyses, the presence of time offsets like these is not a problem, as
long as they are constant.  However, when a new backend instrument is
added, the delay values (e.g., cabling, filters) typically differ from
the previous version.  As described in \S\ref{sec:toa}, this time offset
must be accounted for in order to measure long-timescale
effects.  

A typical approach used in most past timing analyses is to measure
offsets between instruments using pulsar TOAs \citep[e.g.,][]{tw89}.  An
arbitrary offset can be included as an additional term (known as a
``JUMP'' in \tempo/\tempotwo) in a timing model fit (see
\S\ref{sec:timing}).   Although this is a straightforward approach, it
has several drawbacks:  First, in the presence of unmodelled red noise,
this can introduce systematic biases in other parameters \citep[see for
example discussion in][]{chc+11}; this can be mitigated via improved
noise modelling as described in \S\ref{sec:noise} and references
therein.  Next, even if the noise is properly modelled, the offset will
be covariant with other long-term effects, most critically with
low-frequency gravitational waves, reducing sensitivity to these
effects.  A refinement to this is to restrict the offset measurement to
only a shorter, overlapped span of data between the two instruments,
measure it in a separate fit procedure, and hold the resulting value
fixed in the main timing model fit.  This will reduce covariance with
long-term effects, but raises concern that the effect of the offset fit
may not be fully accounted for in the other model parameter
uncertainties.  In both cases, the precision of the measurement is
limited by the relevant TOA uncertainties.  It is sometimes possible to
transfer an offset measurement done using one bright pulsar to other
sources, although there may then be concerns about potential
systematics, for example due to pulse shape evolution with frequency,
calibration inaccuracies, or different instrument configurations used
for different sources.

An alternate approach was used recently by \citet{mhb+13}, wherein a
locally-generated pulsed signal was injected into the common signal path
and used to recover the offsets between different systems with much
higher precision than could be achieved using astronomical signals.
This mitigates all of the problems of TOA-based approaches described
above.  The only drawbacks are that it requires additional
special-purpose hardware be built and installed at the observatory, and
that it can not be applied retroactively -- once an instrument has been
decommissioned it is no longer possible to perform this measurement.  In
contrast, simultaneous (or at least contemporaneous) pulsar data is
often still available in archival data sets long after the relevant
instruments are gone.

We have developed a new method that addresses many of the shortcomings
described above.  This is based on the fact that for observations that
are {\it simultaneous in both time and frequency} (i.e., where a single
signal is split and fed into multiple backend systems), both instruments
see not only the same pulsar signal, but also the same system noise.
This correlated noise can be used to recover a time offset with much
higher precision than is possible from TOA-based methods.  TOA
determination can be viewed as a matched filtering process designed to
recover the template profile shape.  By construction, this filters out a
large fraction of the noise that could otherwise be used for an offset
measurement.  In our method, rather than cross-correlating measured
pulse profiles with a noise-free template, pairs of simultaneous pulse
profiles from each instrument are cross-correlated with each other.  In
contrast with TOA measurement described in \S\ref{sec:toa}, it is
advantageous to average the profiles as little as possible prior to this
step, to preserve more (correlated) noise.  With the profile data at its
original time and frequency resolution, we compute cross-correlations
between all pairs of profiles that overlap in both time and frequency.
The cross-correlations from all simultaneous profile pairs for a given
pulsar, instrument setup, and epoch, are then averaged together (with a
weight proportional to the amount of time-frequency overlap) to form a
final correlation function.  The lag for which this is maximized results
in an offset estimate for that portion of the data.  We use this set of
individual offset measurements to determine an uncertainty on the
average offset, and look for systematic trends as a function of pulsar,
time, or instrument setup.

\begin{table}[tp]
\caption{\label{tab:offsets} Measured instrumental offsets.$^a$}
\begin{center}
\begin{tabular}{c|l|ll}
\hline
Receiver & Cross-corr  & J1713$+$0747 & J1909$-$3744 \\
system   & offset (ns) & JUMP (ns)    & JUMP (ns)    \\
\hline
Arecibo 327    & 785(19) &  -        & -        \\
Arecibo 430    & 789(5)  &  -        & -        \\
Arecibo L-wide & 839(3)  &  820(75)  & -        \\
Arecibo S-wide & 846(6)  &  885(82)  & -        \\
GBT Rcvr\_800  & 897(8)  &  951(124) & 936(42)  \\
GBT Rcvr1\_2   & 693(3)  &  599(86)  & 651(55)  \\
\hline
\multicolumn{4}{l}{$^a$Numbers in parentheses are uncertainties in the last digit quoted.}
\end{tabular}
\end{center}
\end{table}

\begin{figure}[tp]
\centering
\includegraphics[width=\columnwidth]{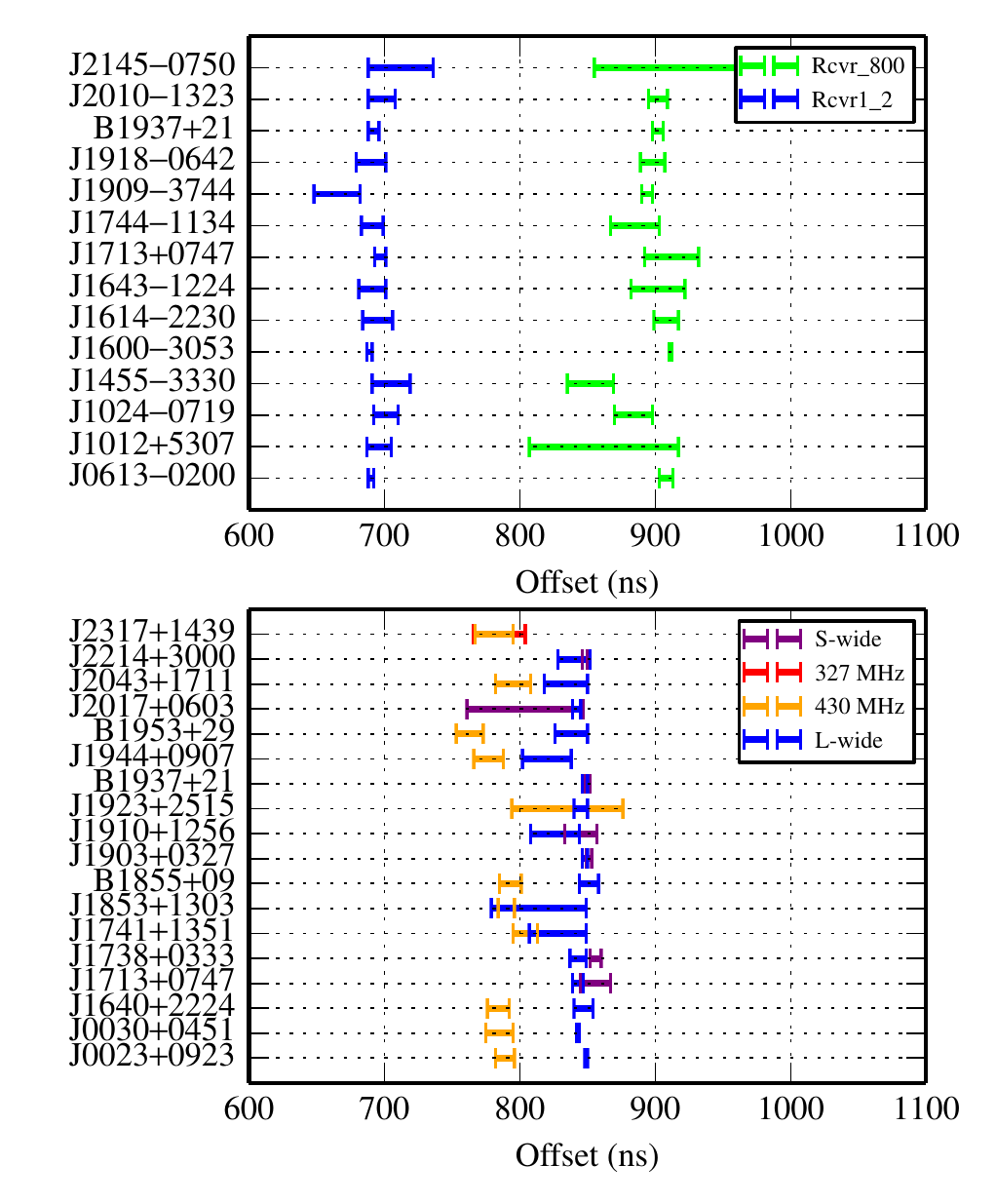}
\caption{Measured instrumental offsets versus pulsar for the GBT (top) and
Arecibo (bottom).}
\label{fig:offsets}
\end{figure}

For this work, we are interested in the offset between
GASP and GUPPI at the GBT and the offset between ASP and PUPPI
at Arecibo.  We analyzed all available simultaneously
collected profiles with overlapping frequency bands from these
pairs of instruments.  The results of this offset analysis are shown in
Table~\ref{tab:offsets} and Figure~\ref{fig:offsets}.  After accounting
for all a priori predictable latencies in the backend systems, there
remains $\sim$700--900~ns additional offset between ASP/GASP and
GUPPI/PUPPI that varies only as a function of signal path and instrument
bandwidth.  Table~\ref{tab:offsets} lists the average value of all data
available for each receiver system, while Figure~\ref{fig:offsets} shows
the same data averaged separately for each pulsar.  At Arecibo, we
obtain consistent results for the 327 and 430~MHz setups, as well as for
the L-wide and S-wide setups.  This result is expected since these pairs
of setups share common analog signal paths and PUPPI bandwidth (100~MHz
and 800~MHz respectively).  At Green Bank, the 820~MHz and 1.4~GHz
receivers have different signal paths to GUPPI, and the instrument is
run at different bandwidth (200~MHz and 800~MHz).  It is likely that
both of these factors contribute to the observed offset difference.  The
sign of these values is such that ASP/GASP pulses arrive later -- the
offsets must be {\it subtracted} from these TOAs to align them with the
GUPPI/PUPPI data.  As can be seen in Figure~\ref{fig:offsets},
consistent values are obtained from all pulsars for a given receiver
system, therefore in our timing analysis we have applied the
overall-average (Table~\ref{tab:offsets}) values to the TOAs.  In our
data set this is provided via a time offset (``-to'') flag on each TOA
line.  

As a check on these results, we performed a standard timing analysis on the
overlapping TOAs of two pulsars, in each case fitting for an offset between
the TOAs from the two different instruments used as part of the timing
solution.  The results are shown in Table~\ref{tab:offsets}, where they are
labeled JUMP (the \tempo parameter used for this offset measurement).  These
values  illustrate that our noise correlation provides both a consistent and
much more precise result.  For most other pulsars, TOA uncertainties are
larger than for the pulsars used in Table~\ref{tab:offsets},
hence the JUMP uncertainties are larger as well.

\section{TOAs in the Low-S/N Limit}
\label{sec:snr}

In the very low signal-to-noise ratio regime, the standard template
matching procedure breaks down, producing underestimated TOA
uncertainties.  In addition, the distribution of TOAs in this regime
becomes significantly non-Gaussian.  Here we derive expressions for the
expected TOA probability distribution, and motivate our choice of S/N
cutoff for TOAs.  The use of a S/N or TOA uncertainty cutoff, or simply
``by-eye'' removal of outlier residuals, is often done in pulsar timing
analyses.  The discussion in this section provides a somewhat more
rigorous and quantitative justification for this practice.  The behavior
of TOA uncertainties in the low-S/N limit was previously explored
empirically by \citet{hbo05a} using simulated data, who reach similar
conclusions to what we present here.

We follow the standard Fourier-domain least squares TOA determination
approach of \citet{tay92} \citep[see also][]{dem07}, writing the
expression for $\chi^2$ as a function of fitted amplitude $a$ and pulse
phase shift $\phi$ as
\begin{equation}
\begin{split}
  \chi^2(a,\phi) &= \sum_k 
      \frac{\left|d_k - at_ke^{-2\pi i k \phi}\right|^2}{\sigma^2} \\
    &= \frac{D^2 + a^2 T^2 - 2aC_{dt}(\phi)}{\sigma^2},
\end{split}
\end{equation}
where these terms come from the discrete Fourier transform of the measured
pulse profile ($d_k$) and template profile ($t_k$):
\begin{eqnarray}
  D^2 &\equiv& \sum_k \left|d_k\right|^2 \\
  T^2 &\equiv& \sum_k \left|t_k\right|^2 \\
  C_{dt}(\phi) &\equiv& \Re \sum_k d_k t_k^* e^{2\pi i k \phi}.
\end{eqnarray}
Here $\sigma^2$ is the noise level in each bin of $d_k$, and the sum is
over pulse harmonics, not including the constant (DC) term.  All pulse
phase information is contained in the $C_{dt}$ term, illustrating why
TOA determination is sometimes described as a cross-correlation between
the data and template profiles -- the minimum $\chi^2$ is always
achieved at the phase shift giving maximum cross-correlation.  With the
assumption of additive Gaussian noise (implicit in a $\chi^2$ fit), the
TOA likelihood function is
\begin{equation}
        p(d|a,\phi) \propto e^{-\frac{1}{2}\chi^2(a,\phi)}
        = \exp\left(
                \frac{2aC_{dt}(\phi)-D^2-a^2T^2}{2\sigma^2}
        \right),
\end{equation}
and with the use of uniform priors on $a$ and $\phi$, the posterior
distribution is simply proportional to the likelihood, $p(a,\phi|d)
\propto  p(d|a,\phi)$.  For TOA determination, $a$ is a nuisance
parameter, which can be analytically marginalized over in the above
expression to get the posterior $\phi$ distribution
\begin{equation}
\label{eqn:toa_pdf}
  p(\phi|d) \propto \exp \left( \frac{C^2_{dt}(\phi)}{2\sigma^2
      T^2}\right).
\end{equation}

\begin{figure}[t]
\centering
\includegraphics[width=\columnwidth]{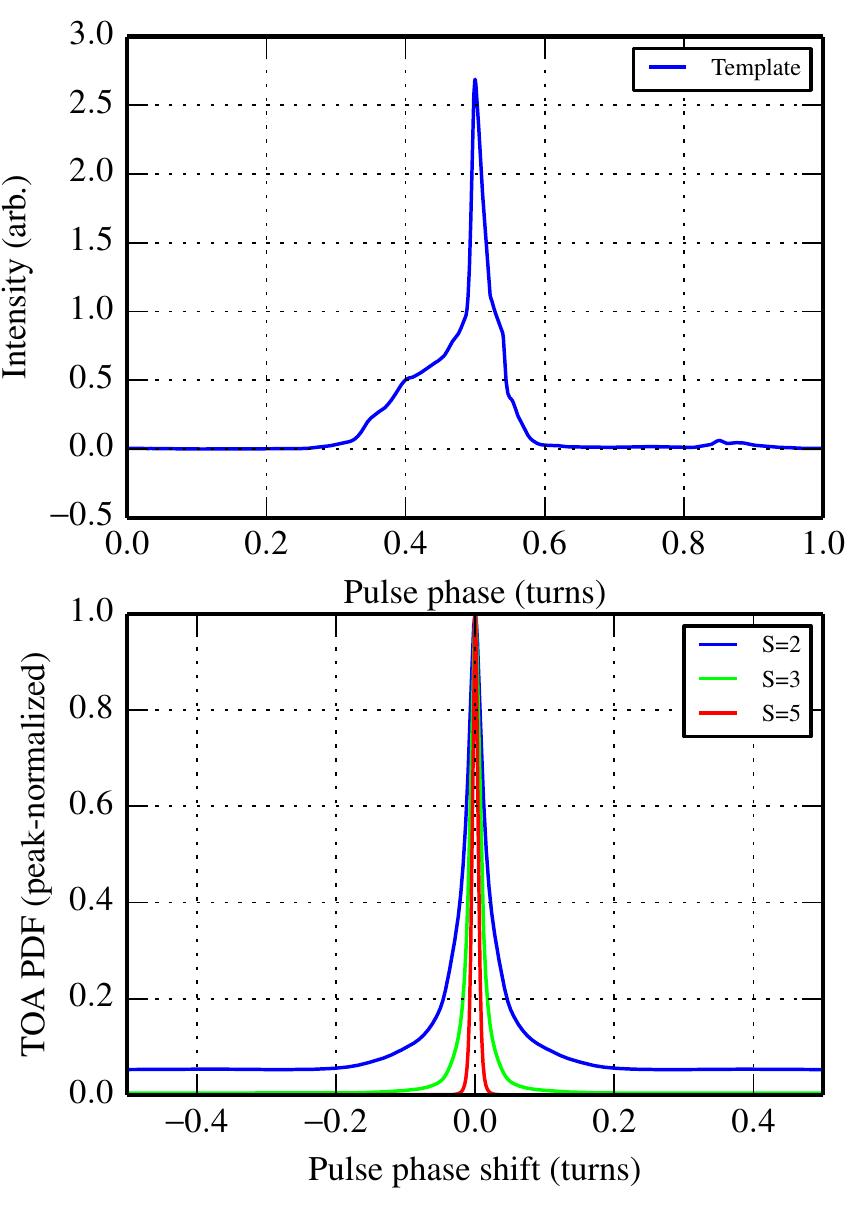}
\caption{{\it Upper panel:} 1.4~GHz template profile for J1455$-$3330.
{\it Lower panel:} Expected pulse phase shift distributions for several
values of S/N ratio, from Eqn.~\ref{eqn:toa_pdf_exp}.  For plot clarity
these are normalized to 1 at $\phi=0$, rather than to constant
integrated area.  This shows the evolution of the distribution from
nearly Gaussian at higher S/N ratio ($S=5$) to clearly non-Gaussian at
low-S/N ($S=2$). \label{fig:toa_pdf}}
\end{figure}

\begin{figure}[t]
\centering
\includegraphics[width=\columnwidth]{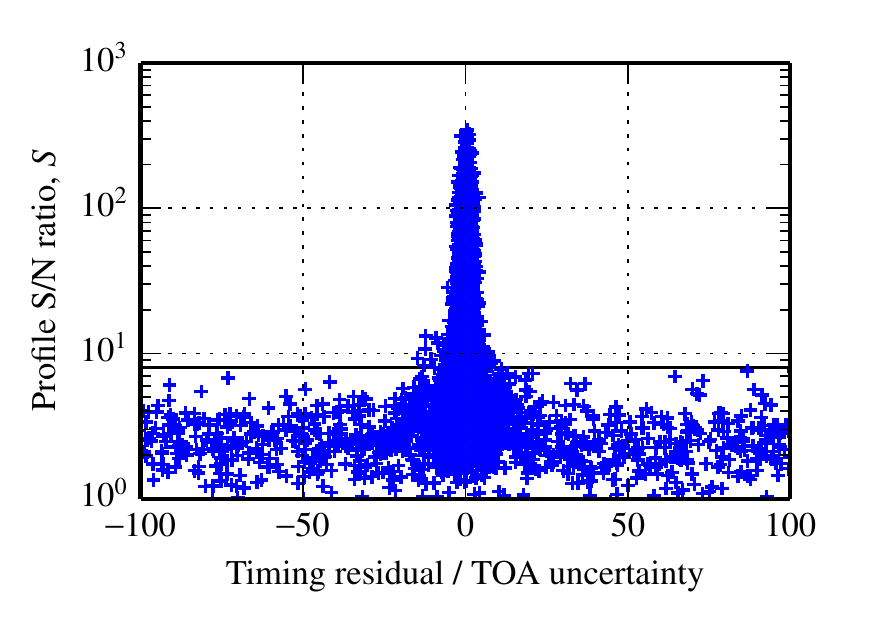}
\caption{Signal-to-noise ratio $S$ versus normalized timing residual
(residual divided by TOA uncertainty) for J1455$-$3330 L-band data.  All
data points below the solid line at $S=8$ were removed from the timing
analysis.}
\label{fig:resid_snr}
\end{figure}

By making the substitution $d_k \to at_k$ -- i.e., the data profile is
simply a scaled copy of the template -- we can explore the expected
shape of these distributions independent of any particular (noisy) data
realization.  In this case Eqn.~\ref{eqn:toa_pdf} becomes
\begin{equation}
\label{eqn:toa_pdf_exp}
  p(\phi) \propto \exp \left( 
          \frac{S^2}{2}
          \frac{C^2_{tt}(\phi)}{T^4}
      \right),
\end{equation}
where $S \equiv aT/\sigma$ defines the signal-to-noise ratio of the
data, and $C_{tt}$ is the template profile's autocorrelation (with
normalization $C_{tt}(0) = T^2$).  For non-detections ($S \to 0$ limit),
$p(\phi)$ becomes a uniform distribution across one turn of phase.  For
high-S/N detections ($S \gtrsim 10$), $p(\phi)$ becomes extremely well
approximated by a Gaussian, with standard deviation given by the usual
template-matching TOA uncertainty formula 
\begin{equation}
\label{eqn:toa_err}
\sigma_\phi = S^{-1} T \left( C_{tt}^{\prime\prime}(0) \right)^{-1/2}.
\end{equation}
In the low-S/N regime between these two limits, the standard uncertainty
formula underestimates the true spread of TOA values and signifcant
non-Gaussianity is present.  We illustrate this in
Figures~\ref{fig:toa_pdf} and \ref{fig:resid_snr} using data from
PSR~J1455$-$3330.  This pulsar provides a clear demonstration of this
effect, because its wide scintillation bandwidth and moderate average
flux result in profiles with a large range of S/N values in our dataset.
If included in a standard $\chi^2$-based timing model fit, the low-S/N
points appear as outliers and have a disproportionately large impact on
the results.  While it would be possible to mitigate this by a
modification of the TOA uncertainties or use of a timing model
likelihood based on Eqn.~\ref{eqn:toa_pdf}, in our data set the amount
of additional information gained from these data points is likely to be
marginal at best.  Instead, for the work presented in this paper we have
simply removed all TOAs with $S<8$.

\section{Derivation of Noise Model Likelihood}
\label{sec:noise_app}

We begin by forming a set of residuals via the standard weighted least squares fitting routine. An $N_{\rm TOA}$ length vector of  residuals can be modeled mathematically as the sum of several deterministic and stochastic sources as follows
\be
\delta\mathbf{t} = M\boldsymbol{\epsilon} + F\mathbf{a} + U\mathbf{j} +\mathbf{n}.
\ee
The first term on the right hand side ($M\boldsymbol{\epsilon}$)
describes small deterministic trends due to timing model subtraction.
Here $M$ is the timing model design matrix and $\boldsymbol{\epsilon}$
is a vector of small timing model parameter offsets. Next, the term
$F\mathbf{a}$ models the red noise via a Fourier
decomposition\footnote{The Fourier basis was chosen to improve
computational efficiency;  it is not a requirement of this noise
modeling method.} -- $F$ is the Fourier design matrix that has columns
of alternating sine and cosine functions for frequencies in the range
$[1/T,n_{\rm mode}/T]$;  $T$ is the total observation time span, $\Delta
f=1/T$, and $n_{\rm mode}$ is the number of frequencies included in the
sum.  The vector $\mathbf{a}$ gives the amplitudes of the Fourier basis
functions \citep[see][for more details]{lah+14,abb+14}.
The term $U\mathbf{j}$ describes noise that is 
uncorrelated in time but completely correlated between TOAs 
obtained at different frequencies measured simultaneously. This
term could be due to pulse phase jitter but could also have other
components not due to jitter. This term characterizes noise that is
completely correlated for all TOAs in a given time bin but completely
\emph{uncorrelated} between time bins. The matrix $U$ is an $N_{\rm
TOA}\times N_{\rm tb}$ matrix that maps TOAs to a given time bin and
$\mathbf{j}$ is the amplitude of the short time-scale fluctuations.
Finally the last term $\mathbf{n}$ describes a Gaussian white noise
process that characterizes time-, and frequency-independent random
noise left in the data. 

Since the white noise is modeled as Gaussian, the likelihood function for the noise is given by
\be
p(\mathbf{n}) = \frac{\exp\left( -\frac{1}{2}\mathbf{n}^TN^{-1}\mathbf{n} \right)}{\sqrt{\det(2\pi N)}},
\ee
where 
\be
N_{ij,k}=E_{k}^2(W_{ij}+Q_{k}^2\delta_{ij}),
\ee
is an $N_{\rm TOA}\times N_{\rm TOA}$ matrix with $E_{k}$ and $Q_{k}$
corresponding to \tempo and \tempotwo's EFAC and EQUAD parameters for each observing backend, respectively, $W={\rm diag}\{\sigma_i^2\}$ is a diagonal matrix of TOA uncertainties, and $\delta_{ij}$ is the Kronecker delta function. The notation is such that the matrix elements $(i,j)$ apply to those TOAs corresponding to the backend observing system labeled by $k$. We can now write the likelihood function of the residuals as
\be
p(\delta\mathbf{t}|\boldsymbol{\epsilon}, \mathbf{a}, \mathbf{j},\boldsymbol{\phi}) = \frac{\exp\left( -\frac{1}{2}\mathbf{r}^TN^{-1}\mathbf{r} \right)}{\sqrt{\det(2\pi N)}},
\ee
where $\boldsymbol{\phi}$ denotes the $E_{k}$ and $Q_{k}$ parameters and 
\be
\mathbf{r} = \delta\mathbf{t}-M\boldsymbol{\epsilon}-F\mathbf{a}-U\mathbf{j}.
\ee
We now wish to impose prior distributions on our short timescale correlated noise and red noise. Both can be modeled as Gaussian processes by imposing the following priors
\begin{align}
p(\mathbf{j}|J_{k}) &= \frac{\exp\left( -\frac{1}{2}\mathbf{j}^{T}\msJ^{-1}\mathbf{j} \right)}{\sqrt{\det(2\pi\msJ)}}\\
p(\mathbf{a}|\rho_{n}) &= \frac{\exp\left( -\frac{1}{2}\mathbf{a}^{T}\varphi^{-1}\mathbf{a} \right)}{\sqrt{\det(2\pi\varphi)}},
\end{align}
where $\msJ_{ij,k} = J_{k}^{2}\delta_{ij}$ is an $N_{\rm tb}\times N_{\rm
tb}$ matrix with diagonal elements, and $J_{k}^{2}$ describes the variance of
the jitter-like correlated noise for each observing backend; it is also
referred to as the ECORR parameter in \tempo and \tempotwo. Furthermore $\varphi_{ij} = {\rm diag}\{10^{\rho_{n}}\}$ is an $2n_{\rm mode}\times 2n_{\rm mode}$ matrix describing the variance of the red noise Fourier coefficients at each frequency. In this framework, the coefficients of the $\varphi$-matrix are related to the power spectral density evaluated at a given frequency. In principle we could use the power spectrum coefficients, $10^{\rho_{n}}$, themselves as free parameters but in this work we parameterize them via a power law
\be
\varphi_{n} \equiv 10^{\rho_{n}} = \frac{1}{T_{\rm span}} A_{\rm red}^2\left( \frac{f_{n}}{f_{\rm yr}} \right)^{\gamma_{\rm red}}
\ee
where $T_{\rm span}$ is the total observation time, $A_{\rm red}$ is the amplitude of the
red noise in $\mu$s~${\rm yr}^{1/2}$, $\gamma_{\rm red}$ is the spectral index of the red noise, $f_{\rm
yr}$ is the reference frequency of 1~${\rm yr}^{-1}$, and $f_{n}$ is the $n$th
Fourier frequency assuming Nyquist sampling. We see that the prior
distributions on jitter-like correlated noise and red noise are themselves
parameterized by some combination of hyper-parameters. 
We can write down the posterior distribution for the residuals
\be
p(\boldsymbol{\epsilon}, \mathbf{a}, \mathbf{j},\boldsymbol{\phi}|\delta\mathbf{t}) \propto
p(\delta\mathbf{t}|\boldsymbol{\epsilon}, \mathbf{a}, \mathbf{j},\boldsymbol{\phi})p(\mathbf{j}|J_{k})p(\mathbf{a}|\rho_{n}).
\ee

For the purposes of estimating the underlying noise characteristics of our
data set, the parameters $\boldsymbol{\epsilon}$, $\mathbf{j}$, and $\mathbf{a}$ are
nuisance parameters that we wish to marginalize over. This can be done in a
sequential fashion as was presented in \cite{abb+14}, but here we take a
different approach. Notice that all timing parameters are linear and can be described with Gaussian prior distributions\footnote{We use uniform priors on the timing model parameter offsets, $\mathbf{\epsilon}$ but this is the same as a Gaussian prior with infinite variance. Technically this prior is not normalizable, but since we are interested in parameter estimation and not Bayesian model selection here, this non-normalizable prior is not a problem.}. We can thus define a combined operator matrix and amplitude vector
\be
T = \bb M  & F  & U \eb, \quad
\mathbf{b} = \bb \boldsymbol{\epsilon} \\ \mathbf{a} \\ \mathbf{j} \eb
\ee
with prior distribution
\be
p(\mathbf{b}|\boldsymbol{\phi}) = \frac{\exp\left( -\frac{1}{2}\mathbf{b}^{T}B^{-1}\mathbf{b} \right)}{\sqrt{\det(2\pi B)}}
\ee
and covariance matrix defined in terms of the block matrix
\be
B = \bb \infty & 0 & 0 \\ 0 & \varphi & 0 \\ 0 & 0 & \msJ \eb,
\ee
where $\infty$ is a diagonal matrix of infinities to describe a uniform
prior on $\boldsymbol{\epsilon}$. The resulting likelihood function is
then
\be
p(\delta\mathbf{t}|\mathbf{b}) = \frac{\exp\left[ -\frac{1}{2}{(\delta\mathbf{t}-T\mathbf{b})}^TN^{-1}{(\delta\mathbf{t}-T\mathbf{b})} \right]}{\sqrt{\det(2\pi N)}}.
\ee
The marginalized posterior distribution is then
\be
\begin{split}
p(\boldsymbol{\phi}|\delta\mathbf{t}) & \propto \int_{-\infty}^{\infty}d\boldsymbol{\epsilon}\,d\mathbf{a}\,d \mathbf{j} \,p(\boldsymbol{\epsilon}, \mathbf{a}, \mathbf{j},\boldsymbol{\phi}|\delta\mathbf{t}) \\
&= \int_{-\infty}^{\infty}d\mathbf{b} p(\delta\mathbf{t}|\mathbf{b})p(\mathbf{b}|\phi)p(\boldsymbol{\phi}) \\
&= \frac{\exp\left[-\frac{1}{2}(\delta\mathbf{t}^{T}N^{-1}\delta\mathbf{t} - \mathbf{d}^{T}\Sigma^{-1}\mathbf{d})  \right]}
{\sqrt{(2\pi)^{N_{\rm TOA}-{\rm dim\,}\mathbf{b}}\det(N)\det(B) \det(\Sigma)}},
\end{split}
\ee
where
\begin{align}
\mathbf{d} &= T^{T}N^{-1}\delta\mathbf{t} \\
\Sigma &= (B^{-1} + T^{T}N^{-1}T).
\end{align}
The maximum likelihood values of $\mathbf{b}$ and their uncertainties can be found as
\begin{align}
\hat{\mathbf{b}} &= \Sigma^{-1}\mathbf{d} \\
{\rm cov}(\mathbf{b}) &= \Sigma^{-1}.
\end{align}
This scheme has the advantage of being computationally efficient in that it
bypasses $O(N_{\rm TOA}^{3})$ matrix operations via rank reduced matrices
\citep{vhv15} resulting in a likelihood evaluation that instead scales as
$O(N_{\rm par}^{3})$, where $N_{\rm par}$ is the sum of the number of timing
parameters, red noise sample frequencies, and observing time bins. For the largest
datasets the computational speed up is a factor of $\sim10^{3}$.

For a given set of hyper-parameters, this allows us to determine the maximum
likelihood timing model parameters  and the maximum likelihood red noise realization present in the data
via the equivalent of a generalized least squares
 fit. 
We can also evaluate the posterior of the hyper-parameters $\boldsymbol{\phi}$
and thus find the maximum likelihood noise parameters including the EFAC, EQUAD,
ECORR, red noise amplitude $A_{\rm red}$ and spectral index $\gamma_{\rm
red}$.
The posterior distributions of the noise parameters are sampled using a Markov Chain Monte
Carlo process in which we sample some parameters in $\log_{10}$ space and
limit them to $\log_{10}J_{k}\in[-8.5, -4]$, with $J_{k}$ in units of seconds,
$\log_{10}A_{\rm red}\in[-7.5,1.5]$ where $A_{\rm red}$ is in units of
$\mu$s ${\rm yr}^{1/2}$, and $\gamma_{\rm red}\in[0, 7]$.

\section{Daily Averaged Residuals}
\label{sec:daily_avg}
For modern wide-band timing campaigns using multi-channel TOAs it becomes useful to visually inspect timing residuals that have been averaged in order to look for long term trends or biases. Here we derive a robust weighted average that will fully account for short timescale correlations introduced by the ECORR in our noise models. This is important since ECORR is meant to model pulse phase jitter, thus when constructing daily averaged residuals, one must include this effect as it results in larger averaged uncertainties on the averaged residuals. In essence this allows for a way to visually determine which pulsars may be dominated by pulse phase jitter.

We begin the derivation by introducing the probability distribution of
the group of residuals that belong to time bin\footnote{In this work, we
have used time bins of size 1 second, thus are only averaging
sets of multi-channel residuals measured simultaneously.} $k$,
\be
p(\delta\mathbf{t}_{k}|\bar{\delta t}_{k}) = \frac{\exp[-\frac{1}{2}(\delta\mathbf{t}_{k}-O\bar{\delta t}_{k})^{T}C_{k}^{-1}(\delta\mathbf{t}_{k}-O\bar{\delta t}_{k})]}{\det(C_{k})},
\ee
where $\delta\mathbf{t}_{k}$, $\bar{\delta t}_{k}$, and $C_{k}$ are the
residuals in time bin $k$, the mean residual in time bin $k$, and the covariance matrix of the residuals in time bin $k$, respectively. Here, $O$ is the design matrix for the mean which in this case is a vector of ones of length $N_{k}$, where $N_{k}$ is the number of residuals in simultaneous time bin $k$. In an identical manner as Appendix \ref{sec:noise_app} we can determine the maximum likelihood estimator and uncertainty for the mean of the probability distribution function (i.e. the daily averaged residual)
\begin{align}
\bar{\delta t}_{k}^{\rm ML} &= (O^{T}C_{k}^{-1}O)^{-1}O^{T}C_{k}^{-1}\delta \mathbf{t}_{k}\\
\bar{\sigma}_{k}^{2} &= (O^{T}C_{k}^{-1}O)^{-1},
\end{align}
where $\bar{\sigma}_{k}$ is the weighted uncertainty on the daily averaged residual. Note that if $C_{k}$ is diagonal with elements corresponding to the TOA uncertainties then we obtain our usual expression for the weighted mean and standard deviation
\begin{align}
\bar{\delta t}_{k}^{\rm ML} &= \frac{\sum_{i=1}^{N_{k}}\delta\mathbf{t}_{i,k}\sigma_{i,k}^{-2}}{\sum_{i=1}^{N_{k}}\sigma_{i,k}^{-2}}\\
\bar{\sigma}_{k}^{2} &= \left( \sum_{i=1}^{N_{k}}\sigma_{i,k}^{-2} \right)^{-1},
\end{align}
where $\sigma_{i,k}$ is the TOA uncertainty for the $i\th$ TOA in simultaneous time bin $k$. We note that the ECORR will add  to the off-diagonal components of $C_{k}$ and can have a large impact depending on the relative strength of ECORR compared to the radiometer noise component.

\bibliographystyle{apj}
\bibliography{journals_apj,nano9y,modrefs,psrrefs,crossrefs}

\newpage

\begin{figure*}[p]
\centering
\includegraphics[scale=1.0]{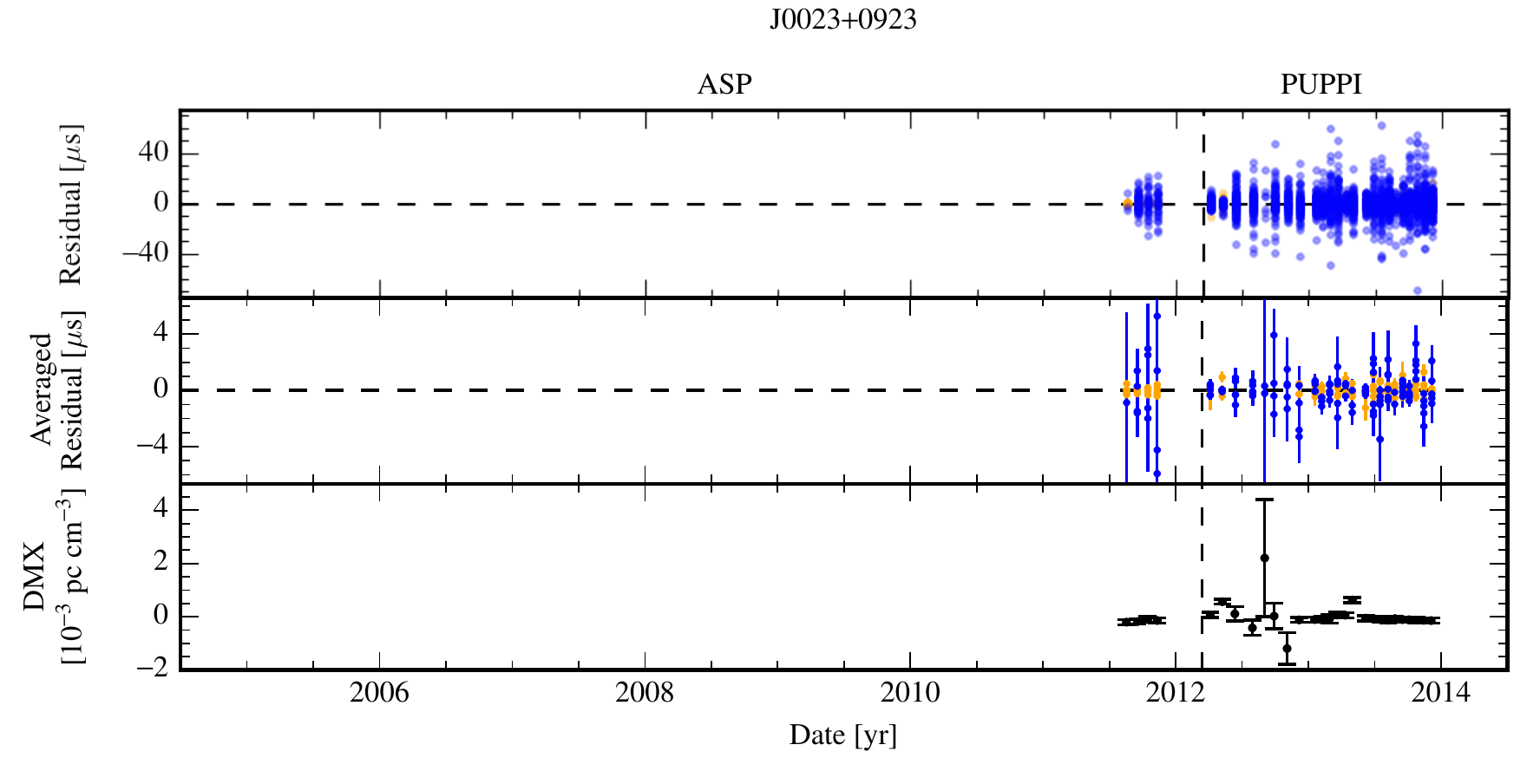}
\caption{Timing summary for PSR J0023+0923. Colors are: Blue: 1.4 GHz, Purple: 2.3 GHz, Green: 820 MHz, Orange: 430 MHz, Red: 327 MHz. In the top panel, individual points are semi-transparent; darker regions arise from the overlap of many points.}
\label{fig:summary-J0023+0923}
\end{figure*}

\begin{figure*}[p]
\centering
\includegraphics[scale=1.0]{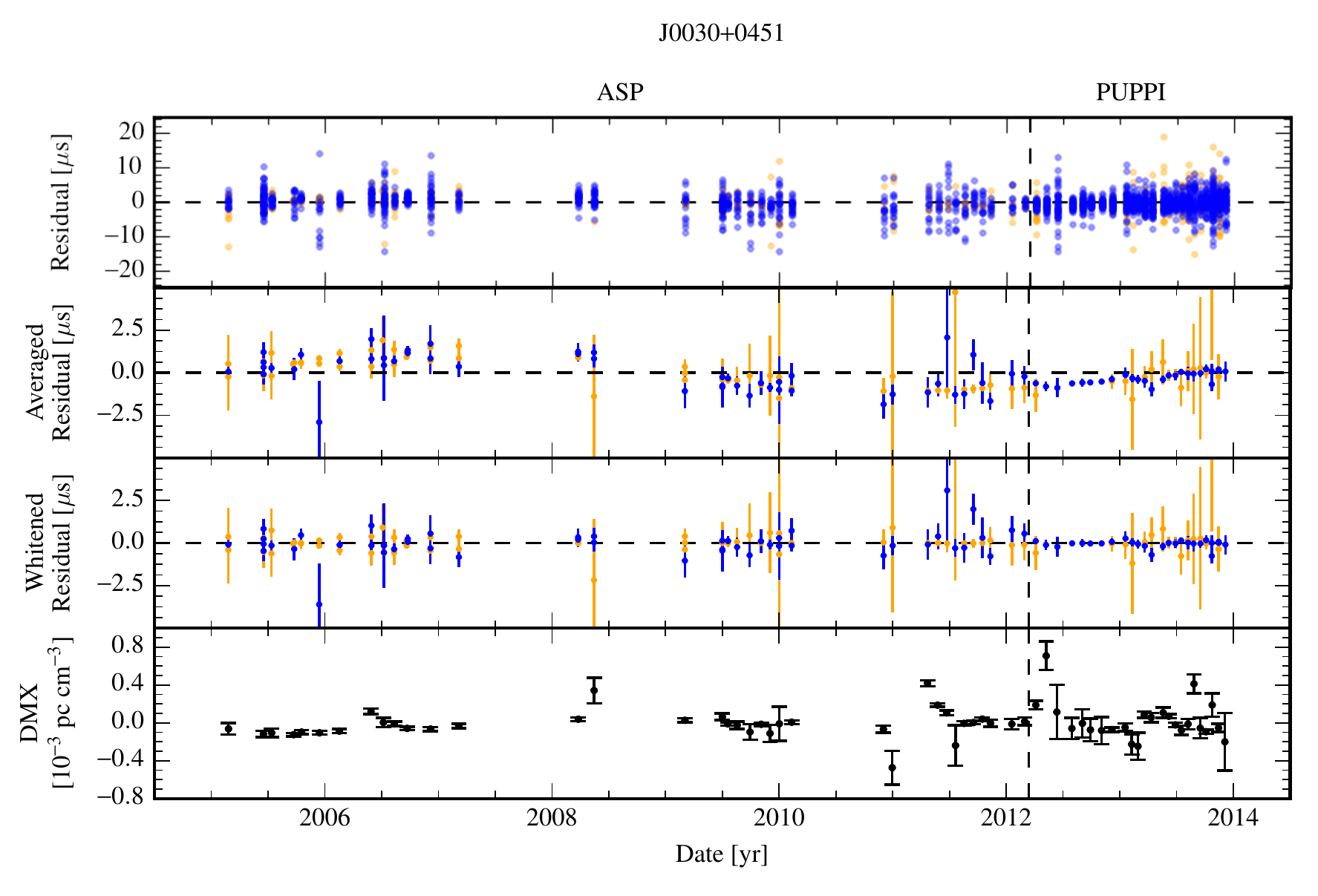}
\caption{Timing summary for PSR J0030+0451. Colors are: Blue: 1.4 GHz, Purple: 2.3 GHz, Green: 820 MHz, Orange: 430 MHz, Red: 327 MHz. In the top panel, individual points are semi-transparent; darker regions arise from the overlap of many points.}
\label{fig:summary-J0030+0451}
\end{figure*}

\begin{figure*}[p]
\centering
\includegraphics[scale=1.0]{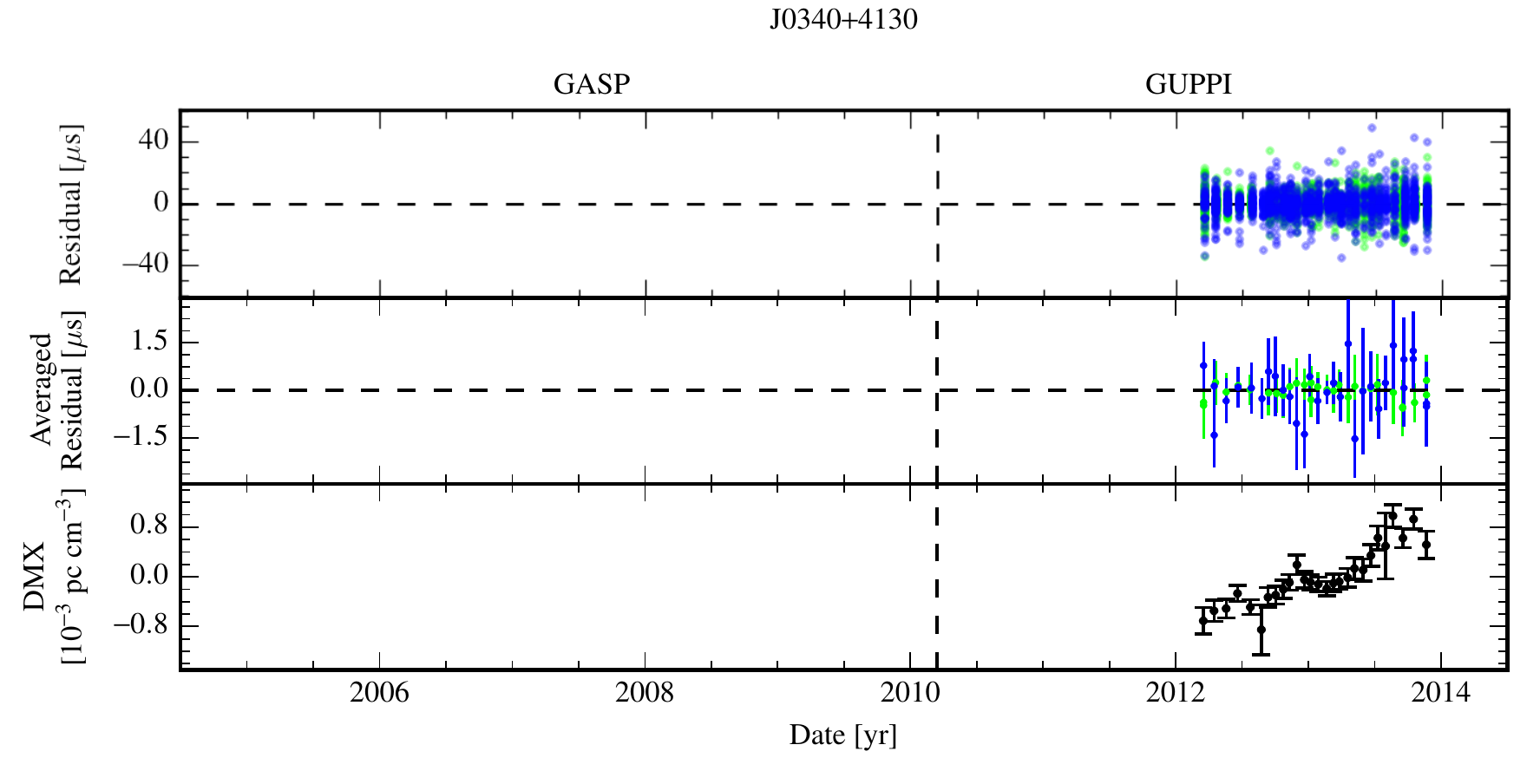}
\caption{Timing summary for PSR J0340+4130. Colors are: Blue: 1.4 GHz, Purple: 2.3 GHz, Green: 820 MHz, Orange: 430 MHz, Red: 327 MHz. In the top panel, individual points are semi-transparent; darker regions arise from the overlap of many points.}
\label{fig:summary-J0340+4130}
\end{figure*}

\begin{figure*}[p]
\centering
\includegraphics[scale=1.0]{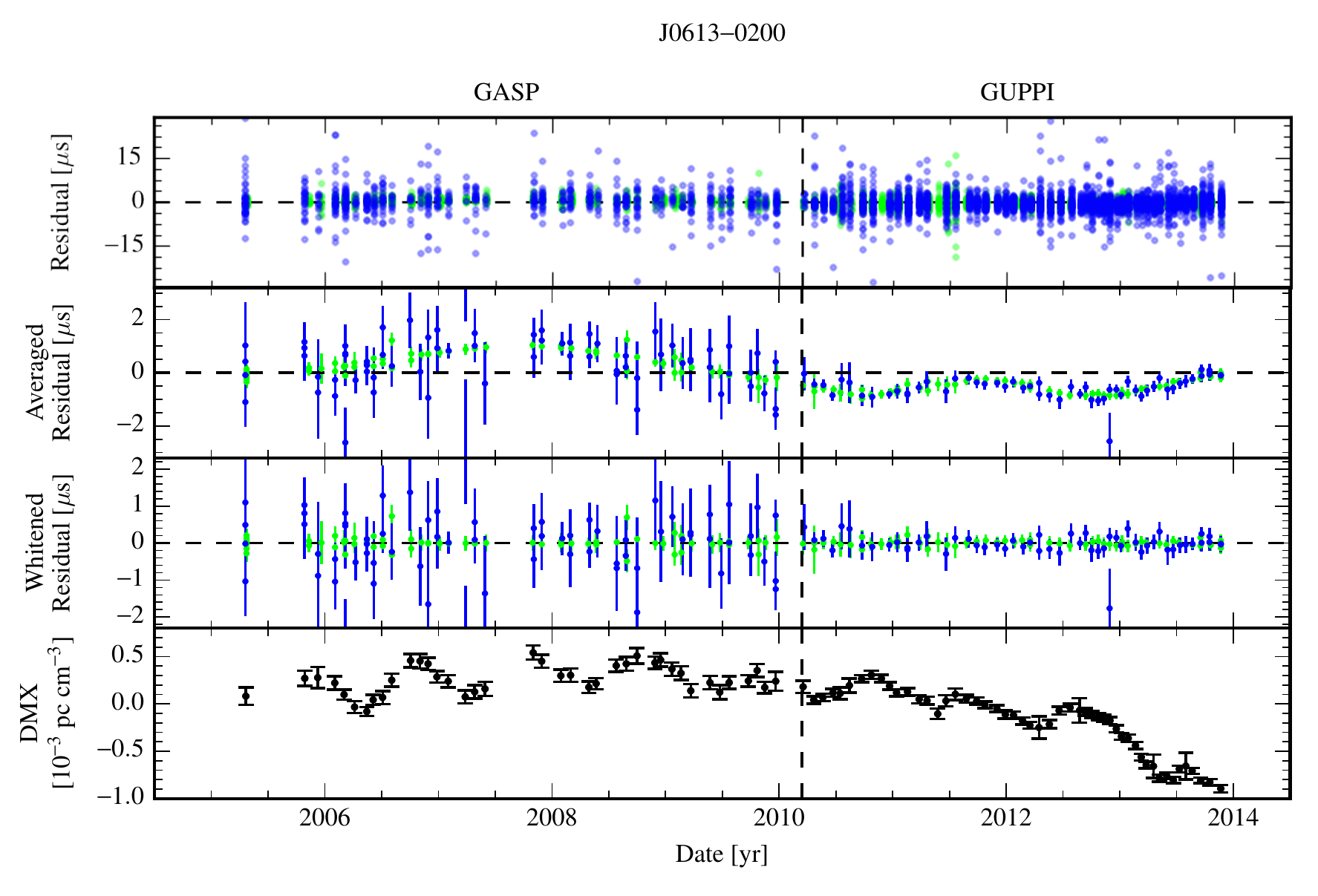}
\caption{Timing summary for PSR J0613-0200. Colors are: Blue: 1.4 GHz, Purple: 2.3 GHz, Green: 820 MHz, Orange: 430 MHz, Red: 327 MHz. In the top panel, individual points are semi-transparent; darker regions arise from the overlap of many points.}
\label{fig:summary-J0613-0200}
\end{figure*}

\begin{figure*}[p]
\centering
\includegraphics[scale=1.0]{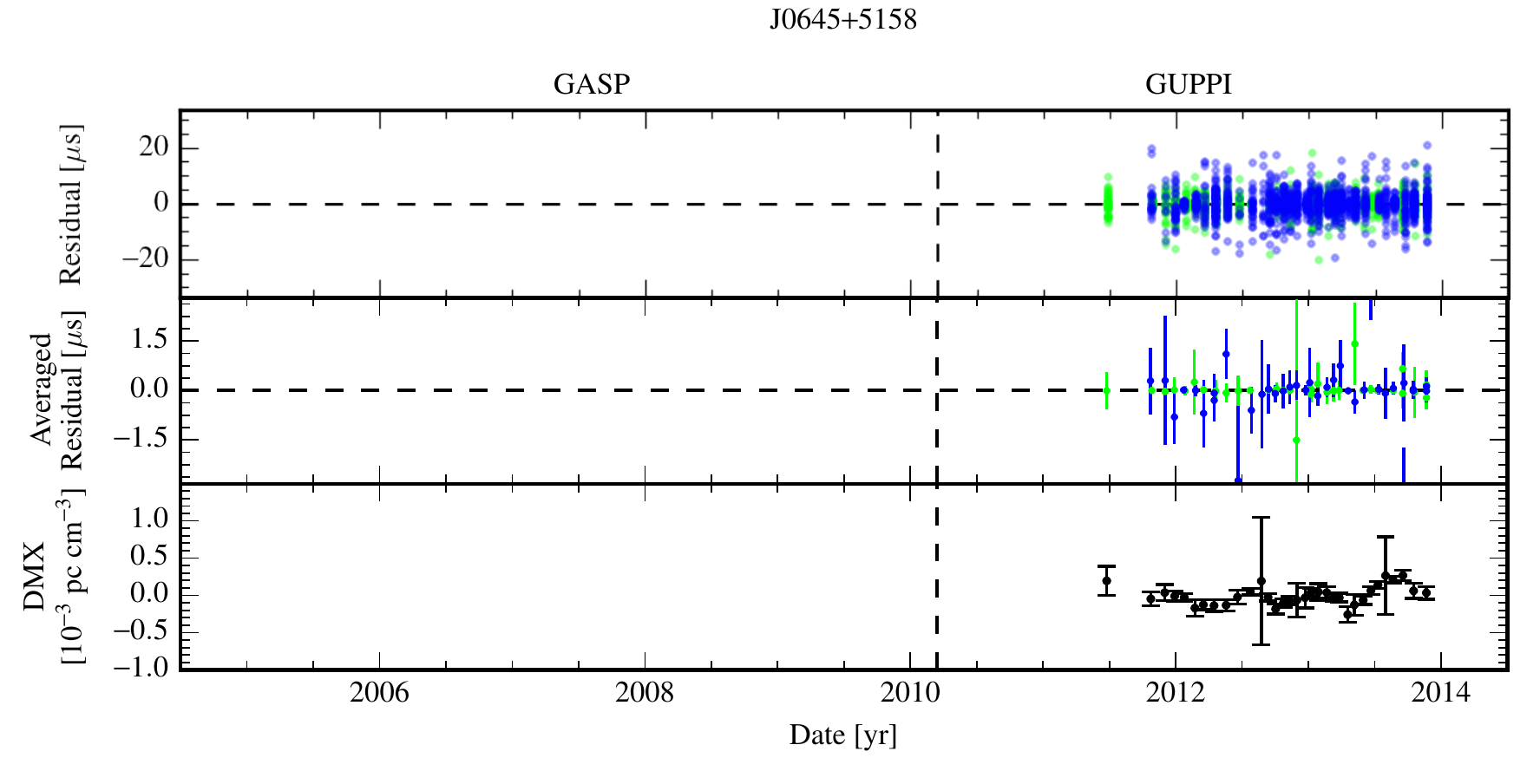}
\caption{Timing summary for PSR J0645+5158. Colors are: Blue: 1.4 GHz, Purple: 2.3 GHz, Green: 820 MHz, Orange: 430 MHz, Red: 327 MHz. In the top panel, individual points are semi-transparent; darker regions arise from the overlap of many points.}
\label{fig:summary-J0645+5158}
\end{figure*}

\begin{figure*}[p]
\centering
\includegraphics[scale=1.0]{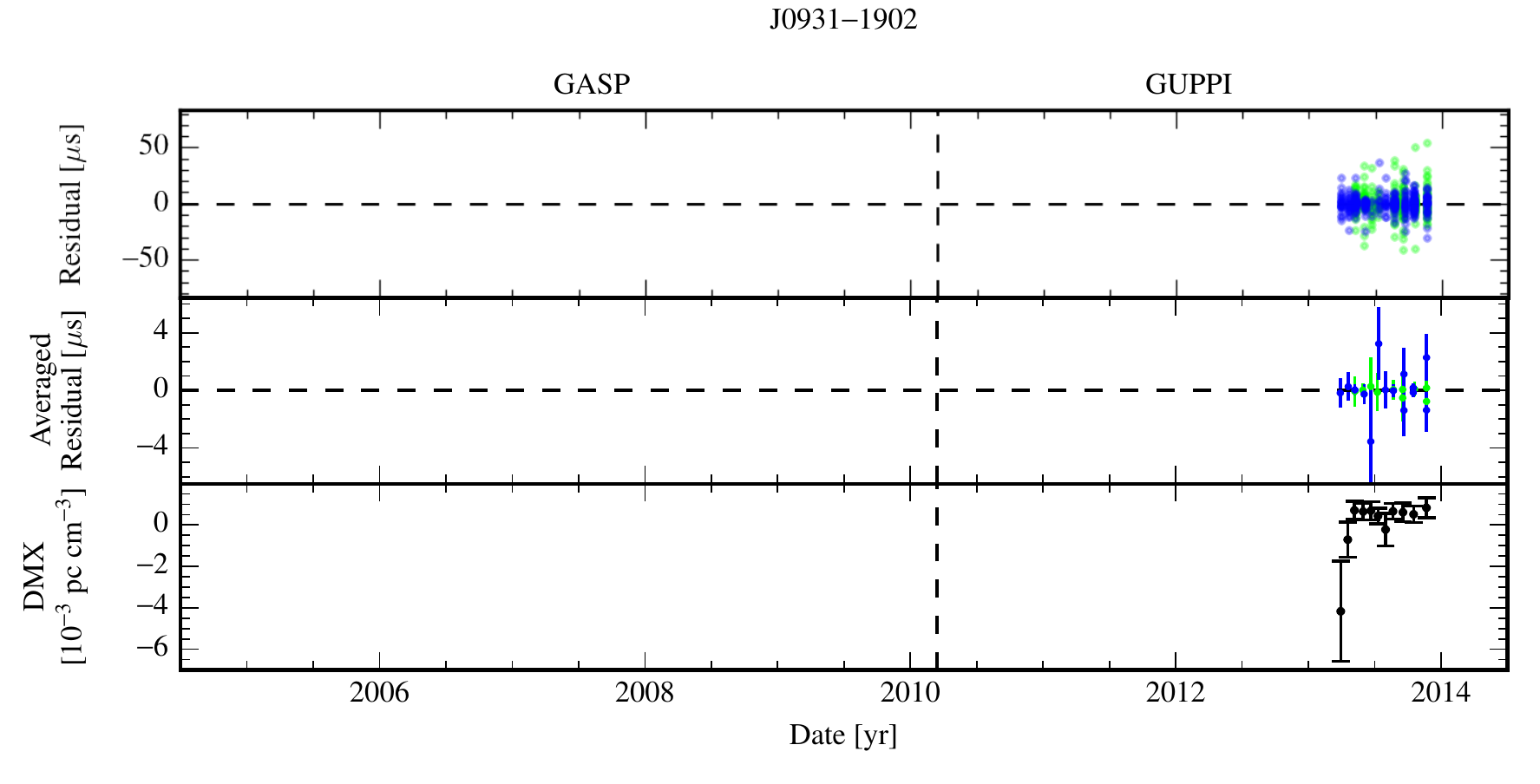}
\caption{Timing summary for PSR J0931-1902. Colors are: Blue: 1.4 GHz, Purple: 2.3 GHz, Green: 820 MHz, Orange: 430 MHz, Red: 327 MHz. In the top panel, individual points are semi-transparent; darker regions arise from the overlap of many points.}
\label{fig:summary-J0931-1902}
\end{figure*}

\begin{figure*}[p]
\centering
\includegraphics[scale=1.0]{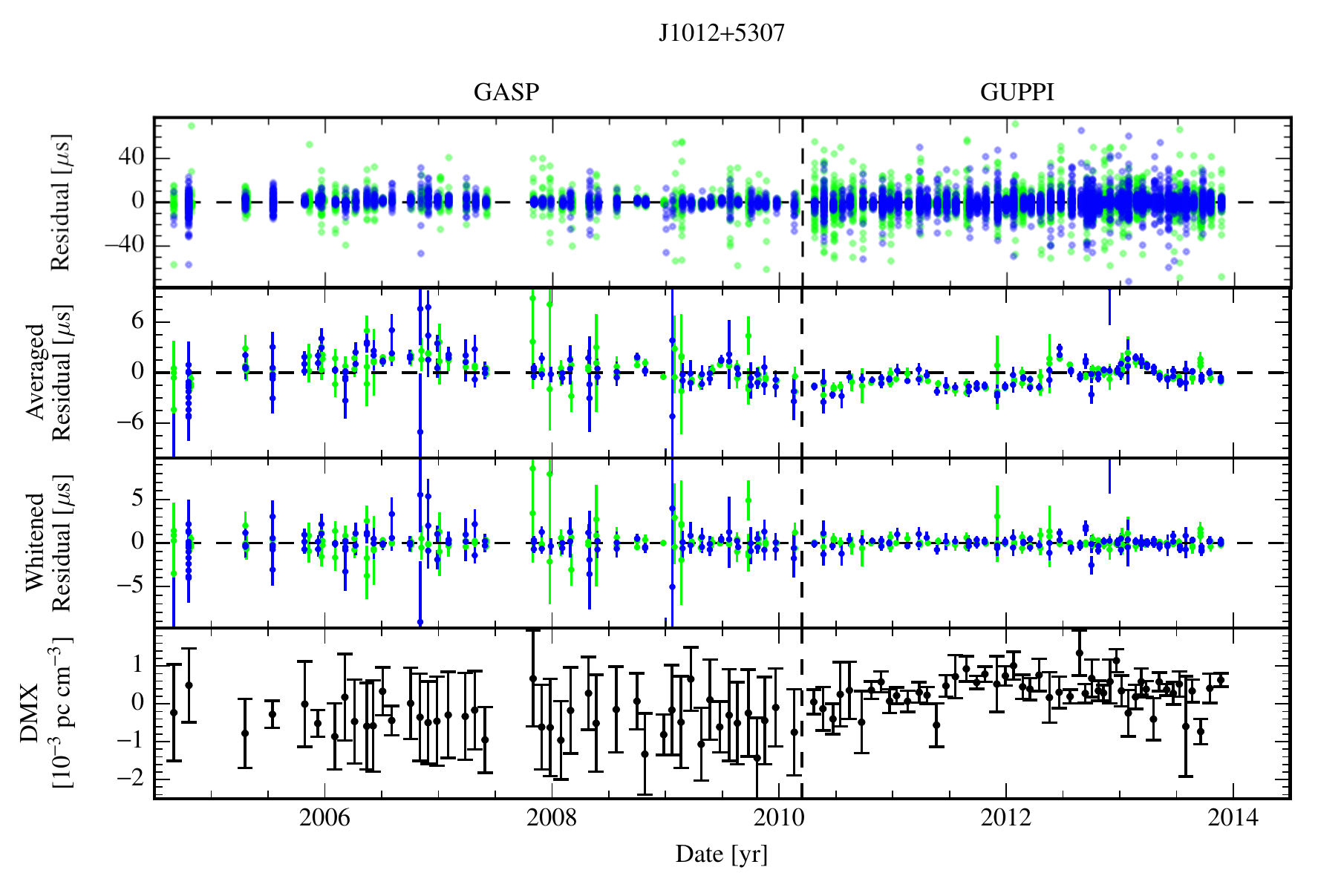}
\caption{Timing summary for PSR J1012+5307. Colors are: Blue: 1.4 GHz, Purple: 2.3 GHz, Green: 820 MHz, Orange: 430 MHz, Red: 327 MHz. In the top panel, individual points are semi-transparent; darker regions arise from the overlap of many points.}
\label{fig:summary-J1012+5307}
\end{figure*}

\begin{figure*}[p]
\centering
\includegraphics[scale=1.0]{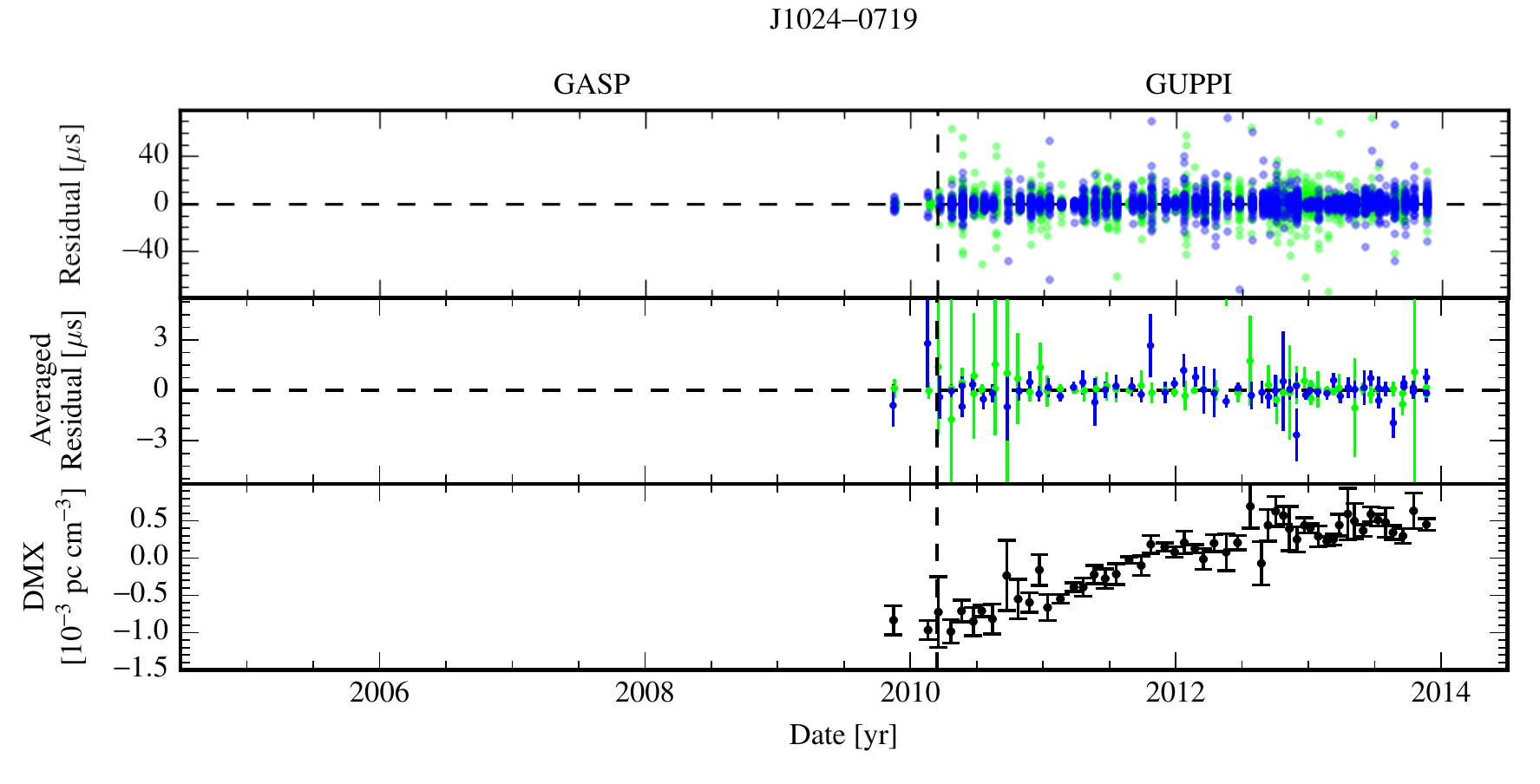}
\caption{Timing summary for PSR J1024-0719. Colors are: Blue: 1.4 GHz, Purple: 2.3 GHz, Green: 820 MHz, Orange: 430 MHz, Red: 327 MHz. In the top panel, individual points are semi-transparent; darker regions arise from the overlap of many points.}
\label{fig:summary-J1024-0719}
\end{figure*}

\begin{figure*}[p]
\centering
\includegraphics[scale=1.0]{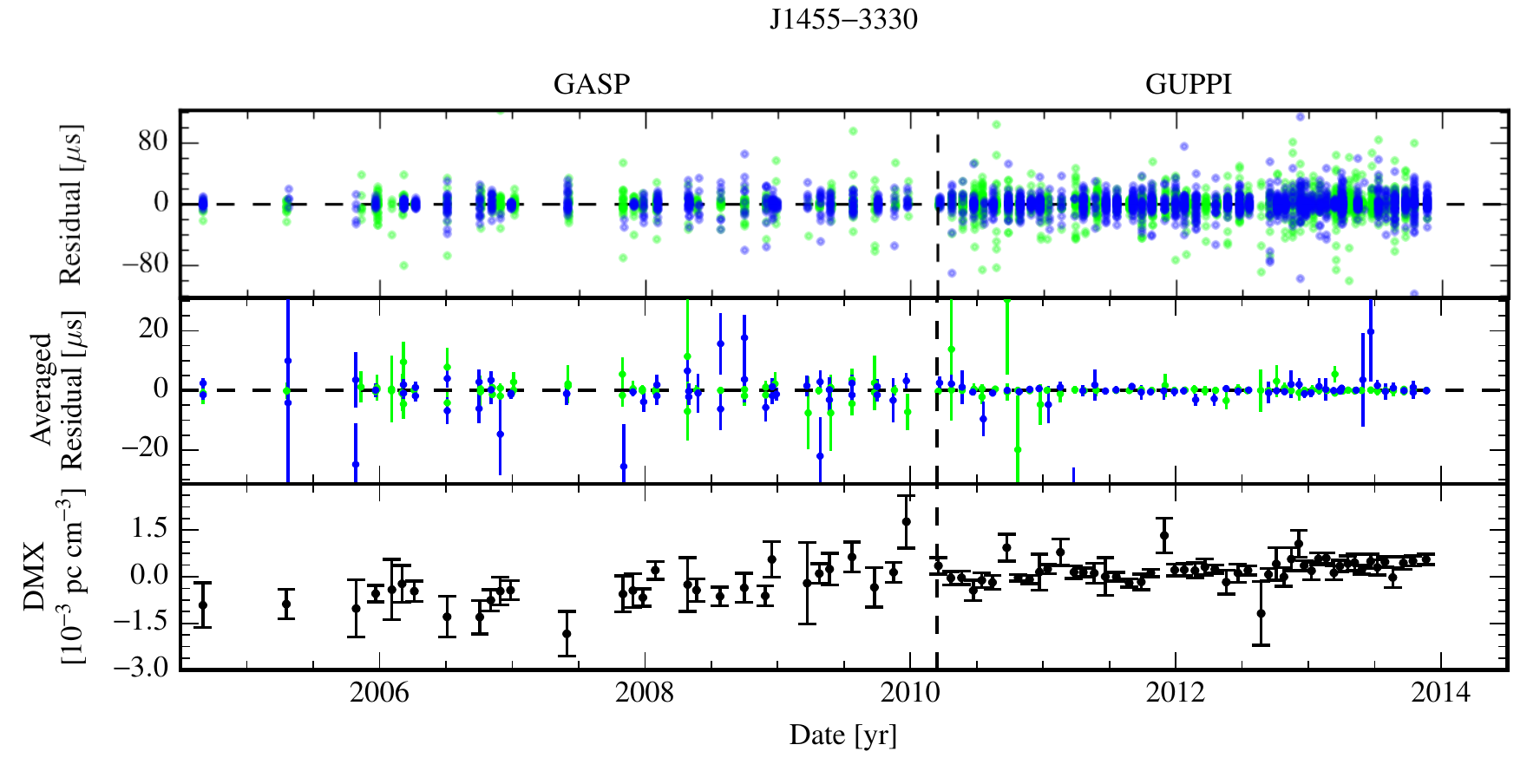}
\caption{Timing summary for PSR J1455-3330. Colors are: Blue: 1.4 GHz, Purple: 2.3 GHz, Green: 820 MHz, Orange: 430 MHz, Red: 327 MHz. In the top panel, individual points are semi-transparent; darker regions arise from the overlap of many points.}
\label{fig:summary-J1455-3330}
\end{figure*}

\begin{figure*}[p]
\centering
\includegraphics[scale=1.0]{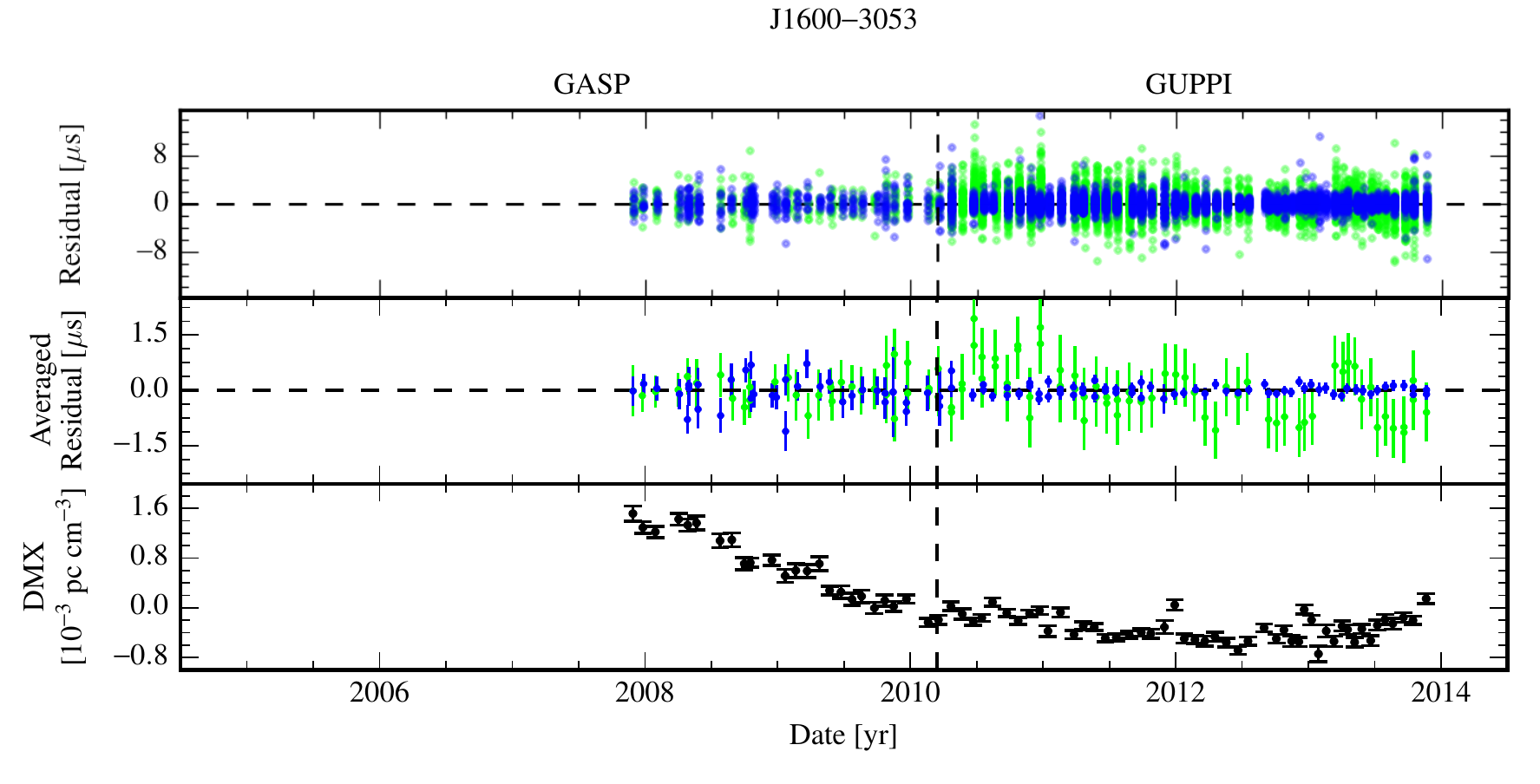}
\caption{Timing summary for PSR J1600-3053. Colors are: Blue: 1.4 GHz, Purple: 2.3 GHz, Green: 820 MHz, Orange: 430 MHz, Red: 327 MHz. In the top panel, individual points are semi-transparent; darker regions arise from the overlap of many points.}
\label{fig:summary-J1600-3053}
\end{figure*}

\begin{figure*}[p]
\centering
\includegraphics[scale=1.0]{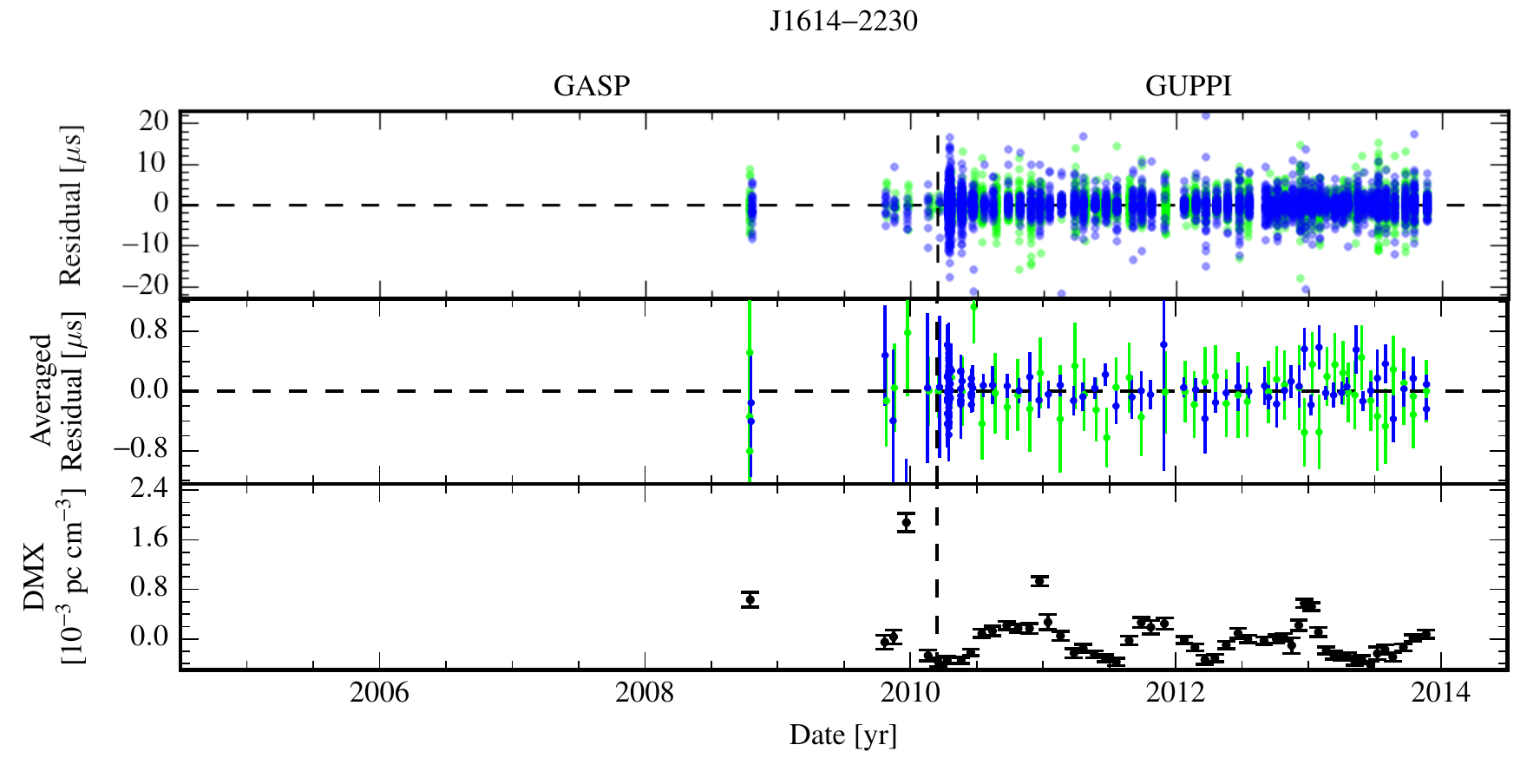}
\caption{Timing summary for PSR J1614-2230. Colors are: Blue: 1.4 GHz, Purple: 2.3 GHz, Green: 820 MHz, Orange: 430 MHz, Red: 327 MHz. In the top panel, individual points are semi-transparent; darker regions arise from the overlap of many points.}
\label{fig:summary-J1614-2230}
\end{figure*}

\begin{figure*}[p]
\centering
\includegraphics[scale=1.0]{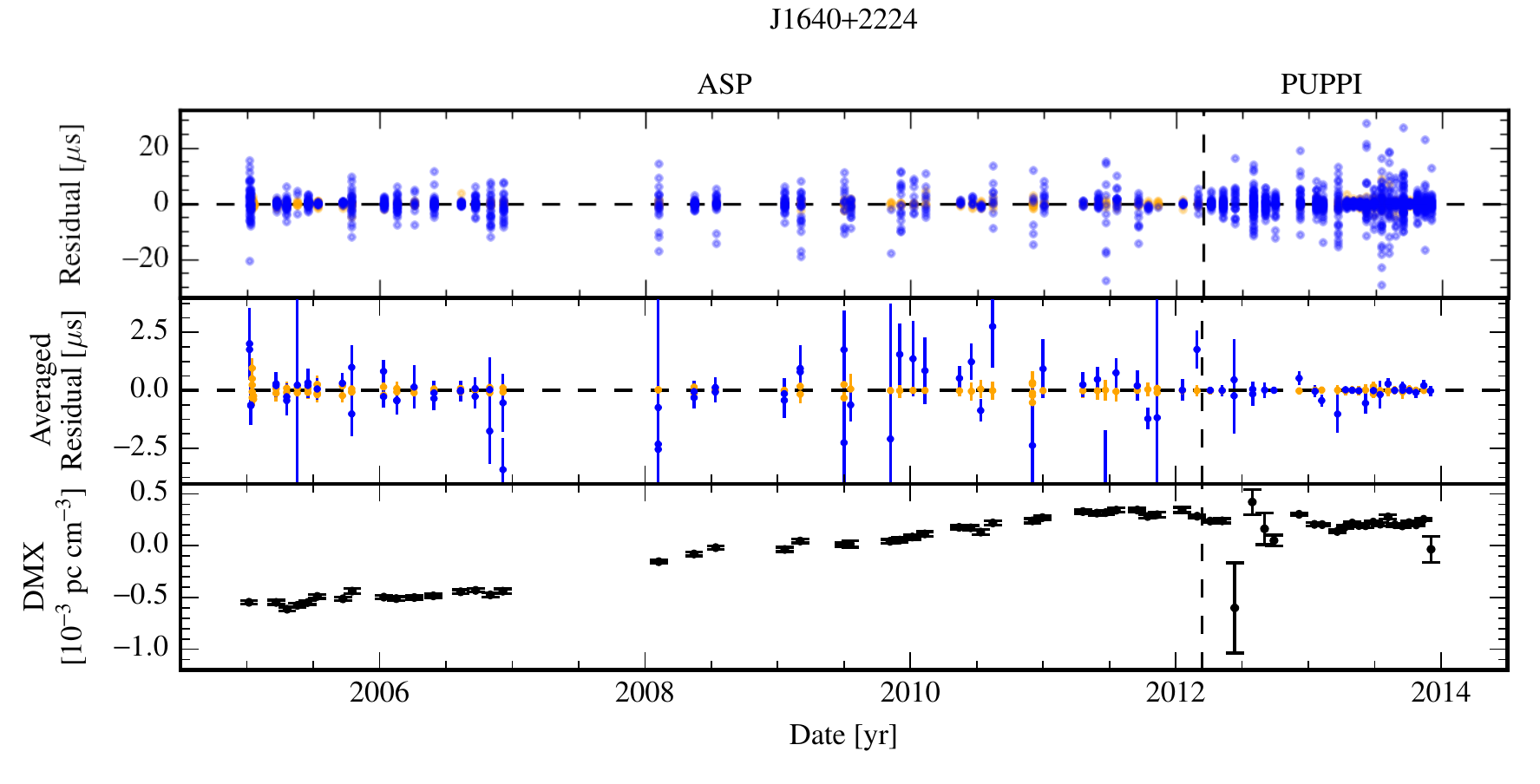}
\caption{Timing summary for PSR J1640+2224. Colors are: Blue: 1.4 GHz, Purple: 2.3 GHz, Green: 820 MHz, Orange: 430 MHz, Red: 327 MHz. In the top panel, individual points are semi-transparent; darker regions arise from the overlap of many points.}
\label{fig:summary-J1640+2224}
\end{figure*}

\begin{figure*}[p]
\centering
\includegraphics[scale=1.0]{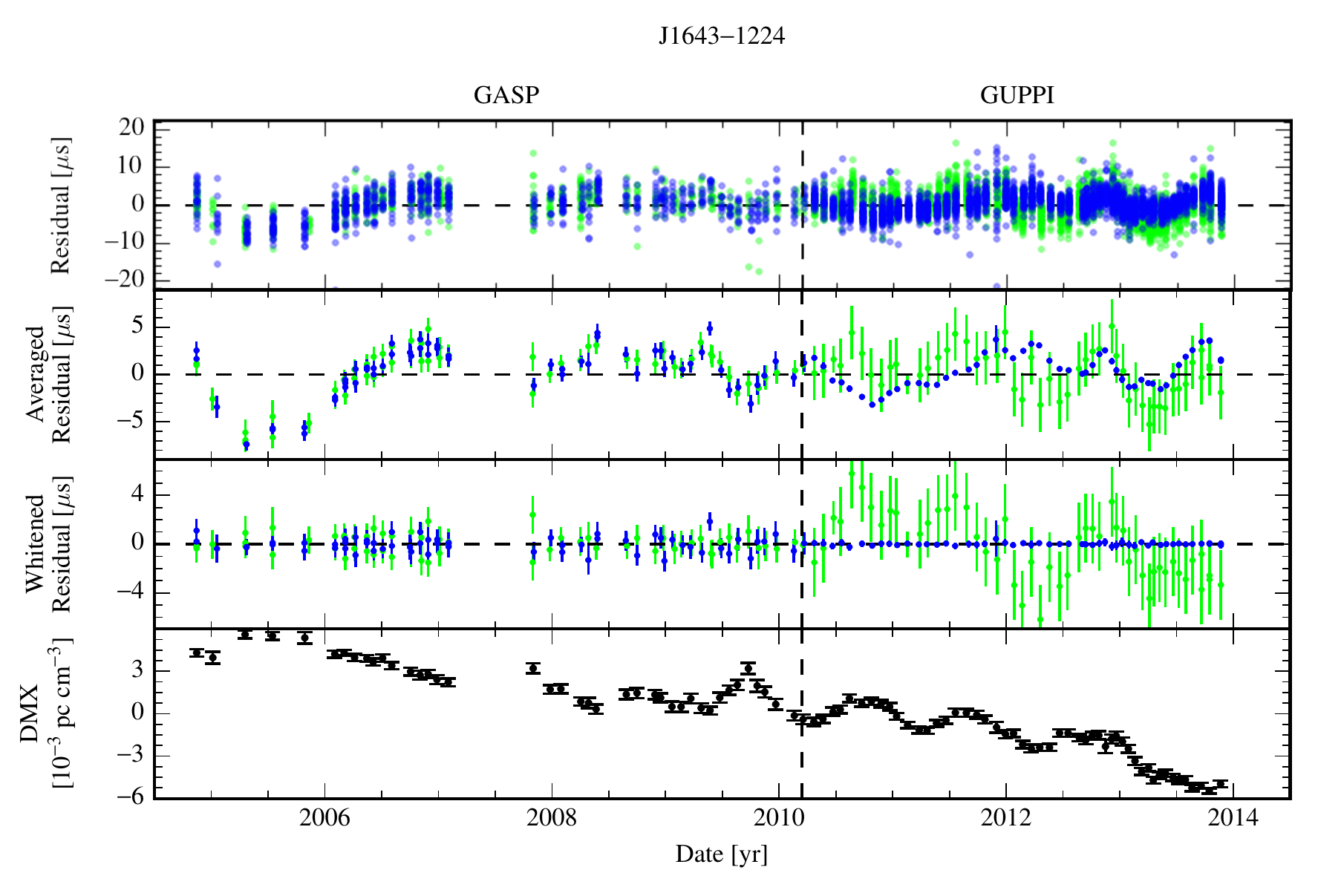}
\caption{Timing summary for PSR J1643-1224. Colors are: Blue: 1.4 GHz, Purple: 2.3 GHz, Green: 820 MHz, Orange: 430 MHz, Red: 327 MHz. In the top panel, individual points are semi-transparent; darker regions arise from the overlap of many points.}
\label{fig:summary-J1643-1224}
\end{figure*}

\begin{figure*}[p]
\centering
\includegraphics[scale=1.0]{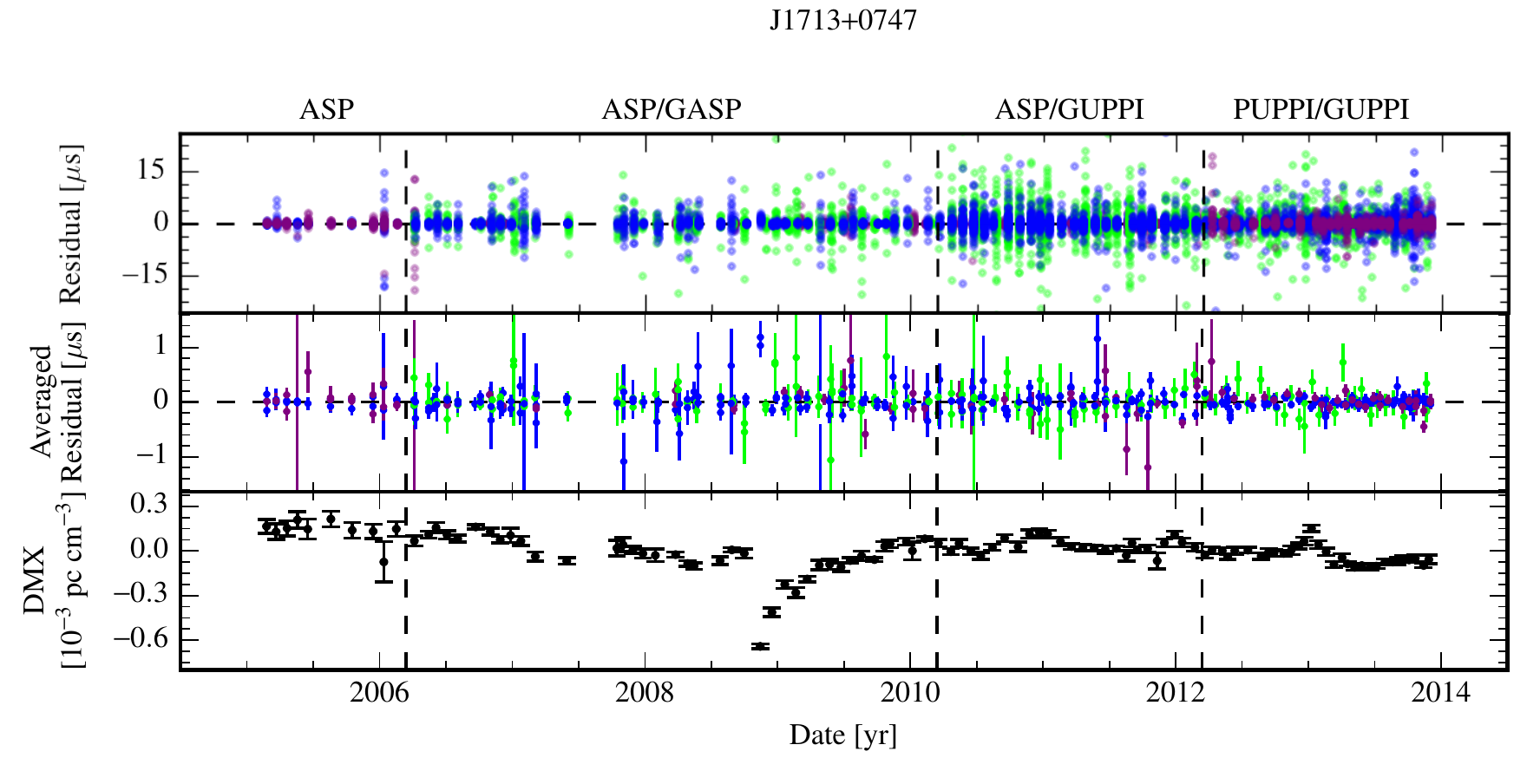}
\caption{Timing summary for PSR J1713+0747. Colors are: Blue: 1.4 GHz, Purple: 2.3 GHz, Green: 820 MHz, Orange: 430 MHz, Red: 327 MHz. In the top panel, individual points are semi-transparent; darker regions arise from the overlap of many points.}
\label{fig:summary-J1713+0747}
\end{figure*}

\begin{figure*}[p]
\centering
\includegraphics[scale=1.0]{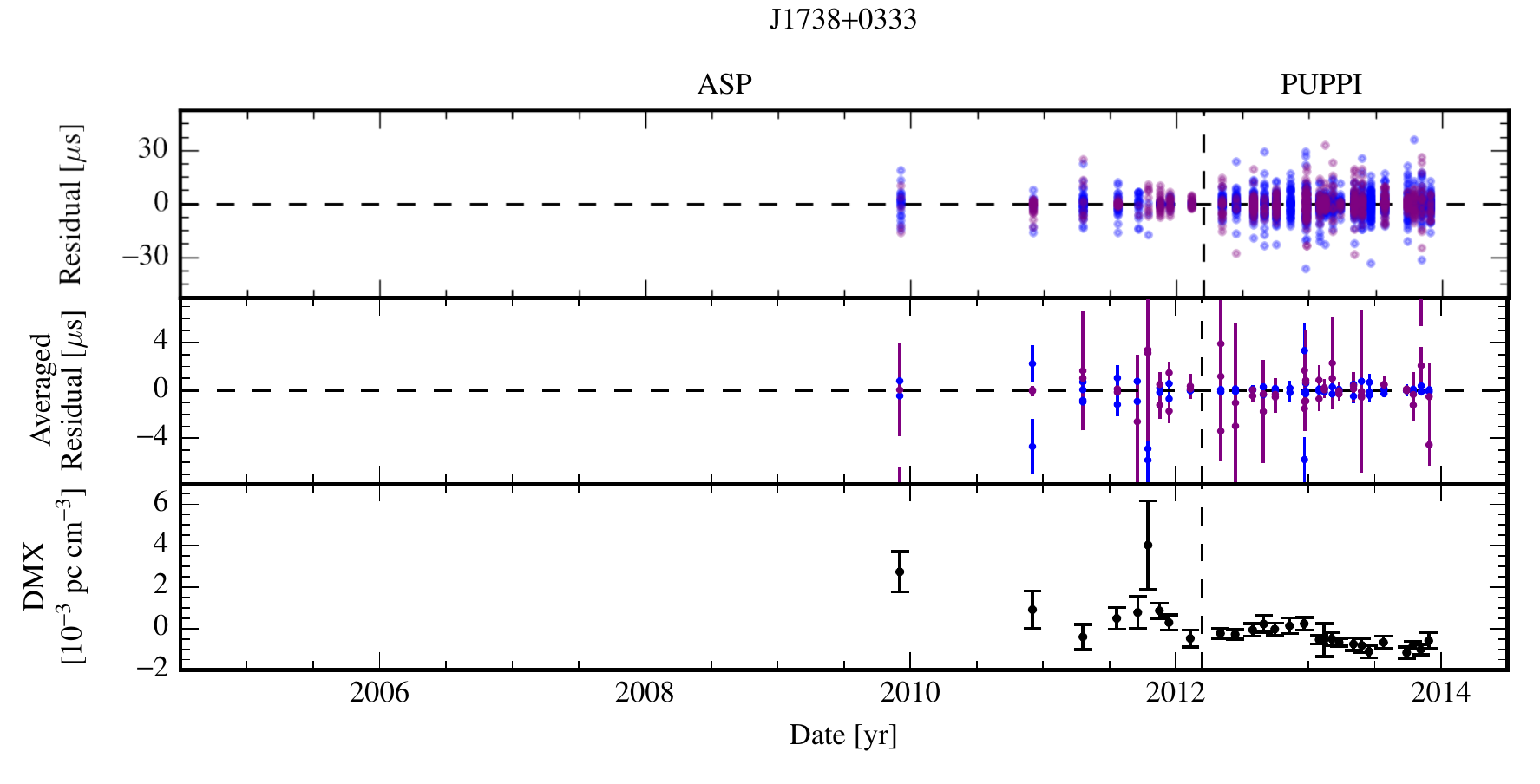}
\caption{Timing summary for PSR J1738+0333. Colors are: Blue: 1.4 GHz, Purple: 2.3 GHz, Green: 820 MHz, Orange: 430 MHz, Red: 327 MHz. In the top panel, individual points are semi-transparent; darker regions arise from the overlap of many points.}
\label{fig:summary-J1738+0333}
\end{figure*}

\begin{figure*}[p]
\centering
\includegraphics[scale=1.0]{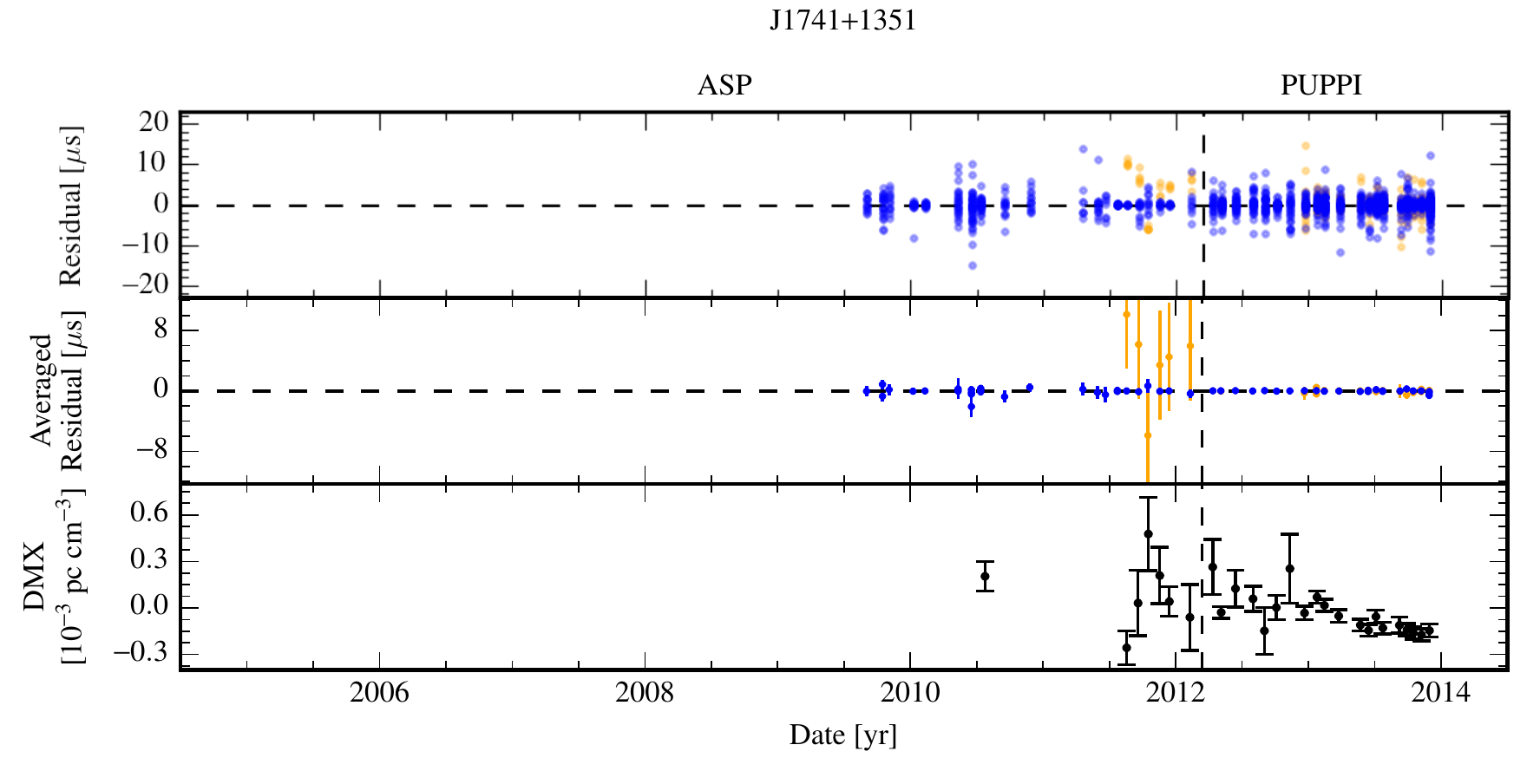}
\caption{Timing summary for PSR J1741+1351. Colors are: Blue: 1.4 GHz, Purple: 2.3 GHz, Green: 820 MHz, Orange: 430 MHz, Red: 327 MHz. In the top panel, individual points are semi-transparent; darker regions arise from the overlap of many points.}
\label{fig:summary-J1741+1351}
\end{figure*}

\begin{figure*}[p]
\centering
\includegraphics[scale=1.0]{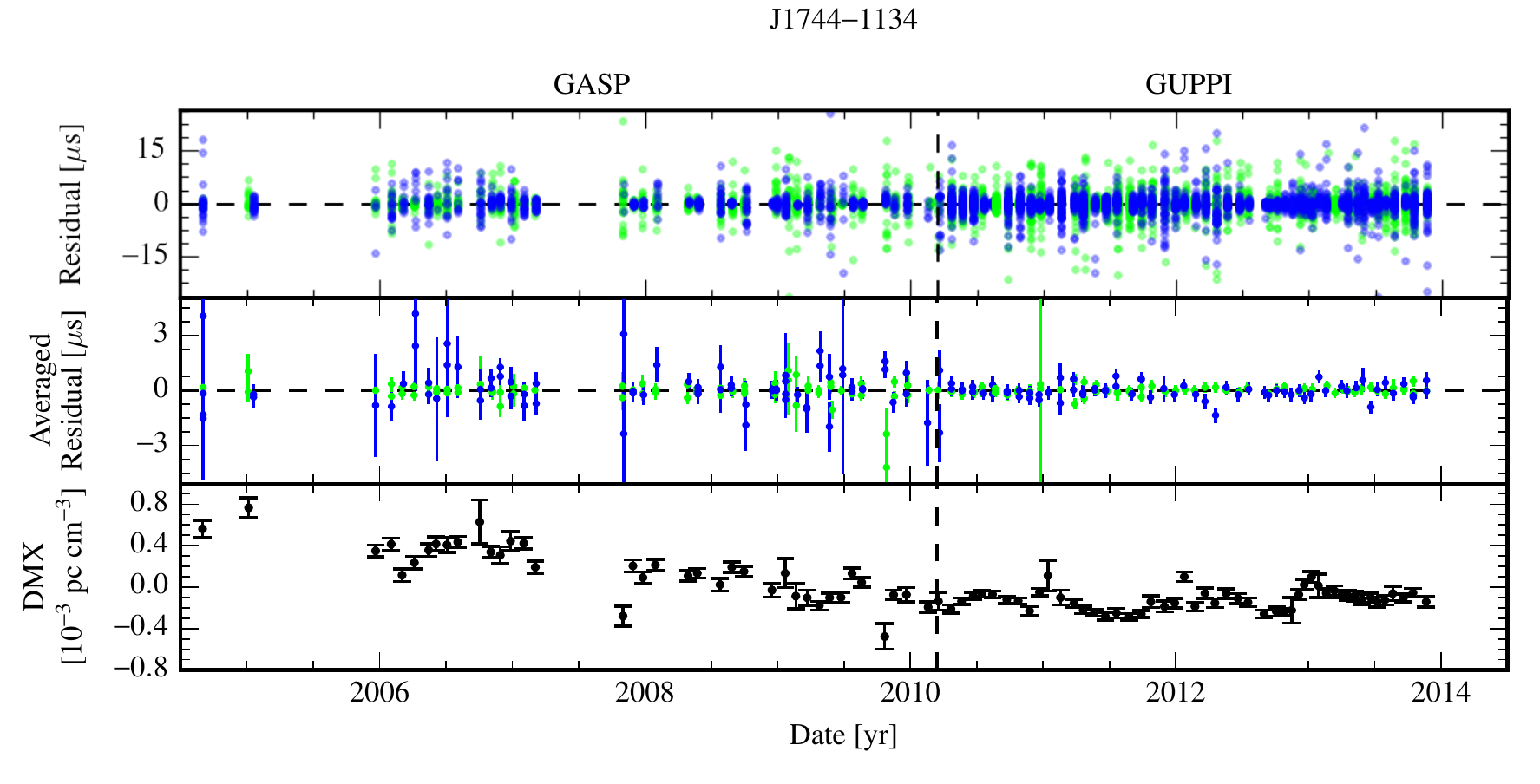}
\caption{Timing summary for PSR J1744-1134. Colors are: Blue: 1.4 GHz, Purple: 2.3 GHz, Green: 820 MHz, Orange: 430 MHz, Red: 327 MHz. In the top panel, individual points are semi-transparent; darker regions arise from the overlap of many points.}
\label{fig:summary-J1744-1134}
\end{figure*}

\begin{figure*}[p]
\centering
\includegraphics[scale=1.0]{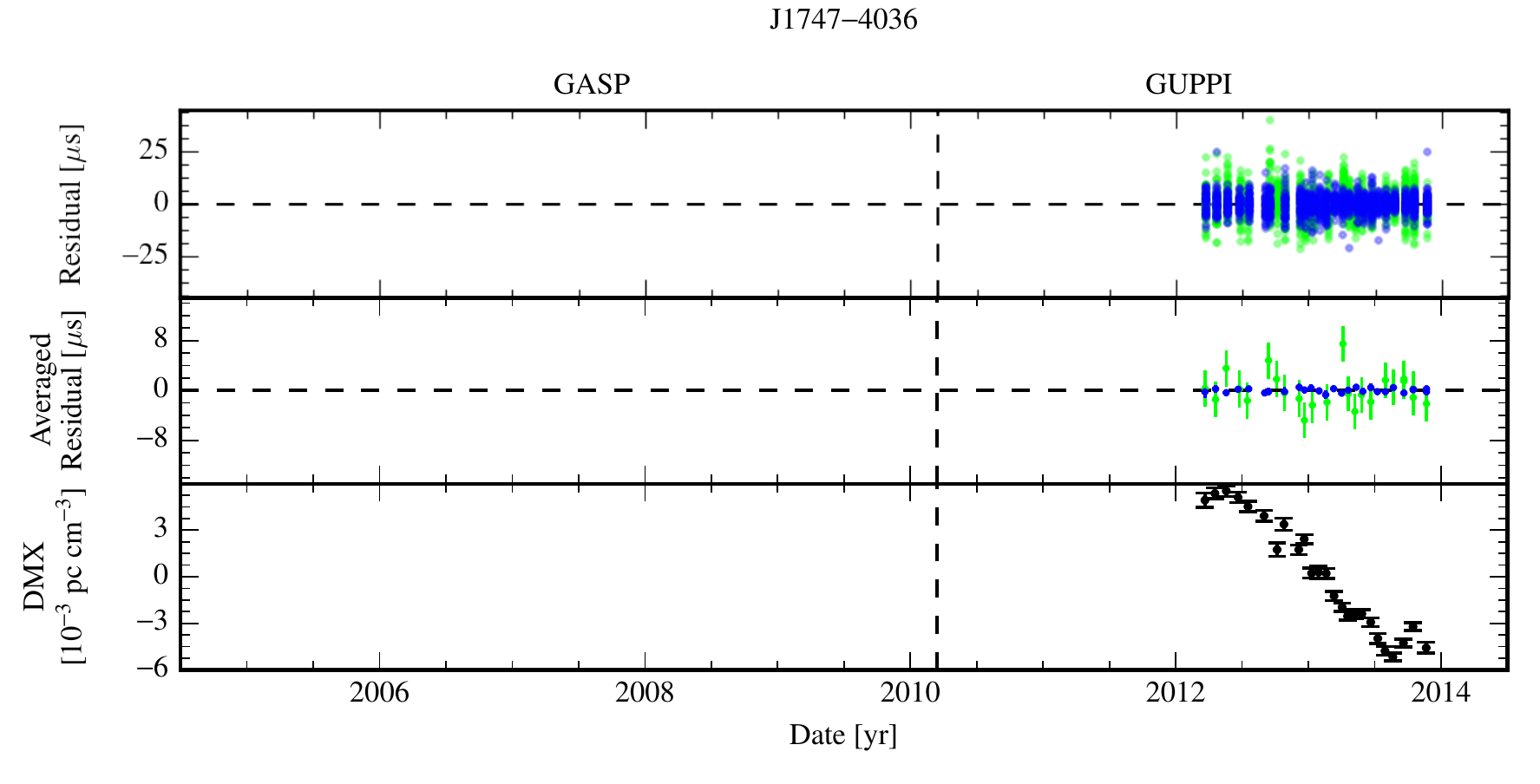}
\caption{Timing summary for PSR J1747-4036. Colors are: Blue: 1.4 GHz, Purple: 2.3 GHz, Green: 820 MHz, Orange: 430 MHz, Red: 327 MHz. In the top panel, individual points are semi-transparent; darker regions arise from the overlap of many points.}
\label{fig:summary-J1747-4036}
\end{figure*}

\begin{figure*}[p]
\centering
\includegraphics[scale=1.0]{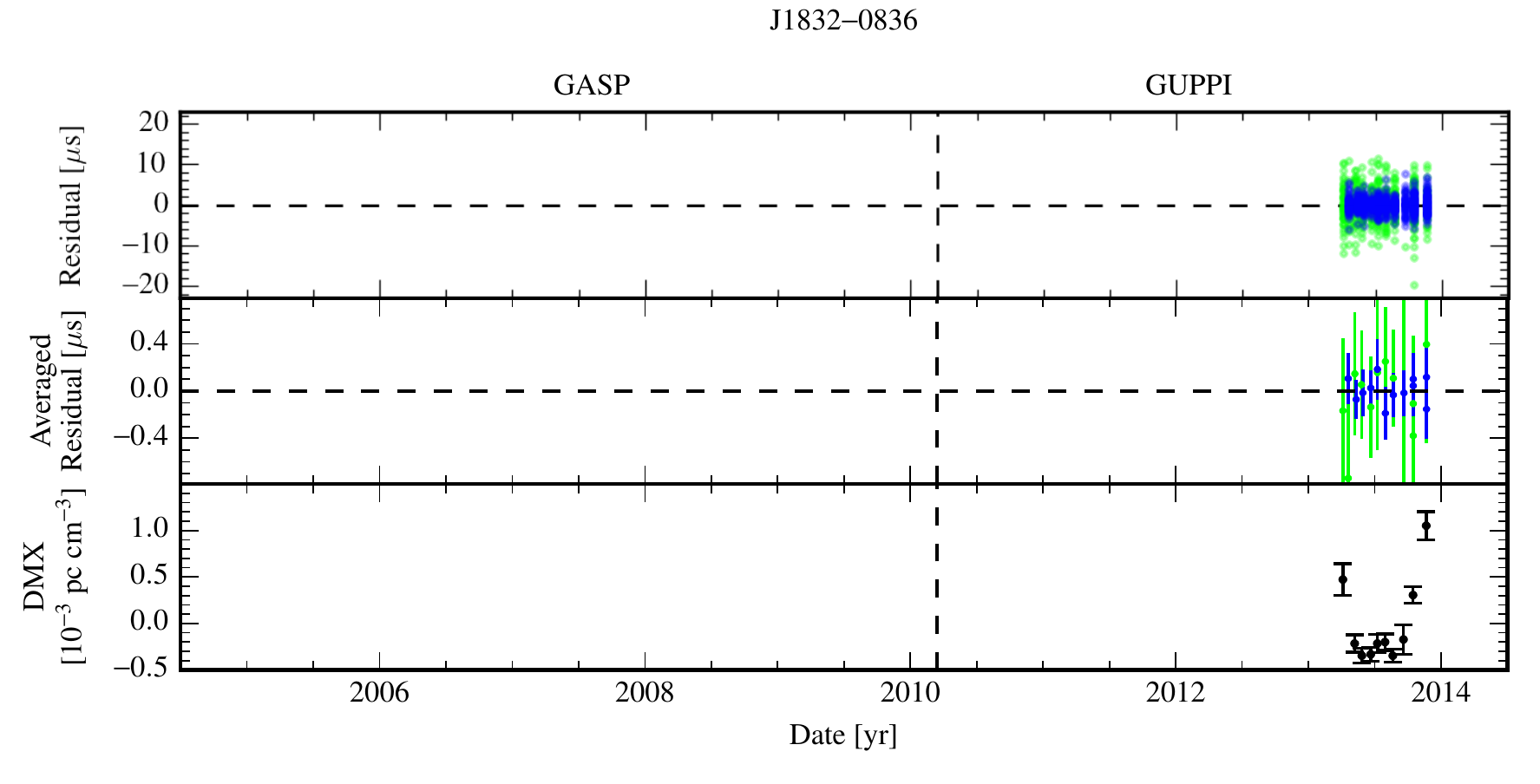}
\caption{Timing summary for PSR J1832-0836. Colors are: Blue: 1.4 GHz, Purple: 2.3 GHz, Green: 820 MHz, Orange: 430 MHz, Red: 327 MHz. In the top panel, individual points are semi-transparent; darker regions arise from the overlap of many points.}
\label{fig:summary-J1832-0836}
\end{figure*}

\begin{figure*}[p]
\centering
\includegraphics[scale=1.0]{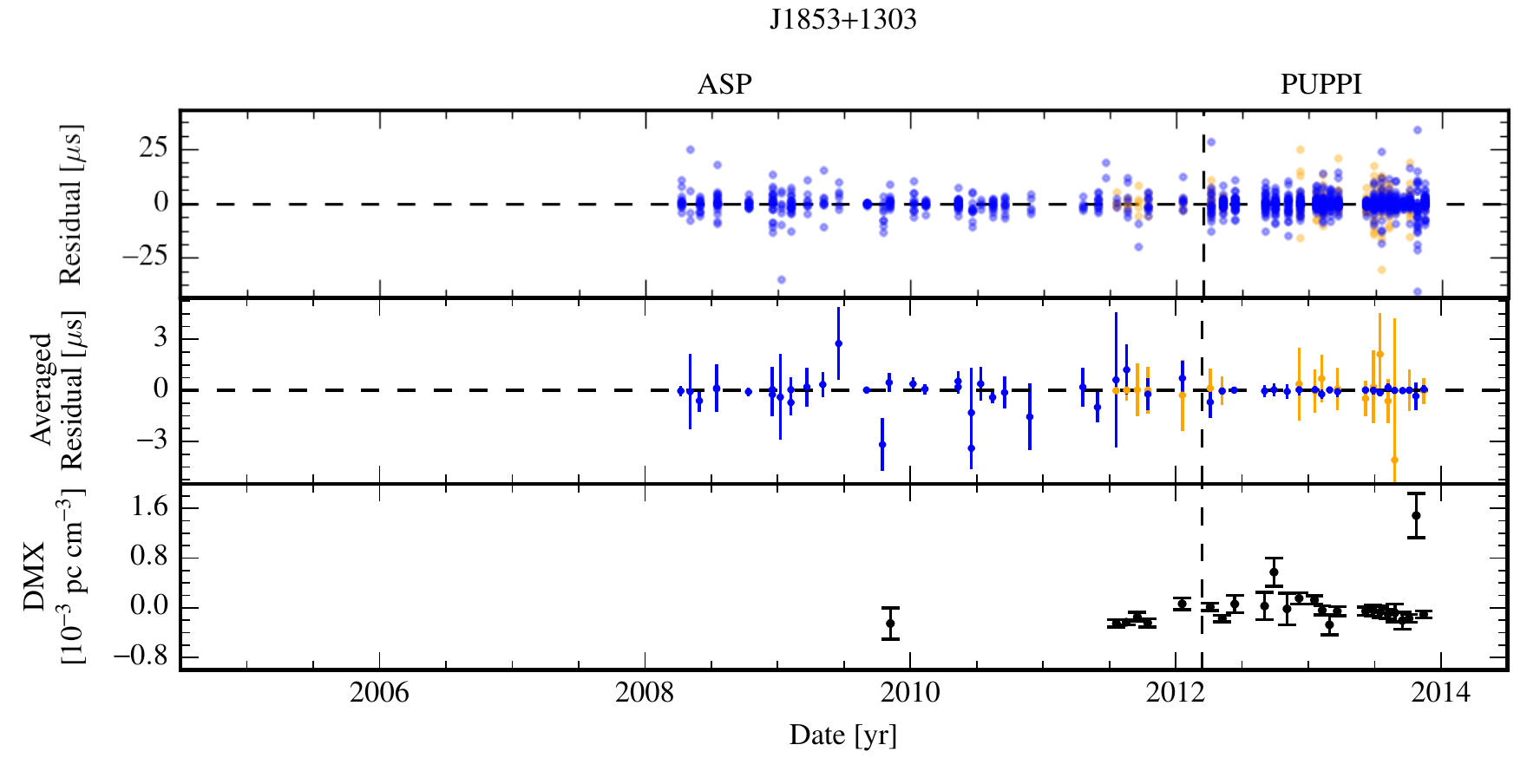}
\caption{Timing summary for PSR J1853+1303. Colors are: Blue: 1.4 GHz, Purple: 2.3 GHz, Green: 820 MHz, Orange: 430 MHz, Red: 327 MHz. In the top panel, individual points are semi-transparent; darker regions arise from the overlap of many points.}
\label{fig:summary-J1853+1303}
\end{figure*}

\begin{figure*}[p]
\centering
\includegraphics[scale=1.0]{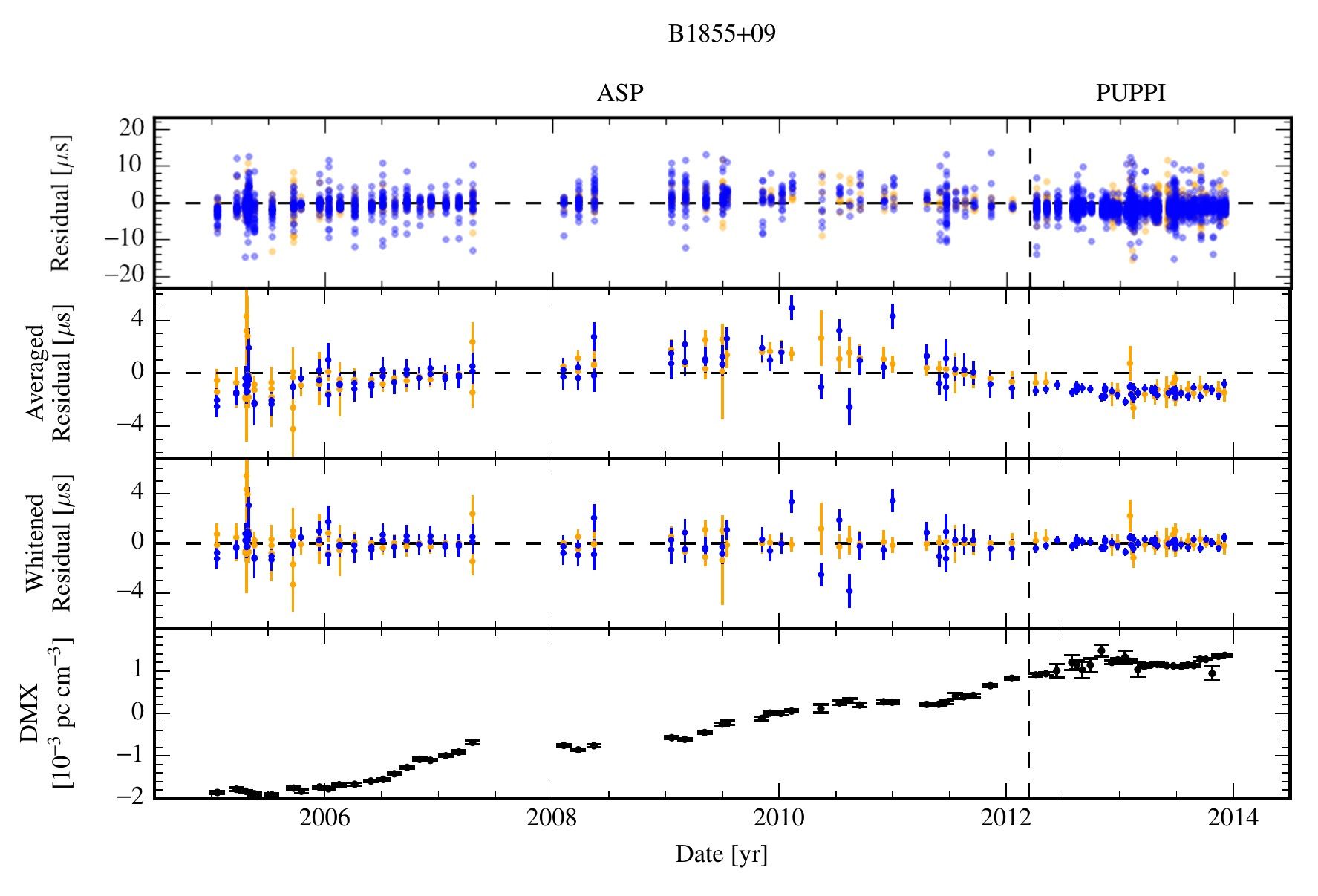}
\caption{Timing summary for PSR B1855+09. Colors are: Blue: 1.4 GHz, Purple: 2.3 GHz, Green: 820 MHz, Orange: 430 MHz, Red: 327 MHz. In the top panel, individual points are semi-transparent; darker regions arise from the overlap of many points.}
\label{fig:summary-B1855+09}
\end{figure*}

\begin{figure*}[p]
\centering
\includegraphics[scale=1.0]{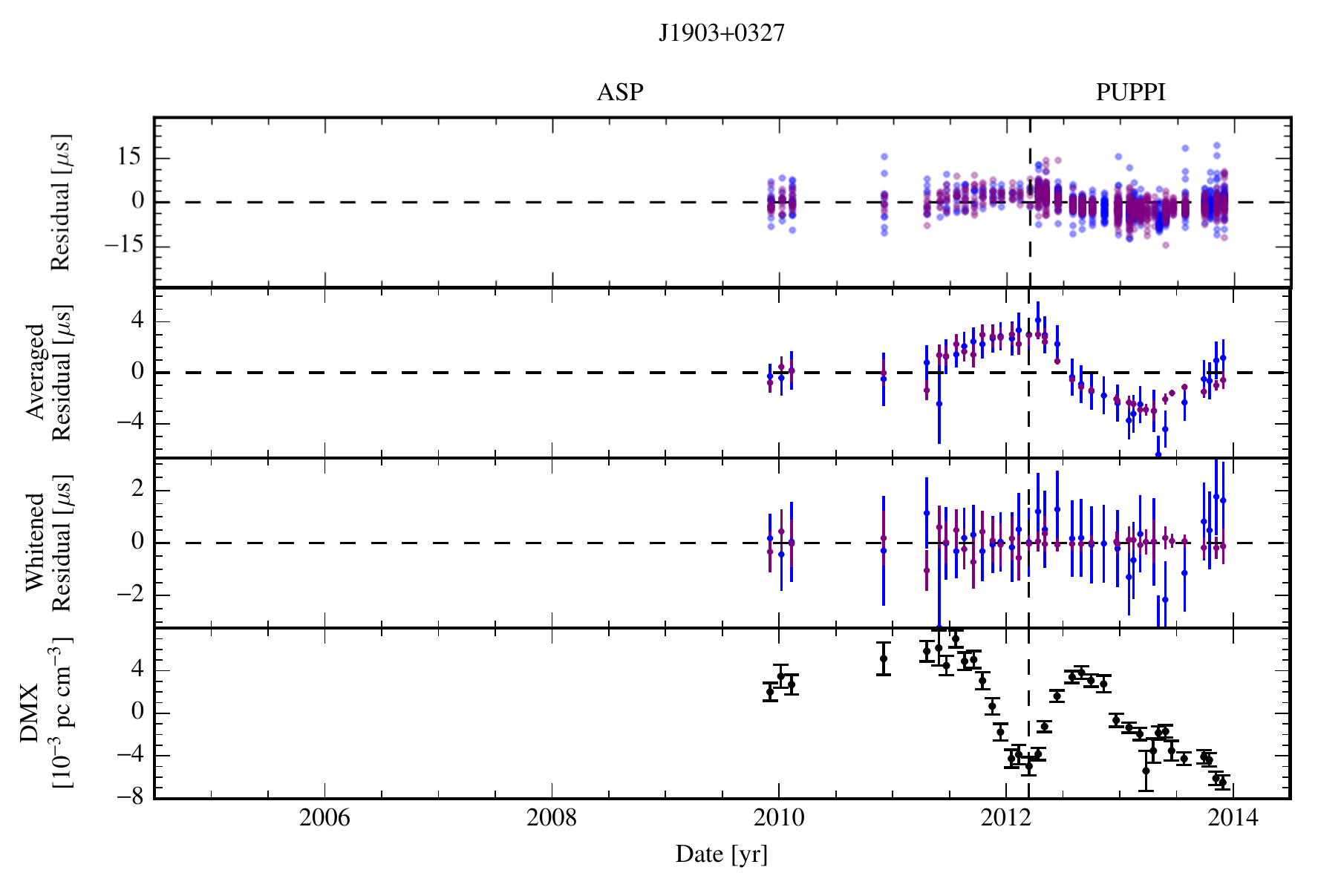}
\caption{Timing summary for PSR J1903+0327. Colors are: Blue: 1.4 GHz, Purple: 2.3 GHz, Green: 820 MHz, Orange: 430 MHz, Red: 327 MHz. In the top panel, individual points are semi-transparent; darker regions arise from the overlap of many points.}
\label{fig:summary-J1903+0327}
\end{figure*}

\begin{figure*}[p]
\centering
\includegraphics[scale=1.0]{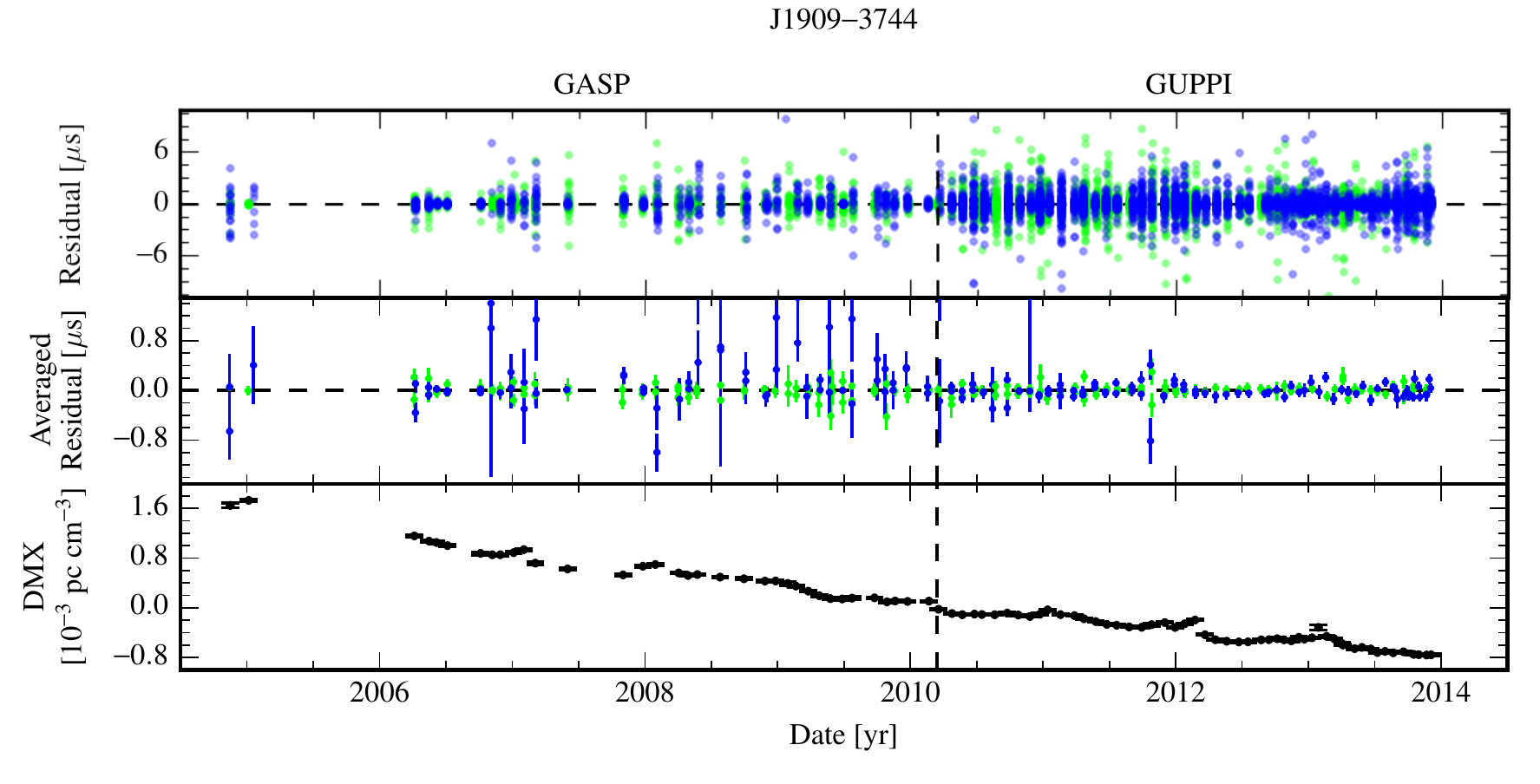}
\caption{Timing summary for PSR J1909-3744. Colors are: Blue: 1.4 GHz, Purple: 2.3 GHz, Green: 820 MHz, Orange: 430 MHz, Red: 327 MHz. In the top panel, individual points are semi-transparent; darker regions arise from the overlap of many points.}
\label{fig:summary-J1909-3744}
\end{figure*}

\begin{figure*}[p]
\centering
\includegraphics[scale=1.0]{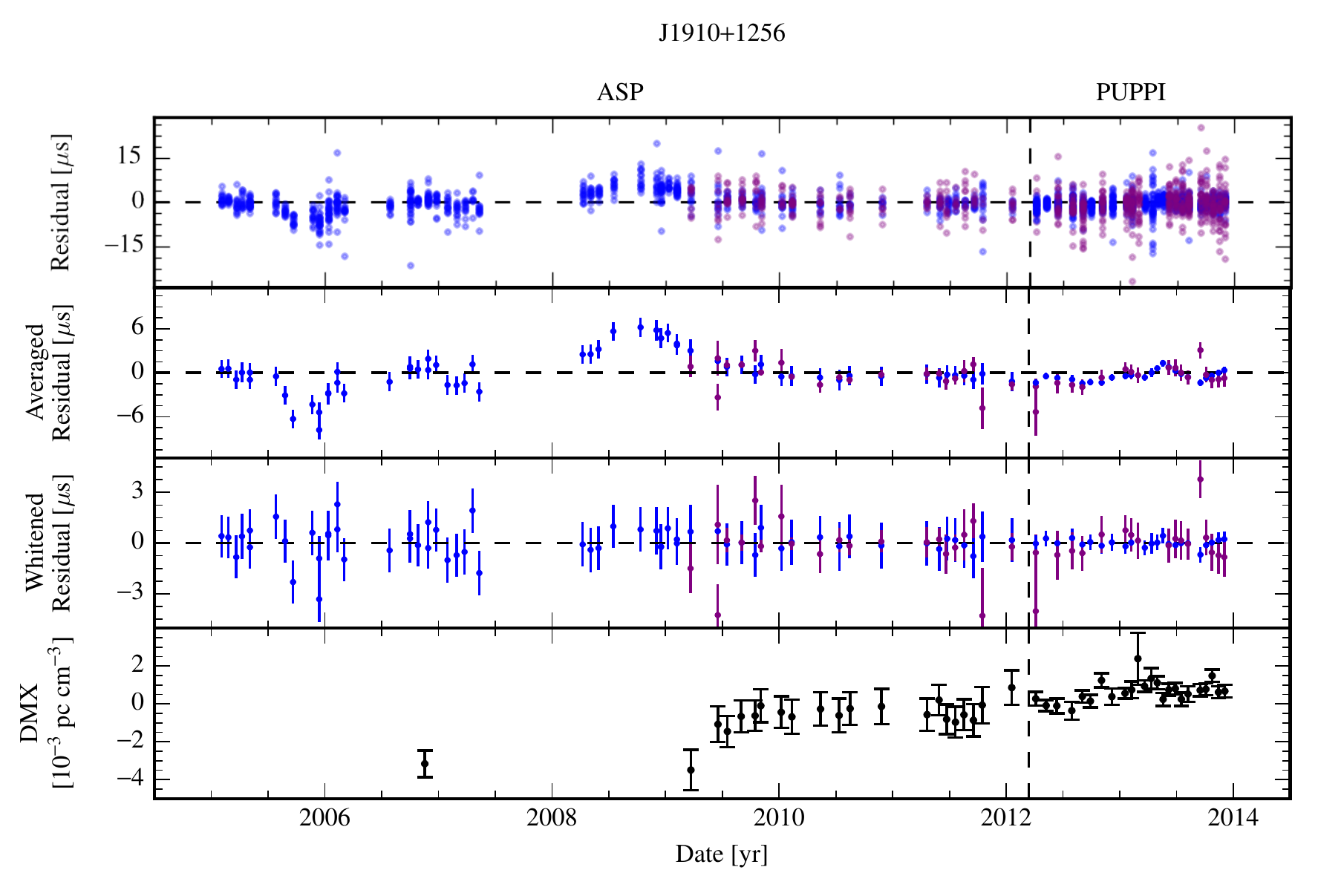}
\caption{Timing summary for PSR J1910+1256. Colors are: Blue: 1.4 GHz, Purple: 2.3 GHz, Green: 820 MHz, Orange: 430 MHz, Red: 327 MHz. In the top panel, individual points are semi-transparent; darker regions arise from the overlap of many points.}
\label{fig:summary-J1910+1256}
\end{figure*}

\begin{figure*}[p]
\centering
\includegraphics[scale=1.0]{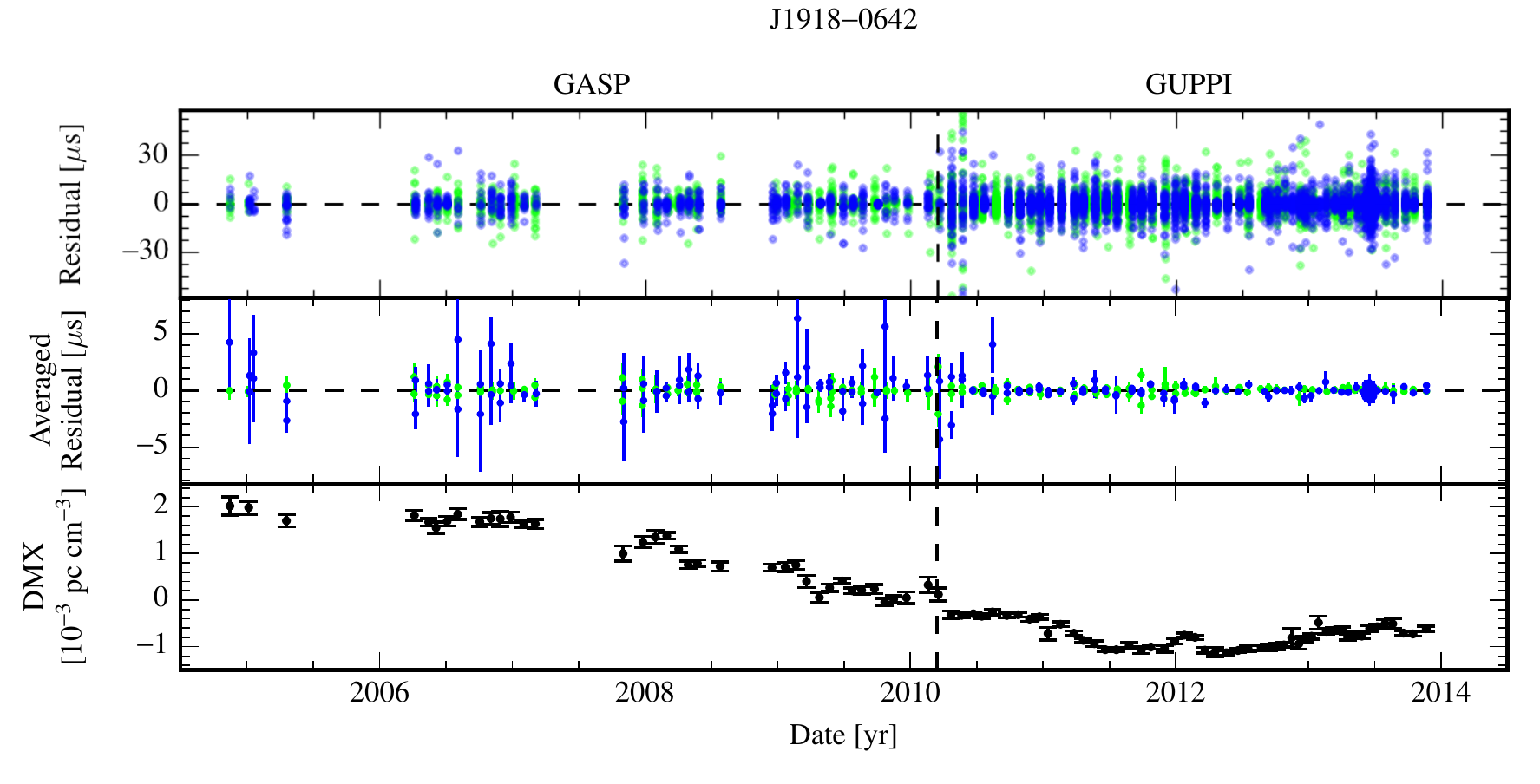}
\caption{Timing summary for PSR J1918-0642. Colors are: Blue: 1.4 GHz, Purple: 2.3 GHz, Green: 820 MHz, Orange: 430 MHz, Red: 327 MHz. In the top panel, individual points are semi-transparent; darker regions arise from the overlap of many points.}
\label{fig:summary-J1918-0642}
\end{figure*}

\begin{figure*}[p]
\centering
\includegraphics[scale=1.0]{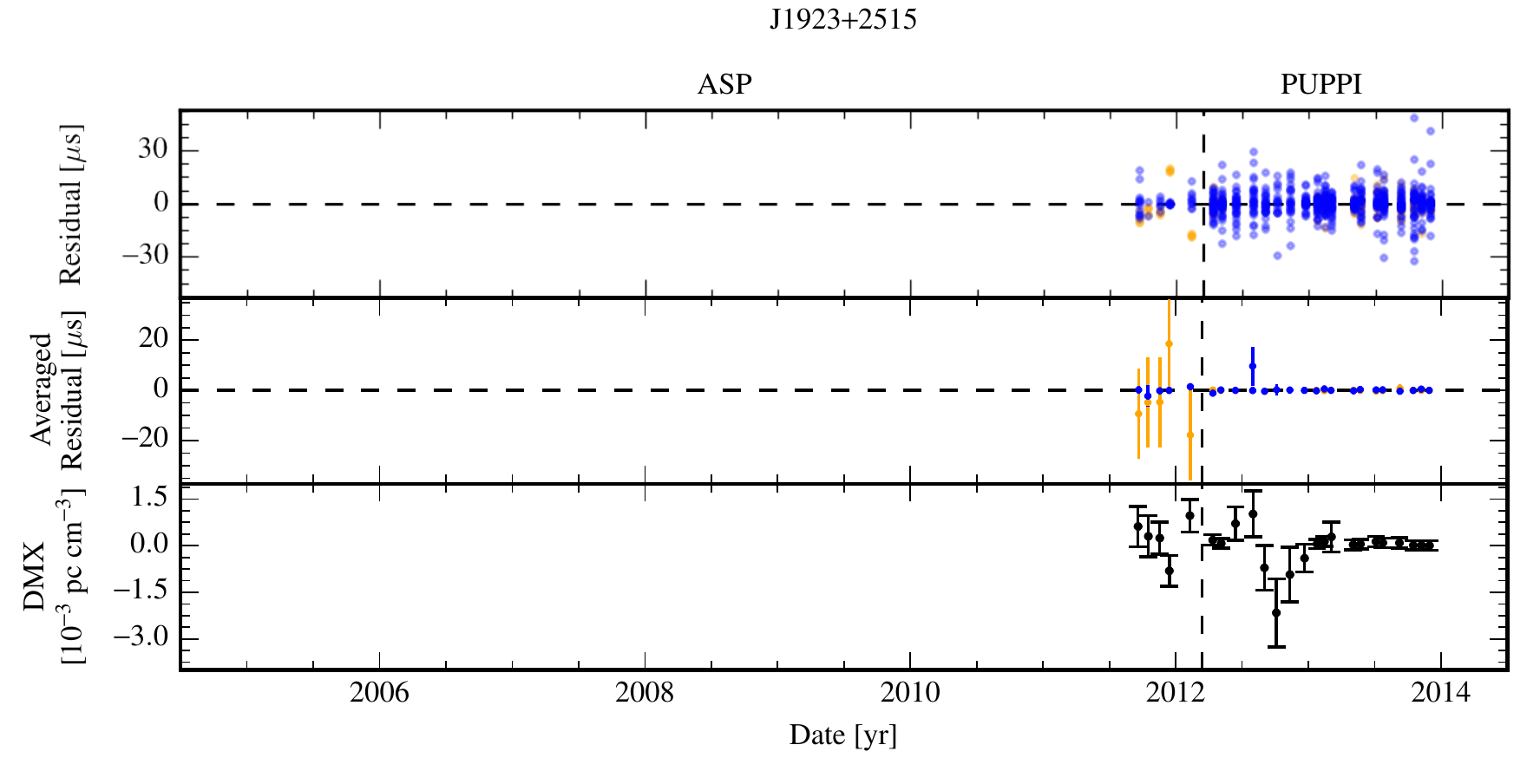}
\caption{Timing summary for PSR J1923+2515. Colors are: Blue: 1.4 GHz, Purple: 2.3 GHz, Green: 820 MHz, Orange: 430 MHz, Red: 327 MHz. In the top panel, individual points are semi-transparent; darker regions arise from the overlap of many points.}
\label{fig:summary-J1923+2515}
\end{figure*}

\begin{figure*}[p]
\centering
\includegraphics[scale=1.0]{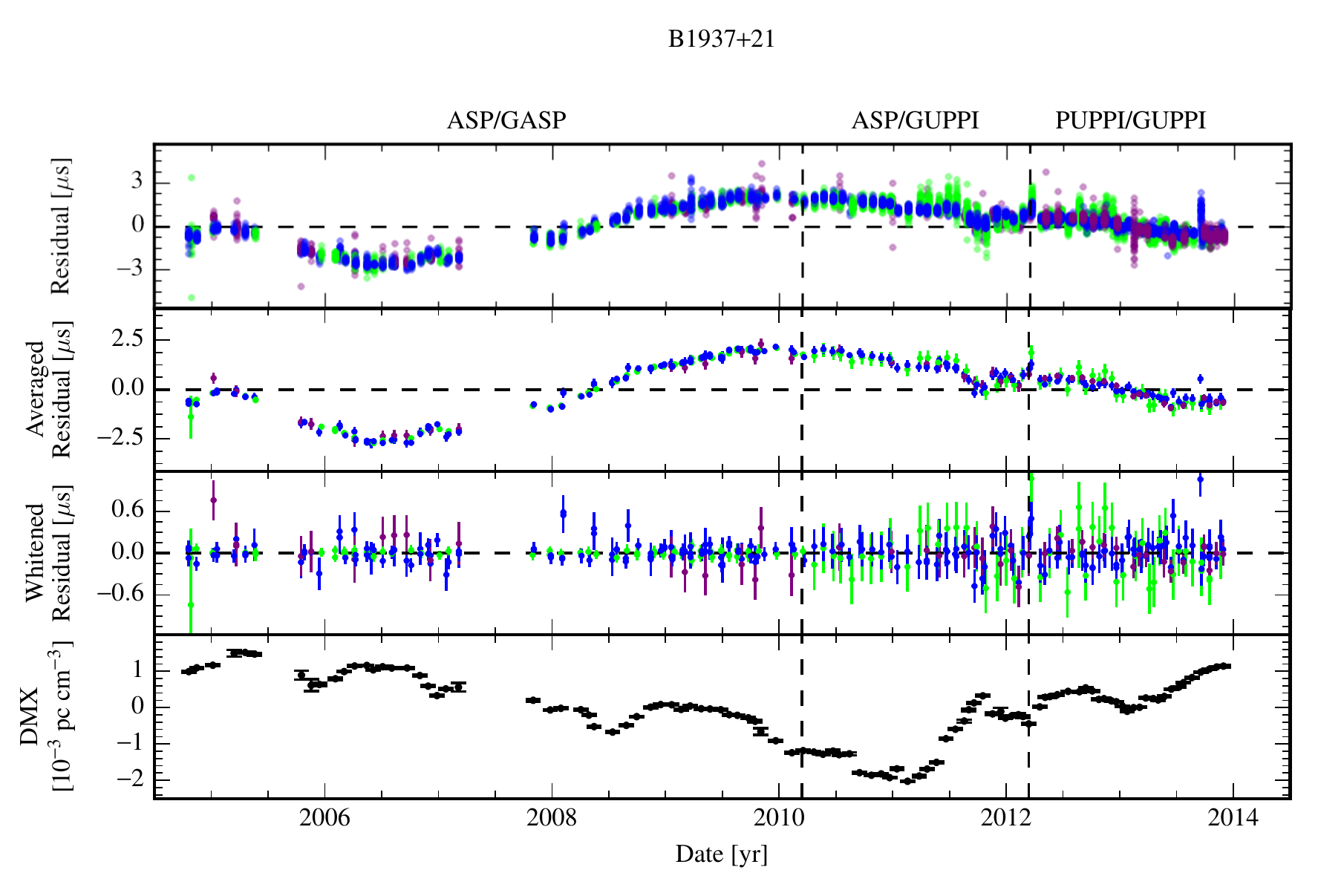}
\caption{Timing summary for PSR B1937+21. Colors are: Blue: 1.4 GHz, Purple: 2.3 GHz, Green: 820 MHz, Orange: 430 MHz, Red: 327 MHz. In the top panel, individual points are semi-transparent; darker regions arise from the overlap of many points.}
\label{fig:summary-B1937+21}
\end{figure*}

\begin{figure*}[p]
\centering
\includegraphics[scale=1.0]{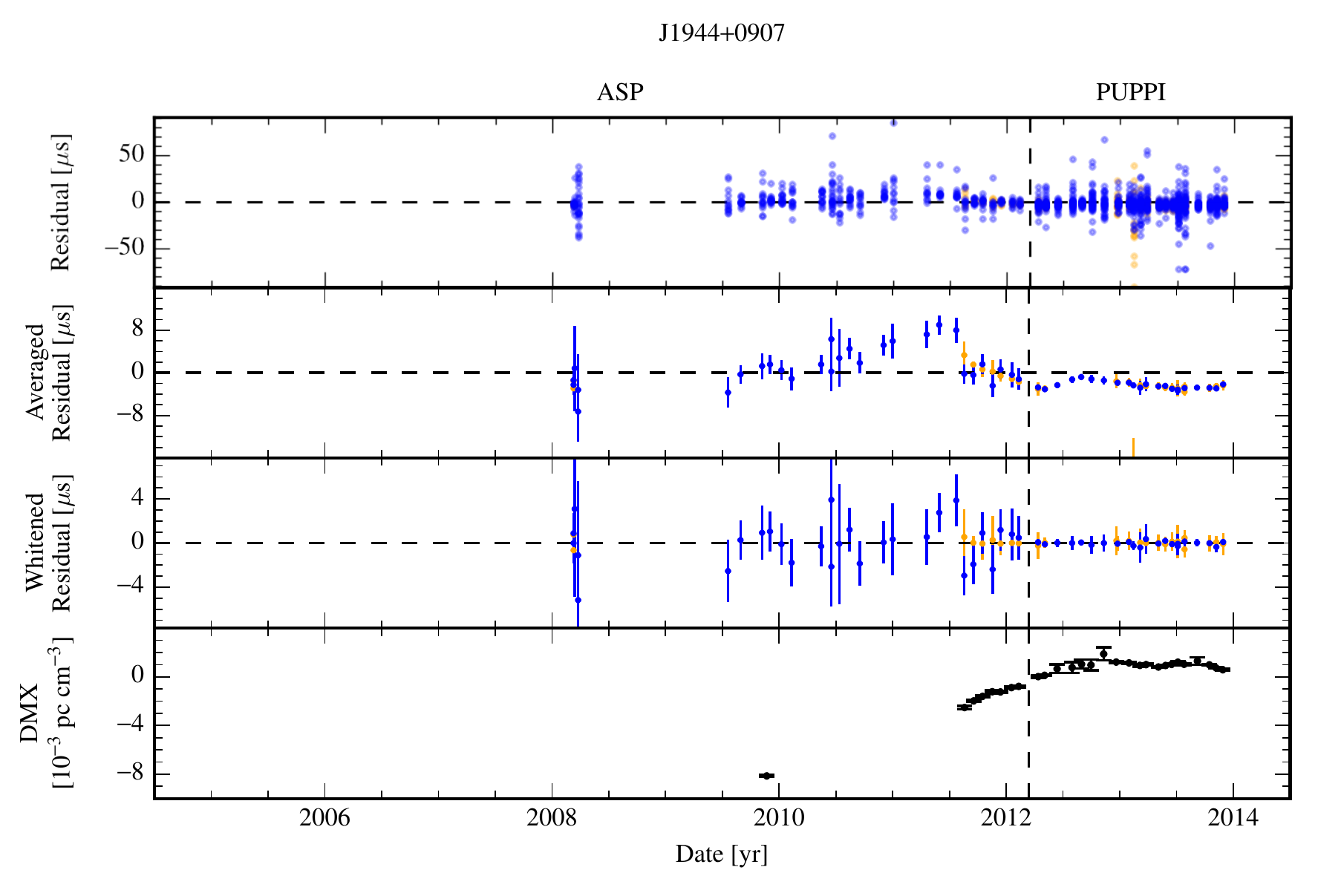}
\caption{Timing summary for PSR J1944+0907. Colors are: Blue: 1.4 GHz, Purple: 2.3 GHz, Green: 820 MHz, Orange: 430 MHz, Red: 327 MHz. In the top panel, individual points are semi-transparent; darker regions arise from the overlap of many points.}
\label{fig:summary-J1944+0907}
\end{figure*}

\begin{figure*}[p]
\centering
\includegraphics[scale=1.0]{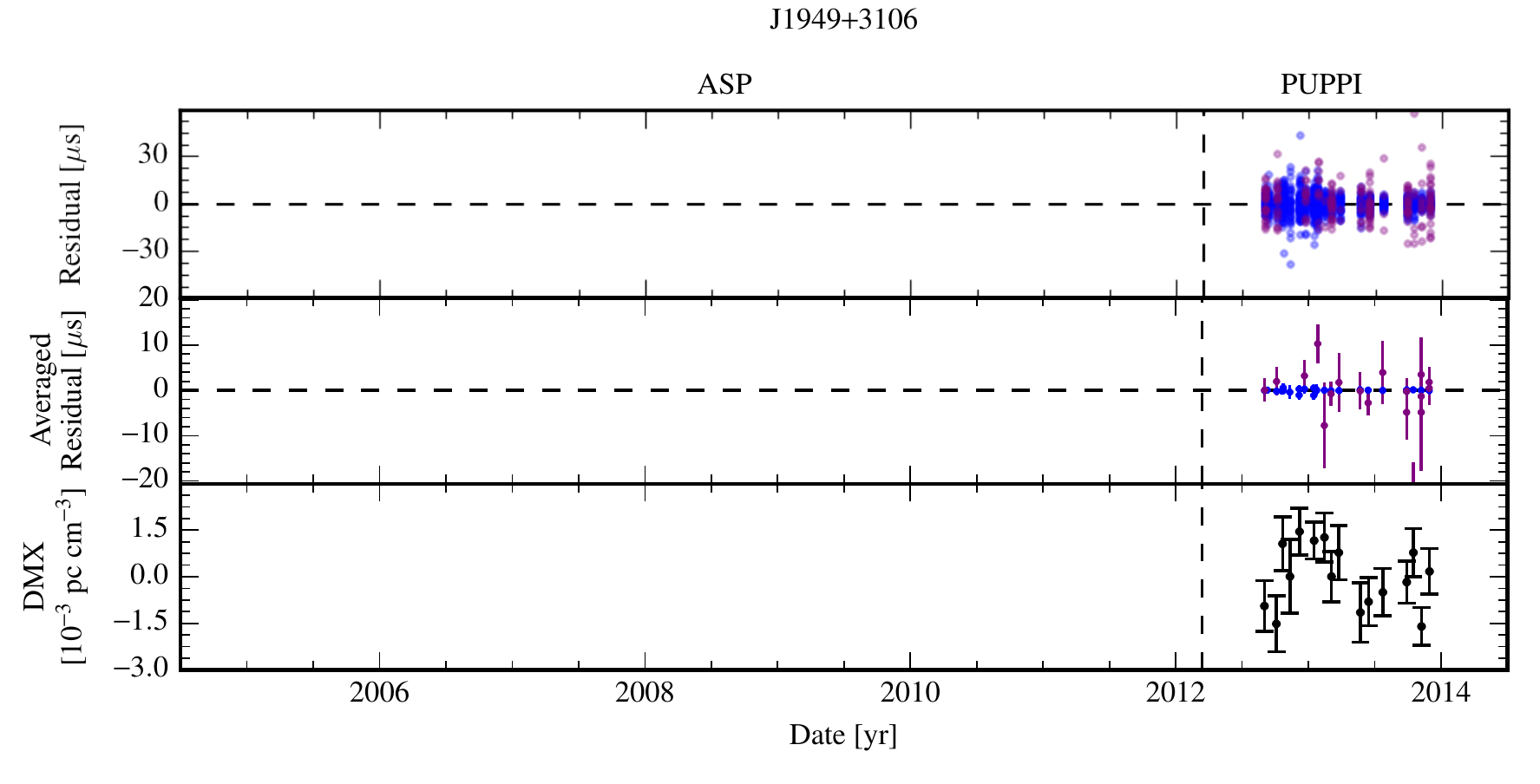}
\caption{Timing summary for PSR J1949+3106. Colors are: Blue: 1.4 GHz, Purple: 2.3 GHz, Green: 820 MHz, Orange: 430 MHz, Red: 327 MHz. In the top panel, individual points are semi-transparent; darker regions arise from the overlap of many points.}
\label{fig:summary-J1949+3106}
\end{figure*}

\begin{figure*}[p]
\centering
\includegraphics[scale=1.0]{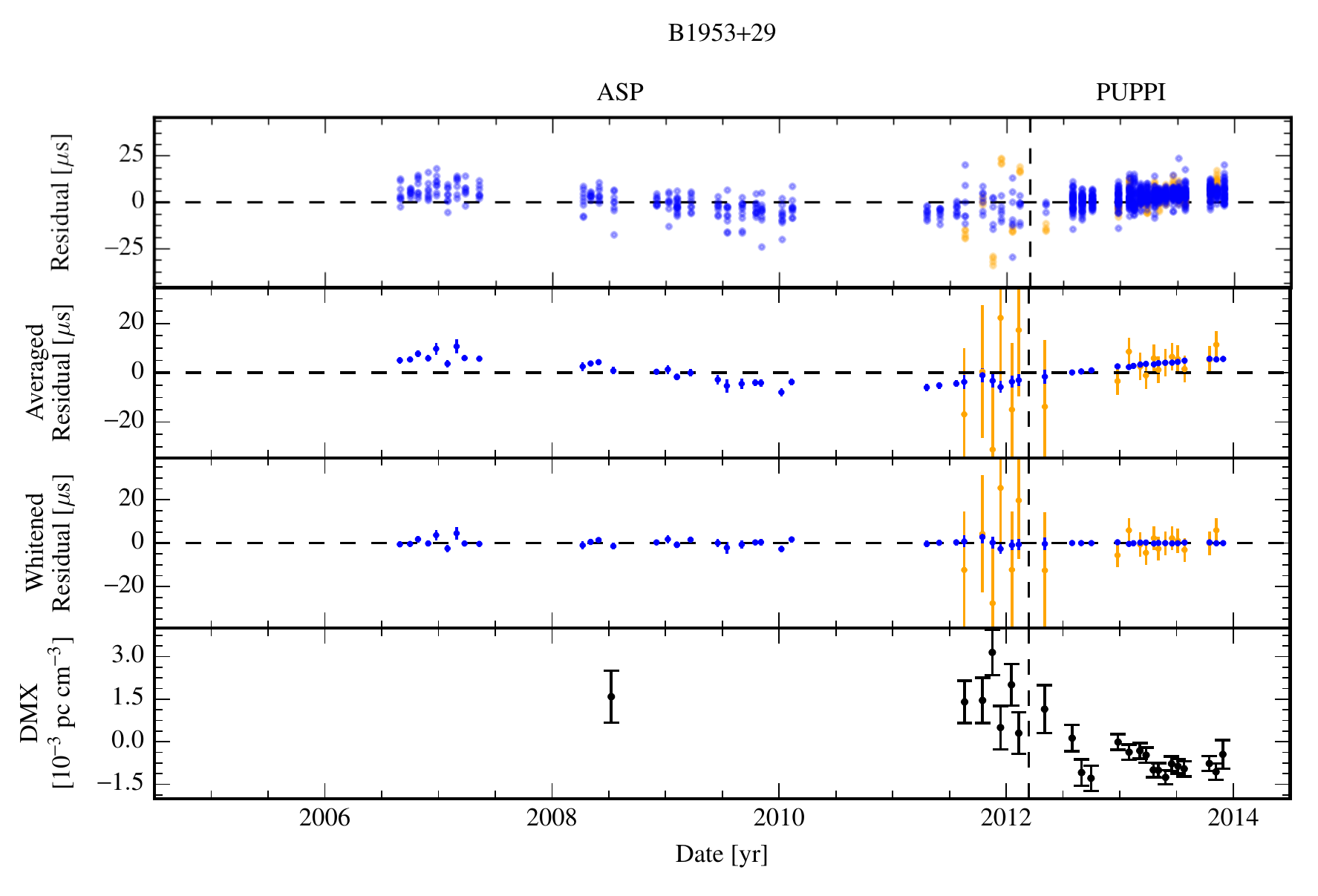}
\caption{Timing summary for PSR B1953+29. Colors are: Blue: 1.4 GHz, Purple: 2.3 GHz, Green: 820 MHz, Orange: 430 MHz, Red: 327 MHz. In the top panel, individual points are semi-transparent; darker regions arise from the overlap of many points.}
\label{fig:summary-B1953+29}
\end{figure*}

\begin{figure*}[p]
\centering
\includegraphics[scale=1.0]{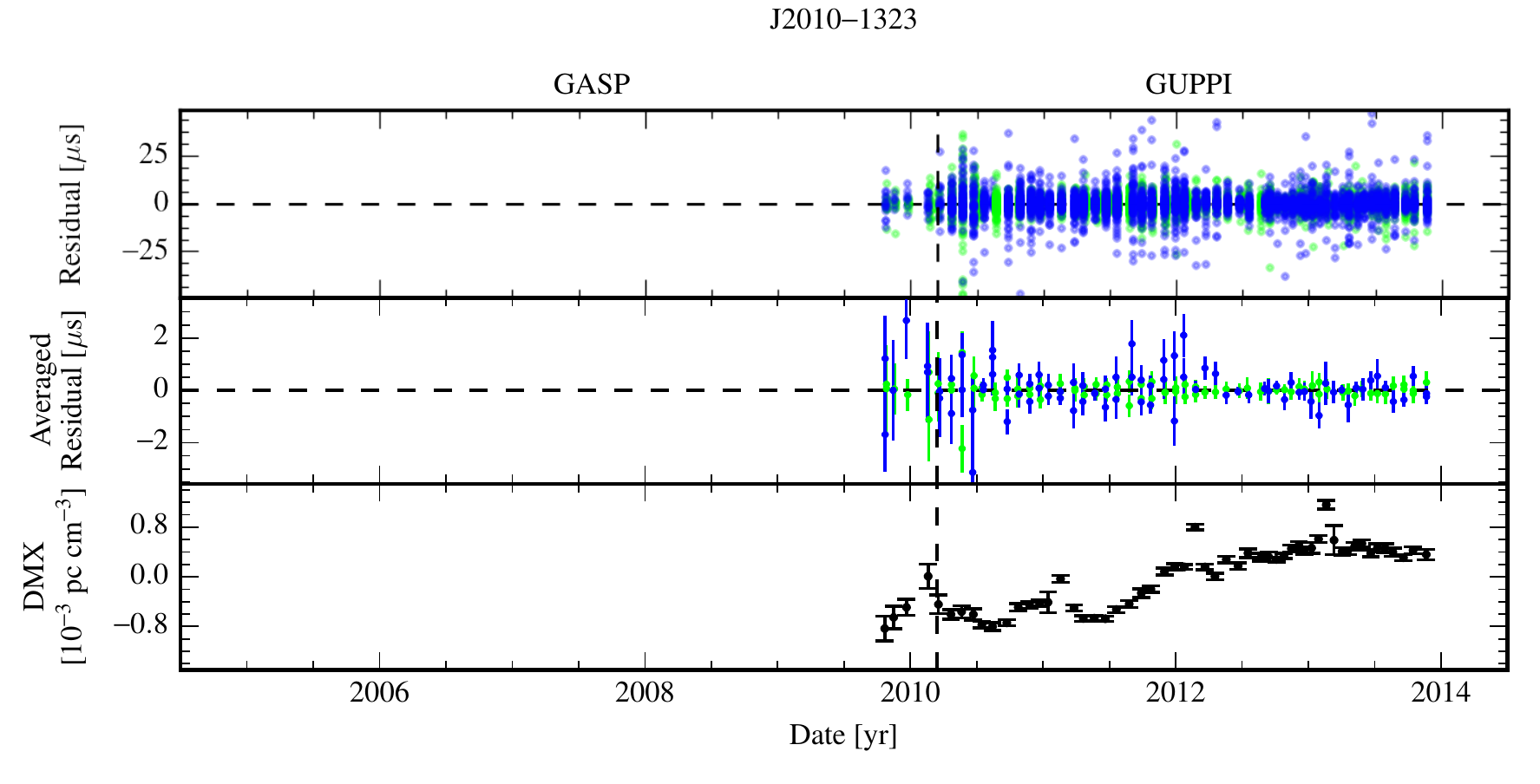}
\caption{Timing summary for PSR J2010-1323. Colors are: Blue: 1.4 GHz, Purple: 2.3 GHz, Green: 820 MHz, Orange: 430 MHz, Red: 327 MHz. In the top panel, individual points are semi-transparent; darker regions arise from the overlap of many points.}
\label{fig:summary-J2010-1323}
\end{figure*}

\begin{figure*}[p]
\centering
\includegraphics[scale=1.0]{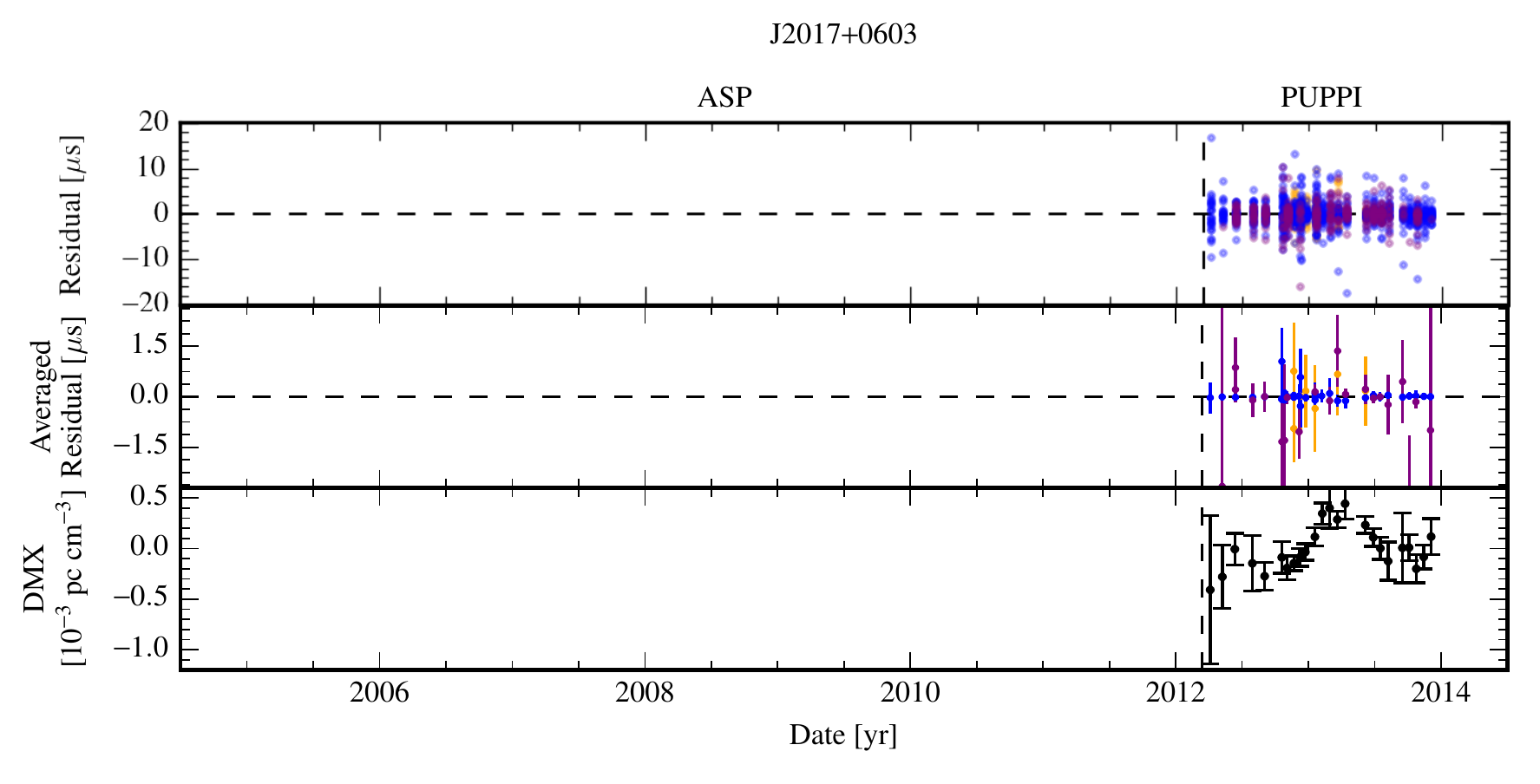}
\caption{Timing summary for PSR J2017+0603. Colors are: Blue: 1.4 GHz, Purple: 2.3 GHz, Green: 820 MHz, Orange: 430 MHz, Red: 327 MHz. In the top panel, individual points are semi-transparent; darker regions arise from the overlap of many points.}
\label{fig:summary-J2017+0603}
\end{figure*}

\begin{figure*}[p]
\centering
\includegraphics[scale=1.0]{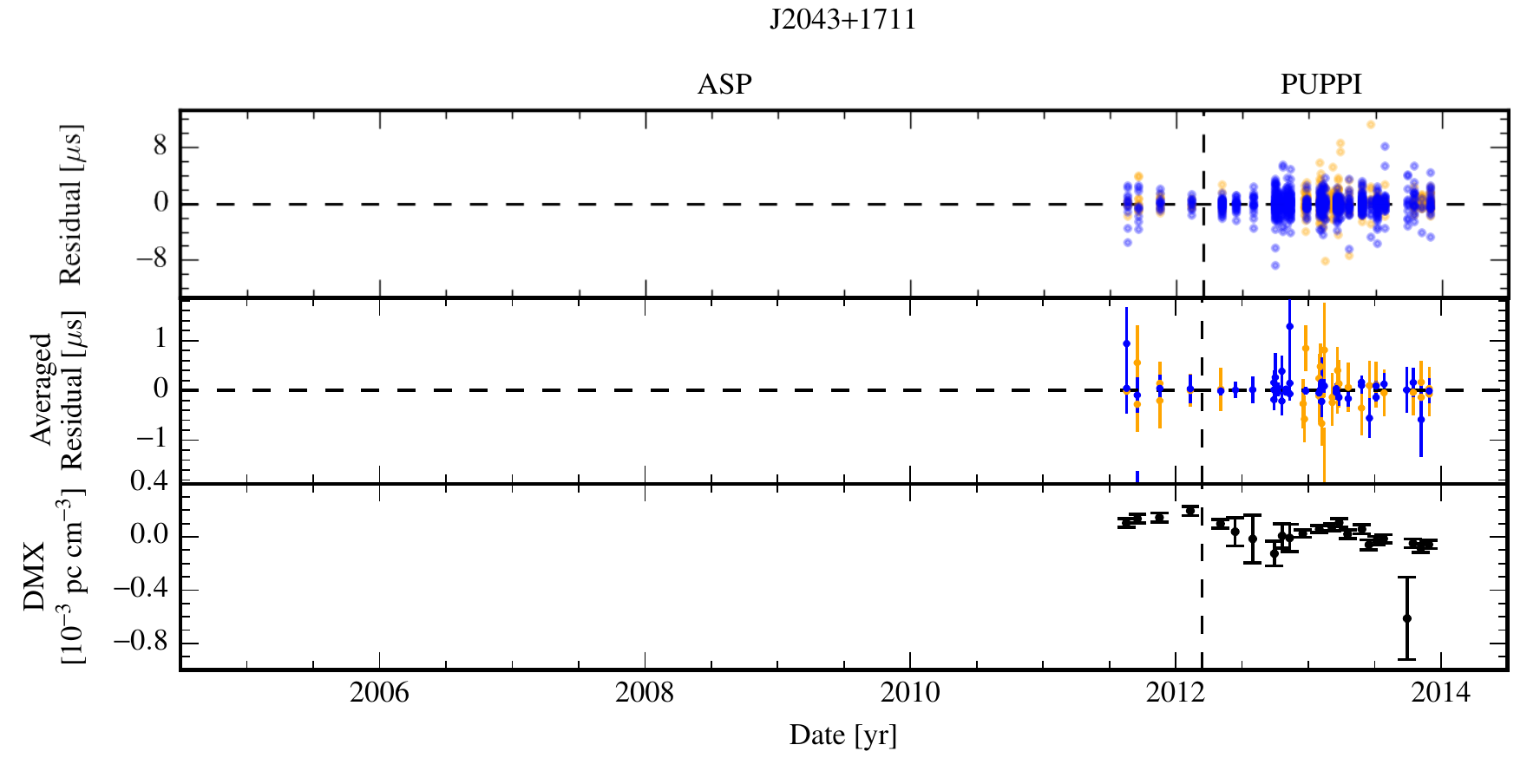}
\caption{Timing summary for PSR J2043+1711. Colors are: Blue: 1.4 GHz, Purple: 2.3 GHz, Green: 820 MHz, Orange: 430 MHz, Red: 327 MHz. In the top panel, individual points are semi-transparent; darker regions arise from the overlap of many points.}
\label{fig:summary-J2043+1711}
\end{figure*}

\begin{figure*}[p]
\centering
\includegraphics[scale=1.0]{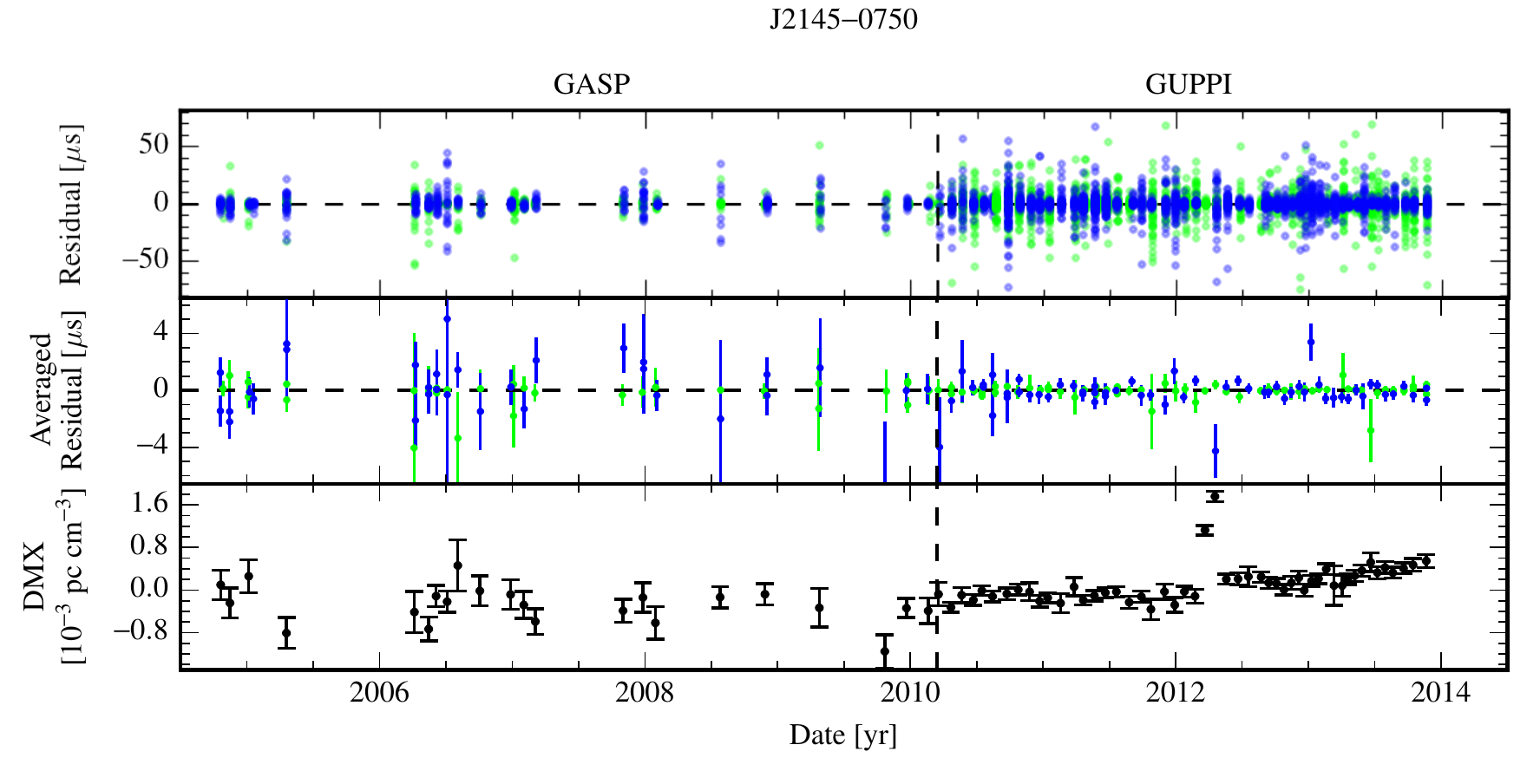}
\caption{Timing summary for PSR J2145-0750. Colors are: Blue: 1.4 GHz, Purple: 2.3 GHz, Green: 820 MHz, Orange: 430 MHz, Red: 327 MHz. In the top panel, individual points are semi-transparent; darker regions arise from the overlap of many points.}
\label{fig:summary-J2145-0750}
\end{figure*}

\begin{figure*}[p]
\centering
\includegraphics[scale=1.0]{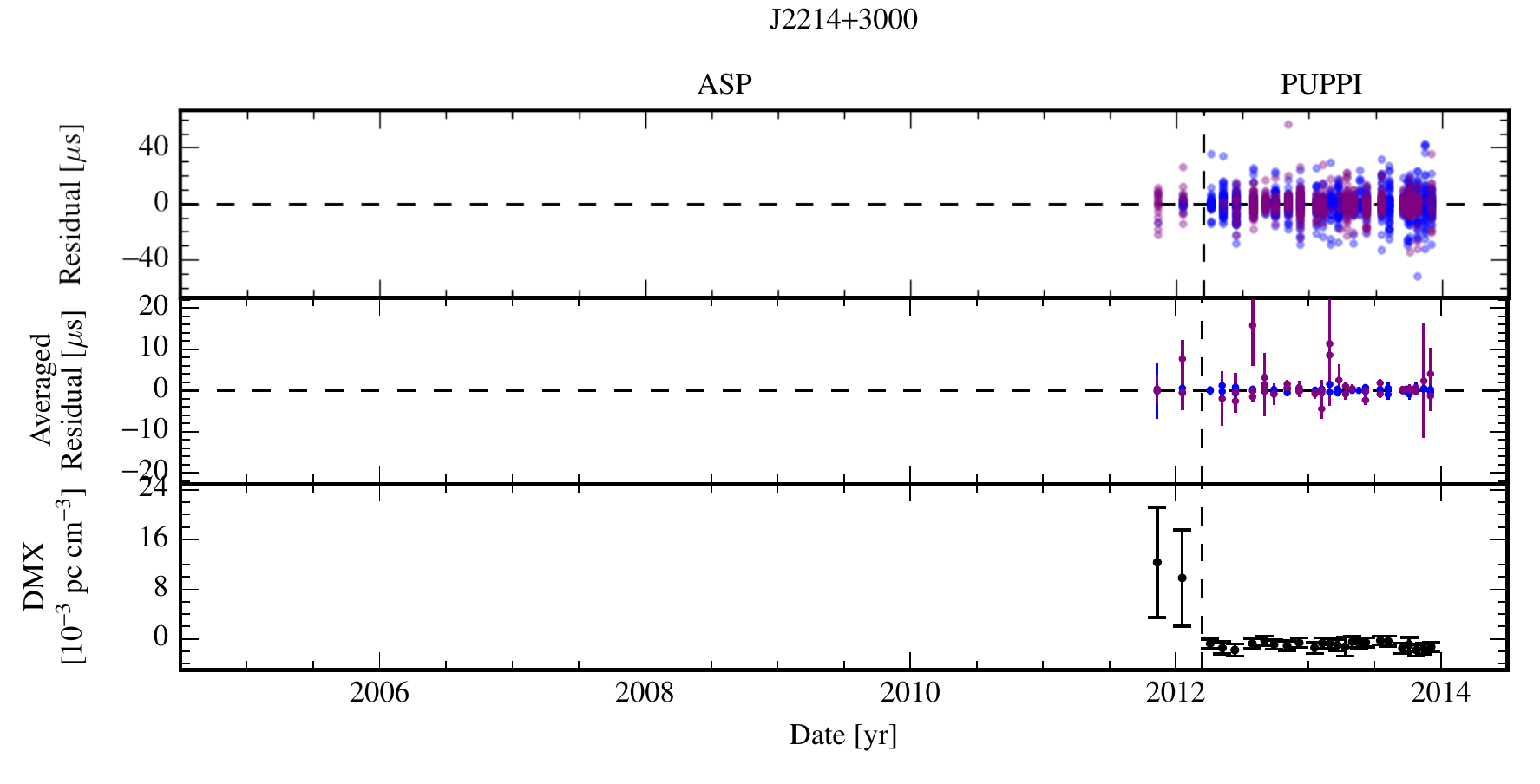}
\caption{Timing summary for PSR J2214+3000. Colors are: Blue: 1.4 GHz, Purple: 2.3 GHz, Green: 820 MHz, Orange: 430 MHz, Red: 327 MHz. In the top panel, individual points are semi-transparent; darker regions arise from the overlap of many points.}
\label{fig:summary-J2214+3000}
\end{figure*}

\begin{figure*}[p]
\centering
\includegraphics[scale=1.0]{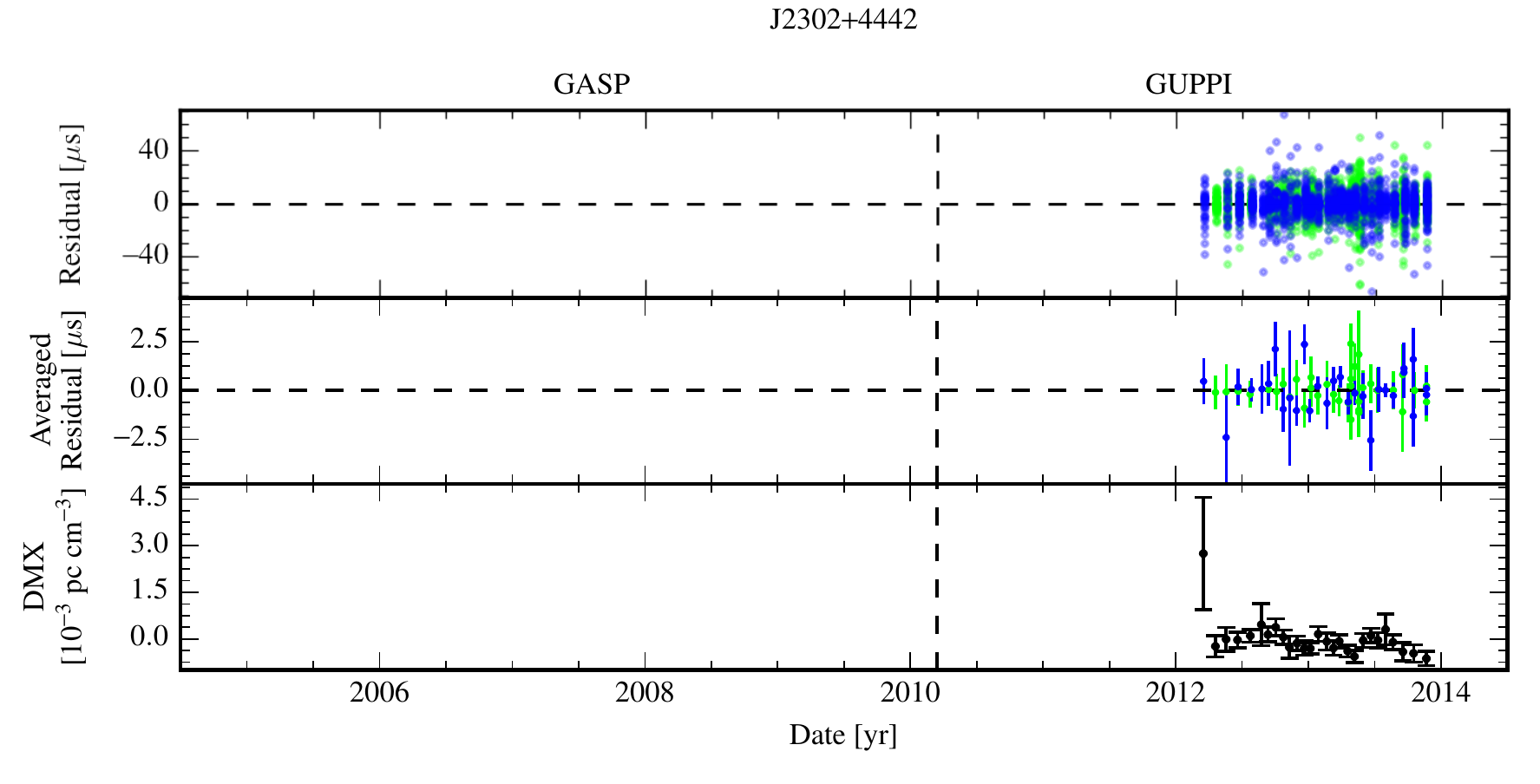}
\caption{Timing summary for PSR J2302+4442. Colors are: Blue: 1.4 GHz, Purple: 2.3 GHz, Green: 820 MHz, Orange: 430 MHz, Red: 327 MHz. In the top panel, individual points are semi-transparent; darker regions arise from the overlap of many points.}
\label{fig:summary-J2302+4442}
\end{figure*}

\begin{figure*}[p]
\centering
\includegraphics[scale=1.0]{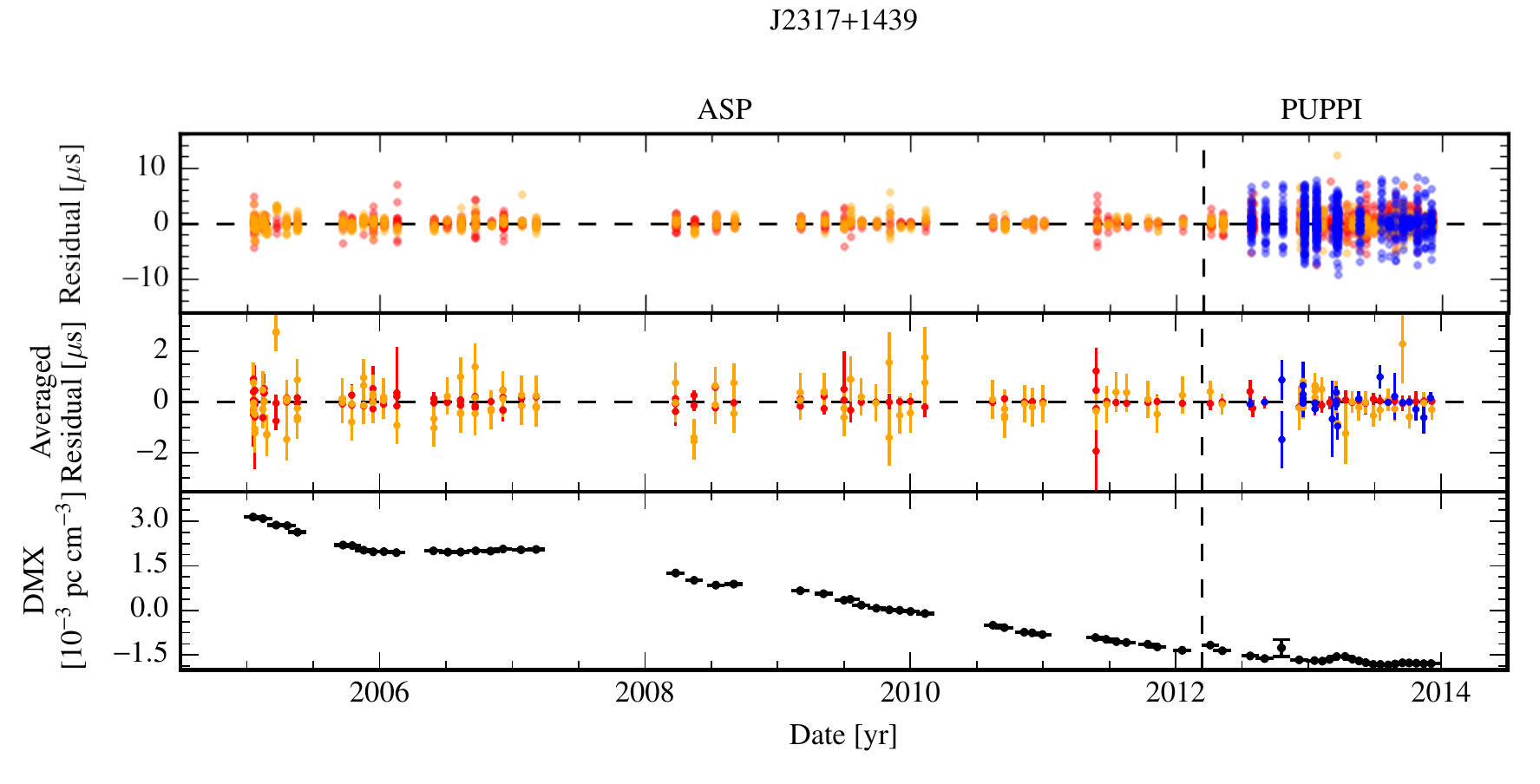}
\caption{Timing summary for PSR J2317+1439. Colors are: Blue: 1.4 GHz, Purple: 2.3 GHz, Green: 820 MHz, Orange: 430 MHz, Red: 327 MHz. In the top panel, individual points are semi-transparent; darker regions arise from the overlap of many points.}
\label{fig:summary-J2317+1439}
\end{figure*}

\end{document}